\documentclass[12pt]{article}
\usepackage{amsfonts,amssymb,graphicx,fullpage}
\usepackage[tight]{subfigure}
\usepackage{rotating}
\usepackage{sidecap}
\pdfoutput=1
\usepackage{ifpdf}
\ifpdf
  \usepackage{color}
  \definecolor{darkblue}{rgb}{0.3,0.3,0.6}
    \definecolor{darkgreen}{rgb}{0,0.6,0}
  \usepackage[pdftitle={Model Building on Z2xZ6'},pdfauthor={Gabriele Honecker, Martin Ripka and Wieland Staessens},pdfsubject={String theory},pdfkeywords={string theory},colorlinks=true,linkcolor=black,urlcolor=darkblue,filecolor=darkblue,citecolor=darkblue,menucolor=darkblue,breaklinks=true,bookmarks=true,anchorcolor=black,unicode=true]{hyperref}
\else
  
\fi
\usepackage[numbers,compress]{natbib}
\usepackage{a4wide}
\usepackage{longtable}
\usepackage{graphicx}
\usepackage{subfigure}
\usepackage[centertags]{amsmath}
\usepackage{multirow}

\newcommand{\bCentering}{\centering}
\newcommand{\bCaption}{\caption}
\newcommand{\sgn}{{\rm sgn}}

\def\muc{\multicolumn}

\def\Z{\mathbb{Z}}

\def\ov{\overline}
\def\N{\mathbf{N}}
\def\Sym{\mathbf{Sym}}
\def\Anti{\mathbf{Anti}}
\def\Adj{\mathbf{Adj}}

\def\ov{\overline}
\def\1{{\bf 1}}
\def\2{{\bf 2}}
\def\3{{\bf 3}}
\def\4{{\bf 4}}
\def\6{{\bf 6}}
\def\OR{\Omega\mathcal{R}}

\def\pp{\uparrow\uparrow}

\def\targ#1#2{\genfrac{[}{]}{0pt}{}{#1}{#2}}

\newcommand{\bCaptionfonts}{\small}
\makeatletter
\long\def\@makecaption#1#2{%
  \vskip\abovecaptionskip
  \sbox\@tempboxa{{\bCaptionfonts #1: #2}}%
  \ifdim \wd\@tempboxa >\hsize
    {\bCaptionfonts #1: #2\par}
  \else
    \hbox to\hsize{\hfil\box\@tempboxa\hfil}%
  \fi
  \vskip\belowcaptionskip}
\makeatother

\makeatletter
\let\ORIGINALlatex@openbib@code=\@openbib@code
\renewcommand{\@openbib@code}{\ORIGINALlatex@openbib@code\setlength{\itemsep}{1ex plus.5ex minus.5ex}\setlength{\parsep}{0pt}}
\makeatother

\def\mathtabfix#1#2#3{\begin{table}[th]\bCentering\resizebox{\linewidth}{!}{$#1$}\bCaption{#3}\label{tab:#2}\end{table}}

\def\mathsidetabfix#1#2#3{\begin{sidewaystable}[H]\bCentering\resizebox{\linewidth}{!}{$#1$}\bCaption{#3}\label{tab:#2}\end{sidewaystable}}

\renewcommand{\arraystretch}{1.3}

\allowdisplaybreaks[4]

\begin{document}
\begin{center}
\begin{flushright}
{\small MZ-TH/12-42\\ 
\today}
\end{flushright}

\vspace{25mm}
{\Large\bf The Importance of Being Rigid: }\\
\vspace{0.1in}{\large \bf D6-Brane Model Building on $T^6/\Z_2 \times \Z_6'$ with Discrete Torsion} 

\vspace{12mm}
{\large Gabriele Honecker${}^{\heartsuit}$, Martin Ripka and Wieland Staessens${}^{\spadesuit}$
}

\vspace{8mm}
{
\it PRISMA Cluster of Excellence \&
Institut f\"ur Physik  (WA THEP), Johannes-Gutenberg-Universit\"at, D-55099 Mainz, Germany
\;$^{\heartsuit}${\tt Gabriele.Honecker@uni-mainz.de},~$^{\spadesuit}${\tt wieland.staessens@uni-mainz.de}}

\vspace{15mm}{\bf Abstract}\\[2ex]\parbox{140mm}{
Model building with rigid D6-branes on the Type IIA orientifold on $T^6/\Z_2 \times \Z_6'$ with discrete torsion
is considered. The systematic search for models of particle physics is significantly reduced by proving new
symmetries among different lattice orientations. Suitable rigid D6-branes without matter in adjoint and
symmetric representations are classified, and $SO(2N)$ and $USp(2N)$ gauge factors on orientifold invariant
D6-branes are distinguished in terms of their discrete Wilson line and displacement parameters.
Constraints on the non-existence of exotic matter prohibit global completions of local MSSM and left-right 
symmetric models, while globally defined supersymmetric Pati-Salam models are found.
For the latter, only one particle generation possesses perturbative Yukawa couplings.
Masses for the mild amount of exotic matter and the role of Abelian symmetries are briefly discussed.
\\
Last but not least, it is shown that for all three two-torus volumes of about the same order of magnitude, 
gauge coupling unification at one-loop can be achieved, while for highly unisotropic choices a low string
scale in the TeV range is compatible with the observed strengths of gauge and gravitational couplings. 
}
\end{center}

\thispagestyle{empty}
\clearpage 

\tableofcontents
\newpage
\setlength{\parskip}{1em plus1ex minus.5ex}
\section{Introduction}\label{S:intro}

Model building in Type IIA string theory orientifolds with D6-branes has proven to be able
to incorporate Standard Model features over the past decade, see e.g. the 
reviews~\cite{Uranga:2003pz,Blumenhagen:2005mu,Dudas:2006bj,Blumenhagen:2006ci,Marchesano:2007de,Lust:2007kw,Cvetic:2011vz,Ibanez:2012zz}.
In particular,  $T^6/\Z_{2N}$ toroidal orbifolds with $2N \in \{6,6'\}$ have provided for stable supersymmetric particle physics
models~\cite{Honecker:2004kb,Honecker:2004np,Bailin:2006vb,Bailin:2006zf,Gmeiner:2007we,Bailin:2007zz,Gmeiner:2007zz,Bailin:2008xx,Gmeiner:2008xq}. 
While the simplest models on the six-torus~\cite{Ibanez:2001nd,MarchesanoBuznego:2003hp}, the $T^6/\Z_2 \times \Z_2$  orbifold without discrete 
torsion~\cite{Cvetic:2001tj,Cvetic:2001nr,Blumenhagen:2004xx,Gmeiner:2005vz} and the $T^6/\Z_2 \times \Z_4$ orbifold~\cite{Honecker:2003vq,Honecker:2003vw,Honecker:2004np,Cvetic:2006by}
automatically incorporate three matter multiplets in the adjoint representation of each $SU(N)$ gauge group
associated to continuous breakings to subgroups by displacements along each two-torus, 
in $T^6/\Z_{2N}$  models two of these multiplets are projected out by the $\Z_2$ symmetry, and additional matter in the adjoint representation
at intersections of orbifold image D6-branes can be avoided for special choices of the discrete Wilson line and displacement parameters along the 
four-torus, where the $\Z_2$ symmetry acts.
The attraction of $T^6/\Z_2 \times \Z_{2M}$ orbifolds with discrete torsion for D6-brane model building relies on the fact that the $\Z_2 \times \Z_2$ 
symmetries project out all three chiral multiplets in  the adjoint representation leading to rigid D6-branes stuck at $\Z_2 \times \Z_2$ singularities, 
which were first discussed for the most simple  case $2M=2$ in~\cite{Blumenhagen:2005tn}, however, to date still
without any globally defined supersymmetric D6-brane configuration 
with particle physics properties like the Standard Model or a GUT gauge group and three generations without chiral exotics. Its T-dual constructions 
to the special choice of untilted tori  in the context of Type IIB orientifolds with vanishing $B$-field have mainly been applied to studies of brane 
recombination and instantons contributing to the effective action, see e.g.~\cite{Angelantonj:2000hi,Angelantonj:2007ts,Angelantonj:2009yj,Angelantonj:2011hs}. 
The framework for model building on $T^6/\Z_2 \times \Z_{2M}$ orbifolds with discrete torsion in particular for $2M \in \{6,6'\}$ has first been 
developed  in~\cite{Forste:2010gw}, demonstrating the potential for particle physics on rigid D6-branes.  

Despite the draw-back of singularities, toroidal orbifolds are advantageous for four dimensional model building due to the power of Conformal 
Field Theory (CFT), which has been used to derive the perturbatively exact holomorphic gauge kinetic functions in Type IIA/$\OR$ orientifolds on the 
six-torus~\cite{Lust:2003ky,Akerblom:2007np,Akerblom:2007uc} and its orbifolds~\cite{Blumenhagen:2007ip,Gmeiner:2009fb,Honecker:2011sm,Honecker:2011hm}, 
the K\"ahler metrics at leading order~\cite{Lust:2004cx,Akerblom:2007uc,Blumenhagen:2007ip,Honecker:2011sm,Honecker:2011hm,Berg:2011ij} and Yukawa 
couplings on the six-torus~\cite{Cremades:2003qj,Cvetic:2003ch,Abel:2003vv,Abel:2003yx,Cremades:2004wa,Lust:2004cx,Abel:2004ue}.
Equally powerful tools of Rational CFT apply to Gepner models, for which large numbers of string vacua with Standard Model or GUT spectra have been
constructed e.g. in~\cite{Dijkstra:2004ym,Dijkstra:2004cc,Anastasopoulos:2010hu,Kiritsis:2009sf}. Last but not least, CFT methods for 
heterotic orbifolds have allowed for scans of large classes of four-dimensional vacua~\cite{Lebedev:2006kn,Lebedev:2008un}, and 
computations of field theoretical quantities have been automatised~\cite{Nilles:2011aj}.
In contrast to the Type IIA string theory constructions of this article, for which the uncharged closed string sector is clearly distinguished from the charged
open string sector, blow-ups of heterotic vacua might or might not extinguish part of the Standard Model gauge group, see e.g.~\cite{Blaszczyk:2009in,Blaszczyk:2010db,Blaszczyk:2011ig,Buchmuller:2012mu}.
On the other hand, large classes of compactifications of the $E_8 \times E_8$ heterotic string on smooth Calabi-Yau manifolds are 
known~\cite{Anderson:2011ns,Anderson:2012yf}. In this framework, it is rather straightforward to find models without bundle moduli, however, at the
cost of working in the supergravity approximation, which hampers the computation of exact field theoretic results.

Applications of string compactifications to scenarios Beyond the Standard Model provide interesting proposals for low-mass string 
resonances at the LHC from intersecting D6-branes~\cite{Antoniadis:1998ig,Anchordoqui:2007da,Lust:2008qc,Anchordoqui:2012wt}, for which 
underlying globally consistent D6-brane constructions are required. Another ubiquitous prediction from string theory is the existence of closed string axions
from the RR sector or the recently discussed option of open string axions~\cite{Berenstein:2012eg} as well as additional Abelian gauge symmetries, which if 
light can act as potential $Z'$ bosons, see e.g.~\cite{Ghilencea:2002da,Berenstein:2008xg,Anchordoqui:2011eg,Cvetic:2011iq}, or as dark photons,
which can even involve kinetic mixing of the usually considered open string gauge bosons, e.g. in~\cite{Grimm:2011dx,Kerstan:2011dy}, with closed string 
vectors~\cite{Camara:2011jg} in analogy to the to date more intensively studied  kinetic mixing in the context of 
heterotic $E_8 \times E_8$ models~\cite{Blumenhagen:2005ga,Blumenhagen:2005pm,Blumenhagen:2005zg,Honecker:2007uw,GrootNibbelink:2007ew,Abel:2008ai,Goodsell:2011wn}.  
While computations for low-mass string resonances have concentrated on the intersecting D6-brane/Type IIA string theory scenario, large volume compactifications
are usually studied in the context of D3/D7-branes in Type IIB models, see e.g.~\cite{Conlon:2005ki,Cicoli:2008va,Cicoli:2012vw}. This article 
provides a first step at closing the gap by showing that highly unisotropic compactifications with some very large two-torus volume and correspondingly 
low string scale are indeed feasible within the context of intersecting D6-branes.

{\bf Outline:}
In section~\ref{S:Geometry+Symmetries+T6Z2Z6p} the global consistency conditions and the derivation of the massless spectra for
intersecting D6-branes on $T^6/(\Z_2 \times \Z_6' \times \OR)$ from~\cite{Forste:2010gw} are briefly reviewed and it is shown for the first time 
that - at the level of the current 
knowledge on consistency conditions, spectra and effective field theory - the four a priori independent factorisable background 
lattices are pairwise identical.
In section~\ref{S:Modelbuilding}, investigations on general model building features including $USp(2N)$ and $SO(2N)$ gauge factors, 
absence of exotic matter and three particle generations are presented.
Section~\ref{S:model_building} contains a systematic survey of attempts to embed the Standard Model and left-right symmetric and Pati-Salam
models in globally consistent D6-brane compactifications. For the first two, local models and obstructions to satisfy the twisted 
RR tadpole cancellation conditions are given. For a globally consistent Pati-Salam model, the full massless spectrum and 
Yukawa and other three-point couplings are derived.
Section~\ref{S:U1s} contains a brief discussion of Abelian gauge symmetries, gauge coupling unification and the size of the string scale.
The conclusions are given in section~\ref{S:Conclusions}.
Finally, technical details are collected in appendices~\ref{A:tables_bulk-cycles} to~\ref{A:6stack_PS}. 

\section{Geometry of IIA on  $T^6/(\Z_2 \times \Z_6' \times \OR)$ and Symmetries}\label{S:Geometry+Symmetries+T6Z2Z6p}

The geometry and consistency conditions of orientifolds of Type IIA string theory on the orbifold $T^6/\Z_2 \times \Z_6'$ 
with discrete torsion are briefly reviewed in section~\ref{Ss:Geometry}, as first presented in~\cite{Forste:2010gw}, and in section~\ref{Ss:identifications}
it is proven for the first time that the four at first seemingly independent background lattices are pairwise related - at least at the level
of our understanding in terms of the combination of the geometry of three-cycles, the massless spectra and CFT results on
one-loop vacuum amplitudes and the related field theoretic quantities.
After having proved the pairwise equivalence of lattices, the investigation of particle physics properties focuses  on the two remaining
inequivalent choices of background orientations from section~\ref{S:Modelbuilding} on.

\subsection{Geometry, Supersymmetry and RR tadpole cancellation}\label{Ss:Geometry}

\begin{figure}[ht]
\begin{center}
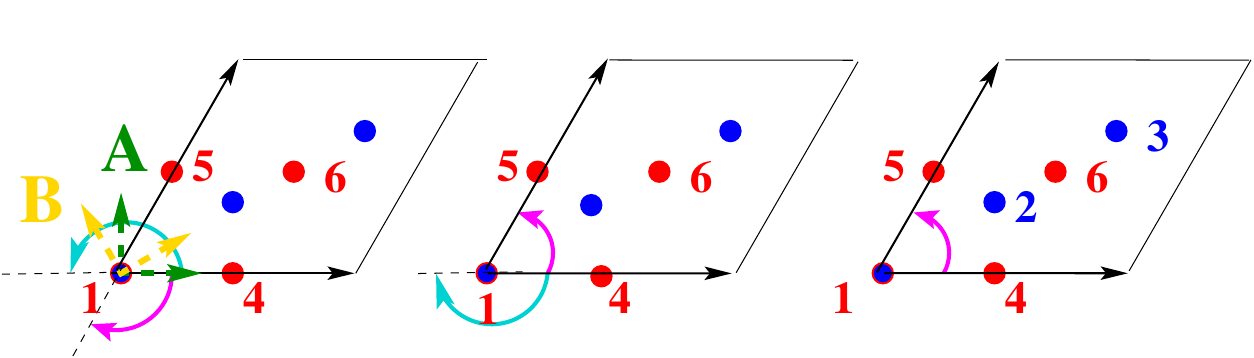
\end{center}
\caption{The $SU(3)^3$ compactification lattice of the $T^6/\Z_2 \times \Z_6'$ orbifold defined in equation~(\ref{Eq:Z2Z6p-def}). The $\Z_2$ fixed points 
$1\stackrel{\omega}{\circlearrowleft},4\stackrel{\omega}{\to}5\stackrel{\omega}{\to}6\stackrel{\omega}{\to}4$ on each two-torus 
are depicted in red, the $\Z_3$ fixed points $1\stackrel{\theta,\omega}{\circlearrowleft},2\stackrel{\theta,\omega}{\leftrightarrow} 3$ in blue. 
The anti-holomorphic involution eq.~(\ref{Eq:Z2Z6p-involution})
is consistent with two choices of orientations {\bf A} (for which $4 \stackrel{\cal R}{\circlearrowleft}, 5 \stackrel{\cal R}{\leftrightarrow} 6$) and {\bf B} 
(for which $4 \stackrel{\cal R}{\leftrightarrow} 5, 6 \stackrel{\cal R}{\circlearrowleft}$) of each $SU(3)$ root lattice.
}
\label{Fig:Z2Z6p-lattice}
\end{figure}

The $T^6/\Z_2 \times \Z_6'$ orbifold is generated by the following action on the complex two-torus coordinates $z^k$ of $T^2_{(k)}$
with $k \in \{1,2,3\}$,
\begin{equation}\label{Eq:Z2Z6p-def}
\theta^m \, \omega^n: \; z^k \to e^{2\pi \, i \, (m v_k + n w_k^{\, \prime})} \; z^k
\qquad
\text{with}
\qquad
\vec{v}= \frac{1}{2}(1,-1,0),
\qquad
\vec{w}^{\, \prime} = \frac{1}{6}(-2,1,1).
\end{equation}
This orbifold possesses three equivalent $\Z_6$ subsectors generated by the shift vectors
$\vec{w}^{\, \prime}$, $\vec{v} + \vec{w}^{\, \prime}$ and $\vec{v} + 2 \vec{w}^{\, \prime}$
with fixed points along the whole six-torus $T^6 \equiv \prod_{k=1}^3 T^2_{(k)}$, one $\Z_3$ subsector generated by $2 \vec{w}^{\, \prime}$ with fixed points
also along the full six-torus $T^6$, as well as three equivalent $\Z_2^{(l)}$ subsectors generated by $\vec{v}$, $3\vec{w}^{\, \prime}$ 
and $\vec{v} + 3 \vec{w}^{\, \prime}$ with fixed points along the four-tori $T^4_{(l)} \equiv \prod_{k \neq l} T^2_{(k)}$  with respectively $l=3,1,2$.
The situation is depicted in figure~\ref{Fig:Z2Z6p-lattice}.
The cristallographic action of the $\Z_6'$ generator $\omega$ on the factorised six-torus requires the shape of an $SU(3)^3$ root lattice,
and the anti-holomorphic involution ${\cal R}$ accompanying the worldsheet parity $\Omega$ in  orientifolds of Type IIA string theory,
\begin{equation}\label{Eq:Z2Z6p-involution}
{\cal R}: z^k \to \ov{z}^k 
,
\end{equation}
is consistent with two possible orientations of each two-torus lattice depicted  in figure~\ref{Fig:Z2Z6p-lattice} as {\bf A} with the one-cycle $\pi_{2k-1}$  
aligned with the $\OR$-invariant plane and {\bf B} with the one-cycle $\pi_{2k-1}+\pi_{2k}$ along the  
$\OR$-invariant direction. 

The sixteen $\Z_2^{(k)}$ invariant points on $T^4_{(k)}$ per sector $k \in \{1,2,3\}$ are 
grouped into a singlet (point (11) on $T^4_{(k)}$) and five triplets ($(p1)$ and $(1p)$ and three $(pq)$'s with $p,q\in \{4,5,6\}$) under 
the $\Z_6'$ action generated by $\omega$, while the $\Z_3$ fixed points form three doublets ($(11r)$ and $(1r1)$ and $(r11)$ 
with $r \in \{2,3\}$) plus five quadruples ($(rs1)$ and $(r1s)$ and $(1rs)$ and two $(rst)$'s with $r,s,t \in \{2,3\}$)
under the $\Z_2^{(k)}$ action.
The three $\Z_6$ fixed points per sector generated by $\omega$, $\theta\omega$ or $\theta\omega^2$ 
coincide with some of the $\Z_3$ fixed points (respectively $(r11)$ or $(1r1)$ or $(11r)$ with $r \in \{1,2,3\}$).

The $\Z_N$ fixed points along $T^6$ or $T^4_{(k)}$ either support two-forms and their dual four-forms, or for the $\Z_2^{(k)}$ fixed points 
instead three-forms dual to three-cycles can arise by tensoring  exceptional divisors $e^{(k)}_{xy}$ at fixed points  $x,y \in \{1,4,5,6\}$ along 
$T^2_{(i)} \times T^2_{(j)} \equiv T^4_{(k)}$  with one-cycles $n^k \, \pi_{2k-1} + m^k \, \pi_{2k}$ on the remaining two-torus $T^2_{(k)}$,
where $(n^k,m^k)$ are the coprime wrapping numbers per two-torus.
The existence of two- and their dual four-forms or instead three-forms at $\Z_2^{(k)}$ singularities depends on the action of another $\Z_2^{(l), \, l \neq k}$
symmetry as follows.

For orbifolds of the type $T^6/\Z_N \times \Z_M$, each generator $\Z_N$ can act with a phase factor $e^{2\pi n \, i/ \gcd(N,M)}$ ($n \in \Z$)
on the $\Z_M$ twisted sector and vice versa~\cite{Vafa:1986wx,Font:1988mk}. For $(N,M)=(2,6')$ this amounts to the choice of a sign factor,
\begin{equation}
\eta = \left\{\begin{array}{cc} 1 & \text{without}
\\ -1 & \text{with}\end{array}  \right. 
\text{ discrete torsion}
,
\end{equation}
for which the relevant Hodge numbers encoding the number of two- and three-cycles are given by
\begin{equation}
\left(\begin{array}{c} h_{11} \\ h_{21} \end{array}\right)
=
\left(\begin{array}{c} h_{11}^U + h_{11}^{\Z_6} + h_{11}^{\Z_3} + h_{11}^{\Z_2}  \\  h_{21}^U + h_{21}^{\Z_6} + h_{21}^{\Z_3} + h_{21}^{\Z_2}   
\end{array}\right)
=\left\{
\begin{array}{cc}
\left(\begin{array}{c} 3+ (3 \times 2) + 9+ (3 \times 6) =36\\  0+ 0+0+0 =0\end{array}\right) & \eta=1
\\
\left(\begin{array}{c} 3+ (3 \times 1) +9+0 =15\\ 0+ 0+0+ (3 \times 5) =15 \end{array}\right) & \eta=-1
\end{array}\right.
.
\end{equation}

The aim of this article is to focus on {\bf  orientifolds} of Type IIA string theory on the $T^6/\Z_2 \times \Z_6'$ orbifold with discrete torsion 
and to explore model building with rigid D6-branes. It was shown in~\cite{Forste:2010gw} that the orientifold O6-planes
are grouped into four orbits under the $\Z_6'$ action containing the $\OR$- and $\OR\Z_2^{(k)}$-invariant planes with $k \in \{1,2,3\}$,
and that a priori there exist four different factorisable $(T^2)^3$ lattice orientations, {\bf AAA}, {\bf AAB}, {\bf ABB} and {\bf BBB}. The 
coprime one-cycle wrapping 
numbers $(n^i,m^i)_{i=1,2,3}$ along $(\pi_{2i-1},\pi_{2i})$ in the notation of figure~\ref{Fig:Z2Z6p-lattice} of each $\OR\theta^p\omega^q$-invariant 
plane as well as the relative angle to the $\OR$-invariant plane are displayed in table~\ref{tab:Z2Z6p-ORonBulkO6plane-torus}.
\mathtabfix{
\begin{array}{|c|c||c|c||c|c|}\hline
 \multicolumn{6}{|c|}{\text{\bf Torus wrapping numbers $(n^i,m^i)$ of O6-planes on } T^6/(\Z_2 \times \Z_6' \times \OR)}
\\\hline\hline
\muc{2}{|c|}{\text{lattice}}& {\color{blue} {\bf AAA}} & {\color{blue} {\bf ABB}} & {\color{red} {\bf AAB}} & {\color{red} {\bf BBB}}
\\\hline\hline
\begin{array}{c}\text{O6-plane}\\\text{orbit} \end{array} & \frac{\text{Angle}}{\pi}  & \multicolumn{4}{|c|}{
(n^1,m^1;n^2,m^2;n^3,m^3) \text{ with }
\begin{array}{c}
{\color{red} (n^3,m^3)=(\text{odd},\text{odd})} \text{ and } n^3>0
\\
{\color{blue} (n^1,n^1)=(\text{odd},\text{odd})} \text{ and } n^1>0
\end{array}
}
\\\hline\hline
\OR & (0,0,0) & (1,0;1,0;1,0) & (1,0;1,1;{\color{red} 1,1})  & (1,0;1,0;{\color{red} 1,1}) &  ({\color{blue}1,1};1,1;1,1)
\\
& (-\frac{2}{3},\frac{1}{3},\frac{1}{3}) & (0,-1;0,1;0,1) & (0,-1;-1,2;-1,2)  & (0,-1;0,1;-1,2) & (1,-2;-1,2;-1,2)
\\
& (\frac{2}{3},-\frac{1}{3},-\frac{1}{3}) & (1,-1;1,-1;{\color{red} -1,1})   & ({\color{blue} 1,-1};2,-1;-2,1) & ({\color{blue} 1,-1};1,-1;-2,1) &(-2,1;2,-1;-2,1)
\\\hline
\OR\Z_2^{(1)} & (0,\frac{1}{2},-\frac{1}{2}) &  (1,0;-1,2;1,-2) & \color{black} (1,0;-1,1;{\color{red} 1,-1})  & (1,0;-1,2;{\color{red} 1,-1}) &
({\color{blue} 1,1};-1,1;{\color{red} 1,-1})
\\
&  (\frac{1}{3},-\frac{1}{6},-\frac{1}{6}) &  (0,1;2,-1;2,-1) & (0,1;1,0;1,0) & (0,1;2,-1;1,0) & (-1,2; 1,0;1,0)
\\
& (-\frac{1}{3},\frac{1}{6},\frac{1}{6}) & ({\color{blue} 1,-1};1,1;{\color{red} 1,1}) & (1,-1; 0,1;0,1) & ({\color{blue} 1,-1};1,1;0,1) & 
(2,-1; 0,1;0,1)
\\\hline
\OR\Z_2^{(2)} &  (\frac{1}{2},0,-\frac{1}{2}) & (-1,2;1,0;1,-2) & (-1,2;1,1;{\color{red} 1,-1}) & (-1,2;1,0;{\color{red} 1,-1}) & ({\color{blue}
 -1,1};1,1;{\color{red} 1,-1})
\\
& (-\frac{1}{6}, \frac{1}{3},-\frac{1}{6}) &  (2,-1;0,1;2,-1) & (2,-1;-1;2;1,0) & (2,-1;0,1;1,0) & (1,0;-1;2;1,0)
\\
& (\frac{1}{6},-\frac{1}{3},\frac{1}{6}) & ({\color{blue} 1,1};1,-1;{\color{red} 1,1})  & ({\color{blue} 1,1}; 2,-1;0,1) &  ({\color{blue} 1,1};
1,-1;0,1) & (0,1;2,-1;0,1)
\\\hline
\OR\Z_2^{(3)} & (\frac{1}{2},-\frac{1}{2},0) &  (-1,2;1,-2;1,0) & (-1,2;1,-1;{\color{red} 1,1}) & (-1,2;1,-2; 1,1) & ({\color{blue} -1,1};1,-1;{\color{red} 1,1})
\\
&  (-\frac{1}{6}, -\frac{1}{6},\frac{1}{3}) & (2,-1;2,-1;0,1) &  (2,-1;1,0;-1,2) & (2,-1;2,-1;-1,2) & (1,0;1,0;-1,2)
\\
& (\frac{1}{6},\frac{1}{6},-\frac{1}{3}) & ({\color{blue} 1,1};1,1;{\color{red} 1,-1})  & ({\color{blue} 1,1};0,1;2,-1) &
({\color{blue} 1,1};1,1;2,-1) & (0,1;0,1;2,-1)
\\\hline
\end{array}
}{Z2Z6p-ORonBulkO6plane-torus}{Angles with respect to the $\OR$-invariant plane and two-torus wrapping numbers $(n^i,m^i)_{i=1,2,3}$ for each of the three components 
at relative angles $\pm \pi (-\frac{2}{3},\frac{1}{3},\frac{1}{3})$ per O6-plane orbit $\OR\theta^p\omega^q$  under the $\Z_6'$ action with respectively $(p,q)=$(even,even), (even,odd), (odd,odd), (odd,even). As argued below in section~\ref{Sss:Geo_bulk}, it is convenient to select the representant per O6-plane orbit with $(n^3,m^3)=\text{(odd,odd)}$ and $n^3>0$ for the 
 {\bf AAA} and {\bf ABB} lattices, while the symmetry between {\bf AAB} and {\bf BBB} orientations is suitably written for  $(n^1,m^1)=\text{(odd,odd)}$ and $n^1>0$.}

It was furthermore shown in~\cite{Blumenhagen:2005tn,Forste:2010gw} that worldsheet consistency conditions of the Klein bottle amplitude lead to the requirement 
\begin{equation}
\eta= \eta_{\OR} \prod_{k=1}^3 \eta_{\OR\Z_2^{(k)}}
,
\end{equation}
where $\eta_{\OR\Z_2^{(k)}} \in \{1,-1\}$ denotes an ordinary or exotic O6-plane orbit with negative or positive RR charge, respectively. 
On the orbifold with discrete torsion ($\eta=-1$),
worldsheet duality thus necessitates an odd number of exotic O6-plane orbits, and a fully supersymmetric open string
spectrum can only be achieved for one exotic O6-plane orbit as detailed below in section~\ref{Sss:Geo_bulk}.

Last but not least, the orientifold projection in the $\Z_2^{(k)}$ twisted sectors of the closed string depends on the choice of the exotic O6-plane
orbit via the sign factors
\begin{equation}\label{Eq:etas_Z2_sectors}
\eta_{(k)} \equiv \eta_{\OR} \eta_{\OR\Z_2^{(k)}}
\qquad
\text{with}
\qquad 
\eta= \prod_{k=1}^3 \eta_{(k)}
,
\end{equation}
and the closed string spectrum of the Type IIA orientifold on $T^6/\Z_2 \times \Z_6'$ with discrete torsion~\cite{Forste:2010gw} 
is reproduced in table~\ref{Tab:ClosedSpectrum_Z2Z6p}
in terms of the Hodge numbers $h_{11}^+$ counting ${\cal N}=1$ supersymmetric vector multiplets, $h_{11}^-$ chiral multiplets containing the K\"ahler moduli and  
$h_{21}$ chiral multiplets containing the complex structure moduli.

\begin{table}
\begin{equation*}
\begin{array}{|c|c||c|c||c|c|}\hline
\muc{6}{|c|}{\text{\bf Closed string spectrum on } T^6/(\Z_2 \times \Z_6' \times \OR)  \; \text{\bf with discete torsion } (\eta=-1)}
\\\hline\hline
{\cal N}=1 \text{ multiplet} & \# & {\bf AAA} & {\bf ABB} & {\bf AAB} & {\bf BBB}
\\\hline\hline
\text{gravity} 
&  \muc{5}{|c|}{ 1 }
\\\hline
\text{dilaton - axion} 
&  \muc{5}{|c|}{ 1 }
\\\hline
\text{vector} & h_{11}^+= & \frac{5 + \sum_{i=1}^3 \eta_{(i)}}{2} & \frac{5 + \eta_{(1)} - \eta_{(2)} - \eta_{(3)}}{2} & \frac{3 + \eta_{(1)} + \eta_{(2)} - \eta_{(3)}}{2} &  \frac{3 - \sum_{i=1}^3 \eta_{(i)}}{2}
\\\hline
\text{K\"ahler moduli} & h_{11}^-= &  \frac{25 - \sum_{i=1}^3 \eta_{(i)}}{2} &   \frac{25 -  \eta_{(1)} + \eta_{(2)} + \eta_{(3)}}{2} &  \frac{27 -  \eta_{(1)} - \eta_{(2)} + \eta_{(3)}}{2} & \frac{27 + \sum_{i=1}^3 \eta_{(i)}}{2}
\\\hline
\text{complex structure} & h_{21}= & \muc{4}{|c|}{ 15} 
\\\hline
\end{array}
\end{equation*}
\caption{${\cal N}=1$ supersymmetric closed string spectrum of Type IIA string theory on \mbox{$T^6/(\Z_2 \times \Z_6' \times \OR)$} with discrete torsion 
as first computed in~\cite{Forste:2010gw}. The spectra on the four different lattice orientations are pairwise related by 
$(\eta_{(1)},\eta_{(2)},\eta_{(3)}) \leftrightarrow  (\eta_{(1)},-\eta_{(2)},-\eta_{(3)})$ for ${\bf AAA} \leftrightarrow {\bf ABB}$ and 
$(\eta_{(1)},\eta_{(2)},\eta_{(3)}) \leftrightarrow  (-\eta_{(1)},-\eta_{(2)},\eta_{(3)})$ for ${\bf AAB} \leftrightarrow {\bf BBB}$
as discussed in the text. }
\label{Tab:ClosedSpectrum_Z2Z6p}
\end{table}

The structure of the closed string spectrum in table~\ref{Tab:ClosedSpectrum_Z2Z6p} gives a first hint on a potential pairwise 
relation among the various lattice orientations
by permuting the O6-plane orbits and thus the sign factors $\eta_{\OR\Z_2^{(k)}}$ and $\eta_{(k)}$,
 \begin{equation}
\begin{array}{cc}
{\bf AAA}  \leftrightarrow {\bf ABB} & {\bf AAB}  \leftrightarrow {\bf BBB}
\\
\left( \eta_{(1)} , \eta_{(2)}, \eta_{(3)} \right) \leftrightarrow \left( \eta_{(1)} ,\,  -\eta_{(2)}, \, -\eta_{(3)} \right) 
\quad&\quad
\left( \eta_{(1)} , \eta_{(2)}, \eta_{(3)} \right)  \leftrightarrow \left(\,  -\eta_{(1)} , \, -\eta_{(2)}, \eta_{(3)} \right) 
\end{array}
.
\end{equation}
At this point, the ansatz for pairwise identifications among background lattices 
is far from unique, but  further evidence will be collected for its validity throughout 
section~\ref{Ss:Geometry}, and in section~\ref{Ss:identifications} a rigorous proof will be presented by means of the bulk and exceptional RR tadpole cancellation and 
supersymmetry conditions, net-chiralities and full massless open string spectra as well as ingredients of one-loop vacuum amplitudes
as used to compute one-loop gauge threshold corrections and the corresponding perturbatively exact holomorphic gauge kinetic functions
as well as K\"ahler metrics at leading order.

\subsubsection{Bulk three-cycles}\label{Sss:Geo_bulk}

The `bulk' part of the unimodular 32-dimensional lattice of three-cycles on the $T^6/\Z_2 \times \Z_6'$ orbifold with discrete torsion ($\eta=-1$)
 consists of $b_3^{\text{bulk}}=2+2 \, h_{21}^U=2$  three-cycles inherited from the underlying factorisable six-torus.
It is spanned by the following two bulk cycles~\cite{Forste:2010gw}
\begin{equation}
\begin{aligned}
\rho_1 = 4 \left( 1 + \omega + \omega^2 \right) \pi_{135} & = 4 \left(\pi_{136} + \pi_{145} + \pi_{235} -\pi_{146} - \pi_{245} - \pi_{236}  \right)
,
\\
\rho_2 = 4 \left( 1 + \omega + \omega^2 \right) \pi_{136} &= 4 \left(\pi_{136} + \pi_{145} + \pi_{235} -\pi_{246} - \pi_{135}  \right)
,
\\
\text{with intersection number } \quad \rho_1 \circ \rho_2 & =4
.
\end{aligned}
\end{equation}
In terms of the bulk wrapping numbers $(X_a,Y_a)$ defined via the (coprime for fixed $i$) toroidal one-cycle wrapping numbers $(n^i_a,m^i_a)_{i=1,2,3}$ by
\begin{equation}\label{Eq:Z2Z6p-Def-XY}
\begin{aligned}
X_a \equiv  n^1_a n^2_a n^3_a  - m^1_a m^2_a m^3_a -\sum_{i\neq j\neq k\neq i} n^i_a m^j_a m^k_a
,
\qquad
Y_a \equiv \!\! \sum_{i\neq j\neq k\neq i} \!\! \left(n^i_a n^j_a m^k_a + n^i_a m^j_a m^k_a  \right)
,
\end{aligned}
\end{equation}
a bulk three cycle can be expressed as
\begin{equation}
\Pi^{\text{bulk}}_a = X_a \, \rho_1 + Y_a \, \rho_2
\qquad 
\text{ with }
\qquad 
\Pi^{\text{bulk}}_a \circ \Pi^{\text{bulk}}_b = 4 \, \left( X_a Y_b - Y_a X_b \right)
.
\end{equation}
The bulk wrapping numbers $(X_{O6},Y_{O6})$ of the O6-plane orbits with the toroidal one-cycle wrapping numbers given in table~\ref{tab:Z2Z6p-ORonBulkO6plane-torus} are collected in table~\ref{Tab:Z2Z6p-ORonBulkO6plane-bulk}.
\begin{table}[h!]
\renewcommand{\arraystretch}{1.3}
  \begin{center}
\begin{equation*}
\begin{array}{|c||c|c||c|c||c|c||c|c|}\hline
 \multicolumn{9}{|c|}{\text{\bf Bulk wrappings of O6-planes on } T^6/\Z_2 \times \Z_6'}
\\\hline\hline
 & \multicolumn{2}{|c|}{\color{blue} \bf AAA}
 & \multicolumn{2}{|c|}{\color{blue}\bf ABB}
 & \multicolumn{2}{|c|}{\color{red} \bf AAB}
 & \multicolumn{2}{|c|}{\color{red}\bf BBB}
\\
{\rm orbit} & X_{O6} & Y_{O6} & X_{O6} & Y_{O6} & X_{O6} & Y_{O6} & X_{O6} & Y_{O6}
\\\hline\hline
\OR & {\color{blue} 1} & {\color{blue} 0}  & 0 & 3 & 1 & 1  & {\color{red} -3} & {\color{red} 6}
\\
\OR\Z_2^{(1)} & 3 & 0 & {\color{blue} 0} & {\color{blue} 1}& 1 & 1  & -1 & 2
\\
\OR\Z_2^{(2)} &  3 & 0 & 0 & 3 &  1 & 1 &  -1 & 2
\\
\OR\Z_2^{(3)} &  3 & 0  & 0 & 3 & {\color{red} 3} & {\color{red} 3} & -1 & 2
\\\hline
\end{array}
\end{equation*}
\end{center}
\caption{
Bulk wrapping numbers for the four O6-plane orbits on $T^6/(\Z_2 \times \Z_6' \times \OR)$ computed from the
one-cycle wrapping numbers in table~\protect\ref{tab:Z2Z6p-ORonBulkO6plane-torus}. The length of a given bulk three-cycle is 
$L_a \equiv \prod_{i=1}^3 L_a^{(i)} =\prod_{i=1}^3  r_i \, \sqrt{ (n^i_a)^2 + n^i_a m^i_a + (m^i_a)^2 }
=r_1 r_2 r_3 \, \sqrt{ X_a^2 + X_a Y_a + Y_a^2 }$ with the two-torus radii $r_i$ defined in figure~\protect\ref{Fig:Z2Z6p-lattice}.
The relation of O6-plane lengths on the different lattices $(3\, L_{\OR})^{\bf AAA} = (L_{\OR\Z_2^{(k),k \in\{1,2,3\}}})^{\bf AAA} = (3 \, L_{\OR\Z_2^{(1)}})^{\bf ABB} 
= (L_{\OR\Z_2^{(l),l \in \{0,2,3\}}})^{\bf ABB}$  and $(L_{\OR\Z_2^{(3)}}/3)^{\bf AAB}=(L_{\OR\Z_2^{(k),k \in\{0,1,2\}}})^{\bf AAB} = (L_{\OR}/3)^{\bf BBB}=(L_{\OR\Z_2^{(l),l \in \{1,2,3\}}} )^{\bf BBB}$
for a given six-torus volume $(\sqrt{3}/2)^3 (r_1r_2r_3)^2$ supports the claim of pairwise background identifications.
}
\label{Tab:Z2Z6p-ORonBulkO6plane-bulk}
\end{table}
It is easy to verify explicitly that the  one-cycle wrapping numbers of the three components of a given $\OR$ or $\OR\Z_2^{(k), \, k \in \{1,2,3\}}$ orbit 
in table~\ref{tab:Z2Z6p-ORonBulkO6plane-torus} are related by
\begin{equation}\label{Eq:1-cycle-orbits}
\left(\begin{array}{cc}
n^1_a & m^1_a \\ n^2_a & m^2_a \\ n^3_a & m^3_a 
\end{array}\right) \stackrel{\omega}{\rightarrow}
\left(\begin{array}{cc}
m^1_a & -(n^1_a+m^1_a) \\ m^2_a & -(n^2_a +m^2_a) \\ m^3_a & -(n^3_a +m^3_a)
\end{array}\right) \stackrel{\omega}{\rightarrow}
\left(\begin{array}{cc}
-(n^1_a+m^1_a) & n^1_a\\ -(n^2_a+m^2_a) & n^2_a \\ -(n^3_a+m^3_a) & n^3_a
\end{array}\right) 
,
\end{equation}
and that the bulk wrapping numbers $(X_a,Y_a)$ defined in equation~(\ref{Eq:Z2Z6p-Def-XY}) are independent of the choice of representant $a$ or $(\omega \, a)$ or $(\omega^2 \, a)$ of a given $\Z_6'$ invariant orbit.
Note here, however, that in comparison to the one-cycle wrapping numbers for $(\omega \, a)$ given in~\cite{Forste:2010gw} an overall sign-flip along 
$T^2_{(2)} \times T^2_{(3)}$ has been performed. As discussed further below in section~\ref{Sss:Geo_exceptional}, this amendment is necessary to ensure 
the independence of a fractional three-cycle - which according to equation~(\ref{Eq:def_frac_cycle}) consists of a linear combination of bulk and exceptional contributions - 
of the orbifold representant upon a given choice of $\Z_2^{(k)}$ eigenvalues.

Using the orientifold projection of the bulk three-cycles below in table~\ref{tab:OR-on-cycles-Z2Z6p} or alternatively the orientifold
projection on the one-cycle wrapping numbers, 
\begin{equation}\label{Eq:OR_on_n+m}
(n^i_{a'},m^i_{a'}) 
= \left\{\begin{array}{cc}
(n^i_a+m^i_a \, , \, - m^i_a) & {\bf A} \\
(m^i_a,n^i_a) & {\bf B}
\end{array}\right.
,
\end{equation}
one arrives at the bulk RR tadpole cancellation and supersymmetry conditions
for the $T^6/(\Z_2 \times \Z_6' \times \OR)$ orientifold with discrete torsion displayed in table~\ref{tab:Bulk-RR+SUSY-Z2Z6p}.
\mathtabfix{
\begin{array}{|c||c|c|c|}\hline
\multicolumn{4}{|c|}{\text{\bf Global bulk consistency conditions on } \; T^6/(\Z_2 \times \Z_6' \times \OR) \; \text{\bf with discrete torsion } (\eta=-1)}
\\\hline\hline
\text{lattice} & \text{\bf Bulk RR tadpole cancellation} & \text{\bf SUSY: necessary}  & \text{\bf SUSY: sufficient} 
\\\hline\hline
{\bf AAA} & \sum_a N_a \left(2 \,X_a+Y_a \right)= 4 \left( \eta_{\OR} + 3 \, \sum_{i=1}^3 \eta_{\OR\Z_2^{(i)}} \right) & Y_a=0 & 2 \, X_a + Y_a > 0
\\\hline
{\bf ABB} & \sum_a N_a \left(X_a+  2\, Y_a \right)= 4 \left( \eta_{\OR\Z_2^{(1)}} + 3 \, \sum_{i=0,2,3} \eta_{\OR\Z_2^{(i)}} \right) & X_a =0 & X_a +2 \, Y_a > 0
\\\hline\hline
{\bf AAB} & \sum_a N_a \left( X_a+Y_a \right)= 4 \left( 3\, \eta_{\OR\Z_2^{(3)}} +  \sum_{i=0}^2 \eta_{\OR\Z_2^{(i)}} \right) & Y_a-X_a=0 & X_a+Y_a > 0
\\\hline
{\bf BBB} & \sum_a N_a \, Y_a = 4 \left( 3\, \eta_{\OR} +  \sum_{i=1}^3 \eta_{\OR\Z_2^{(i)}} \right) & 2 \,X_a + Y_a=0 & Y_a > 0
\\\hline
\end{array}
}{Bulk-RR+SUSY-Z2Z6p}{Model building constraints for the bulk part of fractional D6-branes defined in equation~(\protect\ref{Eq:def_frac_cycle}) as 
first derived in~\cite{Forste:2010gw}. The four lattices are grouped into pairs which are related by a non-supersymmetric rotation by $\pm \pi/3$
along some two-torus $T^2_{(1)}$ or $T^2_{(3)}$ as proven rigorously below in section~\protect\ref{Sss:T6Z2Z6p_identifications}.}
One can read off that the bulk RR tadpole cancellation and supersymmetry conditions are pairwise related among the lattices upon the following 
permutation of bulk wrapping numbers and O6-planes,
\begin{equation}\label{Eq:1st_hint_identifications}
\begin{array}{cc}
\begin{array}{c} {\bf AAA}  \leftrightarrow {\bf ABB}\\ 
 \left(X_a,Y_a \right) =  \left(\ov{X}_a+\ov{Y}_a, - \ov{X}_a \right)   
\\
\eta_{\OR} \leftrightarrow \eta_{\OR\Z_2^{(1)}}\\   \eta_{\OR\Z_2^{(2)}} \leftrightarrow   \eta_{\OR\Z_2^{(3)}} \end{array}
\qquad & \qquad 
\begin{array}{c} {\bf AAB}  \leftrightarrow {\bf BBB} \\
 \left(X_a,Y_a \right)=  \left(\ov{X}_a+\ov{Y}_a, - \ov{X}_a \right) \\
\eta_{\OR} \leftrightarrow \eta_{\OR\Z_2^{(3)}} \\  \eta_{\OR\Z_2^{(1)}} \leftrightarrow  \eta_{\OR\Z_2^{(2)}} \end{array}
\end{array}
.
\end{equation}
At this point, the exchange of the $\OR$-invariant O6-plane orbit with the shortest (${\bf AAA} \leftrightarrow {\bf ABB}$: $k=1$) or longest
(${\bf AAB} \leftrightarrow {\bf BBB}$: $k=3$) $\OR\Z_2^{(k)}$-invariant O6-plane orbit is unique, as can be seen from the corresponding bulk wrapping numbers
in table~\ref{Tab:Z2Z6p-ORonBulkO6plane-bulk}, whereas the  interchange of the remaining two $\OR\Z_2^{(j)}$-invariant O6-plane orbits (with $j=2,3$ for 
${\bf AAA} \leftrightarrow {\bf ABB}$ and $j=1,2$ for ${\bf AAB} \leftrightarrow {\bf BBB}$)  can only be fixed below in 
section~\ref{Sss:T6Z2Z6p_identifications} by considerations of the exceptional wrapping numbers of fractional D6-branes.
Furthermore, merely considering the bulk consistency conditions allows for a second option $(X_a,Y_a)_{\bf AAA} = (\ov{Y}_a,\ov{X}_a)_{\bf ABB}$
and $(X_a,Y_a)_{\bf AAB} =( - \ov{X}_a ,  \, \ov{X}_a + \ov{Y}_a)_{\bf BBB}$. However, this option has to be discarded due to a lack of interpretation in terms
of maps among the one-cycle wrapping numbers $(n^i_a,m^i_a)_{i=1,2,3}$. 

From table~\ref{tab:Bulk-RR+SUSY-Z2Z6p}, one can in particular read off the maximal ranks for globally consistent supersymmetric D6-brane configurations,
\begin{equation}\label{Eq:max-rank}
\text{max. SUSY rank}_{\bf AAA}=\left\{\begin{array}{cc} 16 & \eta_{\OR} =-1 \\ 8 & \text{else} \end{array} \right.
,
\qquad 
\text{max. SUSY rank}_{\bf BBB} = \left\{\begin{array}{cc} 8 & \text{else} \\ 0 & \eta_{\OR} =-1 \end{array} \right.
,
\end{equation}
and analogously for the {\bf ABB} and {\bf AAB} lattices.

The {\bf AAA} lattice with $\OR$ as the exotic O6-plane orbit (or its dual configuration with exotic $\OR\Z_2^{(1)}$ O6-plane orbit on {\bf ABB}) 
is thus the only choice of background lattice orientation admitting {\it a priori} a large hidden gauge group, since the Standard Model branes 
have already rank five for the minimal choice $U(3) \times USp(2) \times U(1)$.
Note also that in the last case $\eta_{\OR} =-1$ on the {\bf BBB} lattice (or equivalently $\eta_{\OR\Z_2^{(3)}}=-1$ on {\bf AAB}), 
all RR tadpoles cancel among the four different O6-plane orbits without the addition
of D6-branes, and the ${\cal N}=1$ supersymmetric massless spectrum consists solely of the closed string states in table~\ref{Tab:ClosedSpectrum_Z2Z6p}
with $(h_{11}^+,h_{11}^-)=(3,12)$ counting massless vectors and K\"ahler moduli.

{\bf Selection of one orbifold and orientifold image:} As argued in~\cite{Forste:2010gw}, a generic bulk three-cycle consists of three torus images 
$(\omega^k \, a)_{k=0,1,2}$ under the orbifold symmetry, and the three orientifold images  $(\omega^k \, a')_{k=0,1,2}$ of these bulk cycles should not 
be counted independently either. Instead an exchange of the role of $\Pi_a \circ \Pi_b$ and $\Pi_a \circ \Pi_{b'}$ will be allowed in the definition 
of Standard Model particles in section~\ref{S:Modelbuilding}.

To single out one representant of a {\it generic} orbifold and orientifold orbit on the {\bf AAA} and {\bf ABB} lattices, one can impose the following conditions on the toroidal one-cycle wrapping numbers:
\begin{enumerate}
\item[1.]
$(n^3_a,m^3_a)=$(odd,odd) selects one orbifold image  $(\omega^k \,  a)_{k \in \{0,1,2\}}$ under the $\omega$ action, and $n^3_a >0$ fixes the orientation of the 
one-cycle along the two-torus $T^2_{(3)}$.
\item[2a.] Choosing the angle \mbox{$-\frac{\pi}{3} < \pi \phi^{(2)}_a < \frac{\pi}{6}$}  on the {\bf A}-type torus $T^2_{(2)}$ and the angle 
\mbox{$-\frac{\pi}{2} < \pi \phi^{(2)}_a < 0$}  on the {\bf B}-type torus $T^2_{(2)}$  singles out a D6-brane compared to its orientifold 
image.\footnote{Note that here the second instead of the first two-torus is used and that the conditions for the {\bf A}-type torus $T^2_{(2)}$ have been
 corrected compared to the original version in~\cite{Forste:2010gw}.} 
For both types of the torus $T^2_{(2)}$, this amounts to demanding $n^2_a > 0$ and $ n^2_a > |m^2_a|$.
\item[2b.] For {\it special} orbifold and orientifold orbits with $n_a^2 = | m_a^2|$, condition 2a does not apply. Treating these cases separately, one can 
distinguish the representant and its orientifold image by imposing $n_a^3 > |m_a^3|$ instead of condition 2a. 
\end{enumerate}
The {\it very special} orbits for which $\pi \phi^{(2)}_a \in \{0, \frac{\pi}{2} \}$ and $\pi \phi^{(3)}_a \in \{0, -\frac{\pi}{2} \}$ are still lacking. 
These cases correspond to D6-brane orbits parallel to some O6-plane orbit, for which the toroidal wrapping numbers are summarised 
in table~\ref{tab:Z2Z6p-ORonBulkO6plane-torus}. 

For the {\bf AAB} and {\bf BBB} lattices, it turns out to be more convenient to impose condition 1. on the first two-torus, i.e. $(n^1_a,m^1_a)=\text{(odd,odd)}$ with $n^1_a >0$
as detailed in appendix~\ref{A:identifications_AAB+BBB}.

Using these criteria for defining a representant for each orbifold and orientifold orbit, an exhaustive list containing all supersymmetric three-cycles
which do not overshoot the bulk RR tadpole cancellation condition of table~\ref{tab:Bulk-RR+SUSY-Z2Z6p}
are given in tables~\ref{Tab:all-cycles-AAA} and~\ref{Tab:all-cycles-BBB} of 
appendix~\ref{A:tables_bulk-cycles} for the {\bf AAA} and {\bf BBB} lattice, respectively.
The lists for the {\bf AAB} and {\bf ABB} lattices are obtained by applying the identifications discussed in section~\ref{Sss:T6Z2Z6p_identifications}
equation~(\ref{Eq:identify_nm_AAA+ABB}) and appendix~\ref{A:identifications_AAB+BBB} equation~(\ref{Eq:identify_nm_AAB+BBB}).

The abundance of three-cycles per given normalised length $\frac{L_a}{r_1r_2r_3}= \sqrt{X_a^2 +X_aY_a+Y_a^2}$ are summarised table~\ref{Tab:Numbers-of-cycles}, 
again supporting the claim of pairwise relations among the four different background lattice orientations.
\begin{table}[h]
\begin{center}
\begin{tabular}{|c||c|c|c||c|c|c||c|c|c||c|c|c|}
\hline \multicolumn{13}{|c|}{\bf Number of SUSY bulk three-cycles on $T^6/(\Z_2 \times \Z_6' \times \OR)$ per lattice} \\
\hline
\hline &\multicolumn{3}{|c||}{\color{red}\bf AAA} & \multicolumn{3}{|c||}{\color{red}\bf ABB} & \multicolumn{3}{|c||}{\color{blue}\bf AAB} &\multicolumn{3}{|c|}{\color{blue}\bf BBB}\\
length&$X_a$& $Y_a$ &\bf number & $X_a$ & $Y_a$ &\bf number & $X_a$ & $Y_a$ &\bf number & $X_a$ & $Y_a$ & \bf number\\
\hline\hline
S&1&0& 2 & 0&1& 2 & 1&1& 6 & -1&2& 6\\
NS&3&0& 6 & 0&3& 6 & 3&3& 2 & -3&6& 2\\
NNS&7&0& 9 & 0& 7& 9 & 7&7& 27 & -7&14& 27\\
NNNS&13&0& 9 & 0&13& 9 & &&  & && \\
\hline
\end{tabular}
\caption{Total number of independent supersymmetric bulk three-cycles on $T^6/(\Z_2 \times \Z_6' \times \OR)$ for each lattice, which do not 
overshoot  the bulk RR tadpole cancellation condition of table~\protect\ref{tab:Bulk-RR+SUSY-Z2Z6p}. These three-cycles are constructed using 
the selection rules given at the end of section~\ref{Sss:Geo_bulk}. 
The cycles are classified in terms of their length for given two-torus radii $r_i$, where `S' stands for shortest, `NS' for next-to-shortest, 
`NNS' for next-to-next-to-shortest, etc.}
\label{Tab:Numbers-of-cycles}
\end{center}
\end{table}

\subsubsection{Exceptional three-cycles}\label{Sss:Geo_exceptional}

In addition to the bulk three-cycles discussed in the previous section, the $T^6/\Z_2 \times \Z_6'$ orbifold with discrete torsion has
 three kinds of local $\mathbb{C}^2_{(k)}/\Z_2^{(k)}$  singularities ($k\in \{1,2,3\}$), at which exceptional divisors with intersection form 
\mbox{$e_{x_1y_1}^{(k)} \circ e_{x_2y_2}^{(k)} =-2 \; \delta_{x_1x_2}\delta_{y_1y_2}$} are supported with fixed point labels $x_l,y_l \in \{1,4,5,6\}$ along $T^2_{(i)} \times T^2_{(j)} \equiv T^4_{(k)}$
as defined in figure~\ref{Fig:Z2Z6p-lattice}. 
On the globally defined $T^6/\Z_2 \times \Z_6'$ orbifold with discrete torsion ($\eta=-1$), these exceptional two-cycles with support at singularities on the four-torus $T^4_{(k)}$ are tensored with a toroidal one-cycle
$\pi_{2k-1}$ or $\pi_{2k}$ on the remaining two-torus $T^2_{(k)}$, and the orbit under the $\Z_6'$ action $\omega$ has to be taken, e.g. in 
the $\Z_2^{(3)}$ twisted sector
\begin{equation}
\sum_{l=0}^5 \omega^l \left(e^{(3)}_{41} \otimes \pi_5 \right) =2 \; \left( e^{(3)}_{41} \otimes \pi_5 + e^{(3)}_{51} \otimes (-\pi_{6})
+ e^{(3)}_{61} \otimes (\pi_6 - \pi_5) \right) \equiv \varepsilon^{(3)}_1
.
\end{equation}
In~\cite{Forste:2010gw}, the three sets of exceptional three-cycles $\varepsilon^{(k)}_{\alpha}$, $\tilde{\varepsilon}^{(k)}_{\alpha}$ were constructed 
explicitly with intersection form
\begin{equation}
\varepsilon^{(k)}_{\alpha} \circ\tilde{\varepsilon}^{(l)}_{\beta} = -4 \; \delta^{kl} \, \delta_{\alpha\beta}
,
\end{equation}
where $k,l \in \{1,2,3\}$ label the three $\Z_2^{(k)}$, $\Z_2^{(l)}$ sectors and \mbox{$\alpha,\beta \in \{1 \ldots 5\}$} the orbits of 
\mbox{$e_{xy}^{(k)} \otimes \pi_{2k-1}$} and  \mbox{$e_{xy}^{(k)} \otimes \pi_{2k}$} under $\omega$ for $\varepsilon^{(k)}_{\alpha}$ and 
$\tilde{\varepsilon}^{(k)}_{\alpha}$, respectively.
The full list of exceptional three-cycles from~\cite{Forste:2010gw} is for later convenience reproduced here in table~\ref{tab:Z2Z6p-fps+excycles}.
\mathtabfix{
\begin{array}{|c|c||c|c|}\hline
\multicolumn{4}{|c|}{\Z_2^{(k)} \; \text{\bf fixed points and exceptional 3-cycles on } \, T^6/\Z_2 \times \Z_6' \; \text{\bf with discrete torsion } (\eta=-1)} 
\\\hline\hline
{\rm f.p.}^{(k)} \otimes (n^{k}_a \pi_{2k-1} + m^{k}_a \pi_{2k}) & {\rm orbit}  & {\rm f.p.}^{(k)} \otimes (n^{k}_a \pi_{2k-1} + m^{k}_a \pi_{2k}) & {\rm orbit} 
\\\hline\hline
11 & - 
& 55 &  -(n^k_a + m^k_a)  \varepsilon^{(k)}_3 + n^k_a \tilde{\varepsilon}^{(k)}_3
\\\cline{1-2}
41 &   n^k_a \; \varepsilon^{(k)}_1 + m^k_a \tilde{\varepsilon}^{(k)}_1
&66 &  m^k_a \varepsilon^{(k)}_3 - (n^k_a + m^k_a)  \tilde{\varepsilon}^{(k)}_3
\\\cline{3-4}
51 & -(n^k_a + m^k_a)  \varepsilon^{(k)}_1 + n^k_a \tilde{\varepsilon}^{(k)}_1
& 45 &    n^k_a \varepsilon^{(k)}_4 + m^k_a \tilde{\varepsilon}^{(k)}_4
\\
61 & m^k_a \varepsilon^{(k)}_1 - (n^k_a + m^k_a)  \tilde{\varepsilon}^{(k)}_1
& 56 &  -(n^k_a + m^k_a)  \varepsilon^{(k)}_4 + n^k_a \tilde{\varepsilon}^{(k)}_4
\\\cline{1-2}
14 &   n^k_a \varepsilon^{(k)}_2 + m^k_a \tilde{\varepsilon}^{(k)}_2
& 64 &  m^k_a \varepsilon^{(k)}_4 - (n^k_a + m^k_a)  \tilde{\varepsilon}^{(k)}_4
\\\cline{3-4}
15 & -(n^k_a + m^k_a)  \varepsilon^{(k)}_2 + n^k_a \tilde{\varepsilon}^{(k)}_2
& 46 &    n^k_a \varepsilon^{(k)}_5 + m^k_a \tilde{\varepsilon}^{(k)}_5
\\
16 &  m^k_a \varepsilon^{(k)}_2 - (n^k_a + m^k_a)  \tilde{\varepsilon}^{(k)}_2
& 54 &  -(n^k_a + m^k_a)  \varepsilon^{(k)}_5 + n^k_a \tilde{\varepsilon}^{(k)}_5
\\\cline{1-2}
44 &   n^k_a \varepsilon^{(k)}_3 + m^k_a \tilde{\varepsilon}^{(k)}_3
& 65 &  m^k_a \varepsilon^{(k)}_5 - (n^k_a + m^k_a)  \tilde{\varepsilon}^{(k)}_5
\\\hline
\end{array}
}{Z2Z6p-fps+excycles}{Complete list of orbits of exceptional divisors $e_{xy}^{(k)}$ at fixed points of $\Z_2^{(k)}$ along $T^4_{(k)}$ times one-cycles $n^{k}_a \pi_{2k-1} + m^{k}_a \pi_{2k}$ 
along $T^2_{(k)}$ under  the $\Z_6'$ orbifold generator $\omega$ as computed in~\cite{Forste:2010gw}.}
By linearly combining bulk and exceptional three-cycles, supersymmetric fractional three-cycles are constructed,
\begin{equation}\label{Eq:def_frac_cycle}
\begin{aligned}
\Pi^{\text{frac}}_a \equiv  \frac{1}{4} \left(\Pi^{\text{bulk}}_a + \sum_{i=1}^3 \Pi^{\Z_2^{(i)}}_a  \right)
= &\frac{1}{4} \left(X_a \, \rho_1 + Y_a \, \rho_2 + \sum_{k=1}^3 \sum_{\alpha=1}^5 \left[x^{(k)}_{\alpha,a} \; \varepsilon^{(k)}_{\alpha}
+  y^{(k)}_{\alpha,a} \; \tilde{\varepsilon}^{(k)}_{\alpha} \right] \right)
\\
\text{with} \qquad 
\chi^{(\N_a,\ov{\N}_b)} \equiv \Pi^{\text{frac}}_a \circ \Pi^{\text{frac}}_b =& \frac{1}{4} \left(X_a Y_b - Y_a X_b - \sum_{k=1}^3 \sum_{\alpha=1}^5 \left[x^{(k)}_{\alpha,a} \,  y^{(k)}_{\alpha,b} 
- y^{(k)}_{\alpha,a} \,  x^{(k)}_{\alpha,b}   \right] \right)
.
\end{aligned}
\end{equation}
For a given bulk three-cycle, the bulk wrapping numbers have been defined in equation~(\ref{Eq:Z2Z6p-Def-XY}).
The exceptional wrapping numbers $(x^{(k)}_{\alpha,a},y^{(k)}_{\alpha,a})$ are obtained by means of the additional
discrete parameters that characterise a fractional three-cycle:
\begin{itemize}
\item
three $\Z_2^{(k)}$ eigenvalues $(-1)^{\tau^{\Z_2^{(k)}}_a}$ with one relation $(-1)^{\tau^{\Z_2^{(3)}}_a}=(-1)^{\tau^{\Z_2^{(1)}}_a+\tau^{\Z_2^{(2)}}_a}$, where
 $\tau^{\Z_2^{(k)}}_a =0,1$ corresponds to $\Z_2^{(k)}$ eigenvalues $+1$, $-1$, respectively,\footnote{In a simplified version, one can view this sign 
factor as encircling a reference fixed point per two-torus clockwise or counter-clockwise. A discrete Wilson line corresponds then to encircling the second 
$\Z_2$ fixed point, which is traversed by the fractional three-cycle along the same two-torus,  in the opposite direction.}
\item
three discrete displacements $(\vec{\sigma}_a)$ with $\sigma^i_a \in \{0,1\}$ parameterising if the bulk three-cycle passes through
the origin on each two-torus $T^2_{(i)}$ or is shifted from it by one-half of some lattice vector,
\item
three discrete Wilson lines $(\vec{\tau}_a)$ with $\tau^i_a \in \{0,1\}$ per two-torus $T^2_{(i)}$, which encode if both exceptional three-cycles at different 
$\Z_2^{(k)}$ fixed points on $T^2_{(i)}$ are added or subtracted from a given bulk three-cycle in equation~(\ref{Eq:def_frac_cycle}).
\end{itemize}
The exceptional contribution  $\Pi^{\Z_2^{(i)}}_a$ to a fractional three-cycle with given bulk three-cycle is thus composed of combinatorics involving 
which exceptional three-cycles are traversed 
for a given set of toroidal wrapping numbers $(n^i_a,m^i_a)_{i=1,2,3}$ and displacements $(\vec{\sigma}_a)$ as displayed in table~\ref{Tab:Reference_Point+Signs}.
\begin{SCtable}
$\begin{array}{|c|ccc|}\hline
\muc{4}{|c|}{\text{\bf Assignment of prefactors $(-1)^{\tau^{\Z_2^{(i)}}_a}$ \!\!\!\!\! or $(-1)^{\tau^{\Z_2^{(i)}}_a \!\!\!+ \tau^i_a}$ }}
\\\hline\hline
(n^i_a,m^i_a) & \text{(odd,odd)} \stackrel{\omega}{\to} & \!\!\!\!\text{(odd,even)} \stackrel{\omega}{\to} & \!\!\!\!\text{(even,odd)}
\\\hline\hline
\sigma^i_a=0 & \left(\begin{array}{c} 1 \\  6\end{array}\right) \to 
& \left(\begin{array}{c} 1 \\ 4 \end{array}\right) \to & \left(\begin{array}{c} 1 \\ 5 \end{array}\right)
\\\hline
\sigma^i_a=1 & \left(\begin{array}{c}  4 \\ 5 \end{array}\right) \to & \left(\begin{array}{c} 5 \\ 6 \end{array}\right)\to  
& \left(\begin{array}{c} 6 \\ 4 \end{array}\right)  
\\\hline
\end{array}$
\caption{Consistent assignment of the reference point (upper entry)  and the second $\Z_2^{(i)}$ fixed point (lower entry) contributing with  sign factor  $(-1)^{\tau^{\Z_2^{(i)}}_a}$ or $(-1)^{\tau^{\Z_2^{(i)}}_a+ \tau^i_a}$ to $\Pi^{\Z_2^{(j),j\neq i}}_a$.
The independence of the sign assignments  on the representant $(\omega^k \,  a)_{k\in \{01,2\}}$ of a given $\Z_6'$ orbit requires the flip from (4,6) to (6,4) in the lower right entry compared to~\cite{Forste:2010gw}.
}
\label{Tab:Reference_Point+Signs}
\end{SCtable}
The resulting $\Z_6'$ invariant orbits given in table~\ref{tab:Z2Z6p-fps+excycles} are dressed with two kinds of sign factors, the global $\Z_2^{(k)}$ eigenvalue 
of $\Pi^{\Z_2^{(k)}}_a$ and the relative signs, the Wilson lines $(\tau_a^i,\tau_a^j)$ among different fixed points within a given $\Z_2^{(k)}$ twisted sector. 
The classification of exceptional wrapping numbers $(x^{(i)}_{\alpha,a},y^{(i)}_{\alpha,a})$ entering equation~(\ref{Eq:def_frac_cycle})
is displayed in short in table~\ref{Tab:xy-entries-Z2Z6p}, and an exhaustive list is for later convenience given in 
table~\ref{tab:exceptional-wrappings-Z2Z6} in appendix~\ref{A:tables_bulk-cycles}.
More details on the generic procedure of constructing the $\Z_2^{(k)}$ twisted sector contributions $\Pi^{\Z_2^{(k)}}_a$ can be found in chapter~2 of~\cite{Forste:2010gw}.
\begin{table}
\begin{equation*}
\begin{array}{|c||c|c|}\hline
\multicolumn{3}{|c|}{\text{\bf Exceptional wrappings $(x^{(i)}_{\alpha,a},y^{(i)}_{\alpha,a})$ expressed via torus wrappings $(n^{i}_a,m^{i}_a)$}}
\\\hline
& {\rm I} & {\rm II}
\\\hline\hline
\text{(a)} & (z^{(i)}_{\alpha,a} \; n^i_a \;, \; z^{(i)}_{\alpha,a} \, m^i_a)
& \left(-z^{(i)}_{0,a} \;  n^i_a+ (z^{(i)}_{\alpha,a}-z^{(i)}_{0,a}) \, m^i_a \; , \, (z^{(i)}_{0,a}-z^{(i)}_{\alpha,a}) \; n^i -z^{(i)}_{\alpha,a} \, m^i_a \right)
\\
\text{(b)} &  (z^{(i)}_{\alpha,a} \; m^i_a \;, \; -z^{(i)}_{\alpha,a} \; (n^i_a+m^i_a) )
& \left( (z^{(i)}_{0,a}-z^{(i)}_{\alpha,a}) \; n^i_a -z^{(i)}_{\alpha,a} \; m^i \; , \; z^{(i)}_{0,a}\; m^i_a+z^{(i)}_{\alpha,a} \; n^i_a \right)
\\
\text{(c)} & (-z^{(i)}_{\alpha,a} \; (n^i_a+m^i_a)  \; , \; z^{(i)}_{\alpha,a} \; n^i_a)
&\left( z^{(i)}_{\alpha,a} \; n^i_a+z^{(i)}_{0,a} \; m^i_a \; , \; -z^{(i)}_{0,a}\; n^i_a +(z^{(i)}_{\alpha,a} -z^{(i)}_{0,a} ) \; m^i_a \right)
\\\hline
\end{array}
\end{equation*}
\caption{The exceptional wrapping numbers of type I stem from a single $\Z_2^{(i)}$ fixed point, while wrapping numbers of type II arise from a linear combination 
of contributions from two
$\Z_2^{(i)}$ fixed points. Any three-cycle $\Pi^{\Z_2^{(i)}}$ receives contributions from exactly three $\alpha \in \{1 \ldots 5\}$ with at most one of the shape II. The sign factors 
are given by $z^{(i)}_{0,a} \equiv (-1)^{\tau^{\Z_2^{(i)}}_a}$ and \mbox{$z^{(i)}_{\alpha,a}  \in \{ (-1)^{\tau^{\Z_2^{(i)}}_a+\tau^j_a} , (-1)^{\tau^{\Z_2^{(i)}}_a+\tau^k_a} , (-1)^{\tau^{\Z_2^{(i)}}_a+\tau^j_a + \tau^k_a}  \}$}
with $(ijk)$ some permutation of $(123)$. The full list of coefficients in dependence of even- or oddness of wrapping numbers and displacements $(\sigma^j_a;\sigma^k_a)$ is given
in table~\protect\ref{tab:exceptional-wrappings-Z2Z6} of appendix~\protect\ref{A:tables_bulk-cycles}.
It incorporates the sign corrections for $(n^{j}_a,m^{j}_a)=\text{(even,odd)}$ compared to~\cite{Forste:2010gw} detailed in the text.}
\label{Tab:xy-entries-Z2Z6p}
\end{table}

It is important to notice that two amendments were performed here. The first one consists in the overall sign factor of the one-cycle components 
$(n^2_{(\omega \, a)}, m^2_{(\omega \, a)}; n^3_{(\omega \, a)}, m^3_{(\omega \, a)}) $ 
of $(\omega \, a)$ in equation~(\ref{Eq:1-cycle-orbits}), which is required to preserve the $\Z_2^{(k)}$ eigenvalue for each choice of the orbifold 
representant.\footnote{This problem did not occur for the $T^6/\Z_6$ orbifold~\cite{Honecker:2004kb,Gmeiner:2007we} since for the $\Z_2^{(1)} \subset \Z_6$ sector, 
only the overall sign of 
$(n^1_{(\omega \, a)}, m^1_{(\omega \, a)})$ enters the definition of $(x^{(1)}_{\alpha,a},y^{(1)}_{\alpha,a})$. 
For the massless spectrum of the Type IIA orientifold example on $T^6/\Z_2 \times \Z_6'$ with discrete torsion given in~\cite{Forste:2010gw}, it does, however, 
modify the spectrum in the $a_m(\omega^k a_n)_{k=1,2}$ sectors. Instead of some of the bifundamental representations $(m \neq n)$, two adjoint representations 
for each $m=n$ arise, while the remaining bifundamental representations are $[(\2,\1,\1,\ov{\2}) + (\1,\2,\ov{\2},\1) + h.c.]$ with
a different permutation of charge entries from those given in ~\cite{Forste:2010gw}.}
The second correction consists in the choice of the reference fixed point for $(n^i_a,m^i_a)=\text{(even,odd)}$ in the presence of a displacement $\sigma^i_a=1$
and discrete Wilson line $\tau^i_a=1$ in table~\ref{Tab:Reference_Point+Signs}, which is in agreement with the analogous discussion for the $T^6/\Z_6'$ orbifold in appendix~A.1 of~\cite{Gmeiner:2009fb}.

The {\bf orientifold} projection on exceptional three-cycles is composed of the projection on the exceptional divisors at $\mathbb{C}^2_{(k)}/\Z_2^{(k)}$ 
singularities, 
\begin{equation}
\OR : e^{(k)}_{xy} \to -\eta_{(k)} \; e^{(k)}_{x'y'}
,
\end{equation}
times the permutation of one-cycles along $T^2_{(k)}$ according to equation~(\ref{Eq:OR_on_n+m}). 
The sign $\eta_{(k)} = \pm 1$ depends on the choice of the exotic O6-plane orbit as defined in equation~(\ref{Eq:etas_Z2_sectors}), 
and the $\Z_2^{(k)}$ fixed points $x,y \stackrel{\cal R}{\to} x',y'$ transform as detailed in the caption of figure~\ref{Fig:Z2Z6p-lattice}. 
The result of the orientifold projection for the $\Z_6'$ invariant orbits of all three-cycles on $T^6/(\Z_2 \times \Z_6' \times \OR)$ with discrete torsion is collected in table~\ref{tab:OR-on-cycles-Z2Z6p}.
\mathtabfix{
\begin{array}{|c||c|c||c|c||c|c|}\hline
\multicolumn{7}{|c|}{\text{\bf Orientifold images of three-cycles on} \; T^6/(\Z_2 \times \Z_6' \times \OR) \; \text{\bf with discrete torsion $(\eta=-1)$}}
\\\hline\hline
\text{lattice} & \OR (\rho_1) & \OR (\rho_2) & \OR (\varepsilon^{(k)}_{\alpha} ) & \OR ( \tilde{\varepsilon}^{(k)}_{\alpha} ) & \alpha,\alpha' & k
\\\hline\hline
{\bf AAA} & \rho_1 & \rho_1 - \rho_2  &  -\eta_{(k)} \; \varepsilon^{(k)}_{\alpha'}  & \eta_{(k)} \; \left( \tilde{\varepsilon}^{(k)}_{\alpha'} - \varepsilon^{(k)}_{\alpha'} \right)
&  \alpha=\alpha'=1,2,3 ; \;  \alpha \leftrightarrow \alpha' = 4,5  &  k=1,2,3 
\\\hline\hline
{\bf ABB} & \rho_2-\rho_1 & \rho_2 &  \eta_{(1)} \;  \left( \varepsilon^{(1)}_{\alpha'} - \tilde{\varepsilon}^{(1)}_{\alpha'} \right) &   -\eta_{(1)} \; \tilde{\varepsilon}^{(1)}_{\alpha'} 
&  \alpha=\alpha'=1,2,3  ; \; \alpha \leftrightarrow \alpha' = 4,5   & k=1
\\\cline{4-7}
& & &   -\eta_{(k)} \; \tilde{\varepsilon}^{(k)}_{\alpha'} &   -\eta_{(k)} \; \varepsilon^{(k)}_{\alpha'} 
 &  \alpha=\alpha'=1,5 ; \;  \alpha \leftrightarrow \alpha'=3,4  & k=2,3
\\\cline{4-6}
& & &  \eta_{(k)} \; \varepsilon^{(k)}_{2}  & \eta_{(k)} \;  \left( \varepsilon^{(k)}_{2} - \tilde{\varepsilon}^{(k)}_{2} \right) & \alpha=\alpha'=2  & 
\\\hline\hline
{\bf AAB} & \rho_2 & \rho_1 &  -\eta_{(k)} \; \varepsilon^{(k)}_{\alpha'}  & \eta_{(k)} \; \left( \tilde{\varepsilon}^{(k)}_{\alpha'} - \varepsilon^{(k)}_{\alpha'} \right)
 &   \alpha=\alpha'=1,5 ; \;  \alpha \leftrightarrow \alpha'=3,4 & k=1,2
\\\cline{4-6}
&  & &  \eta_{(k)} \;  \left( \varepsilon^{(k)}_{2} - \tilde{\varepsilon}^{(k)}_{2} \right) &   -\eta_{(k)} \; \tilde{\varepsilon}^{(k)}_{2}  & \alpha=\alpha'=2  &
\\\cline{4-7}
& & &   -\eta_{(3)} \; \tilde{\varepsilon}^{(3)}_{\alpha'} &   -\eta_{(3)} \; \varepsilon^{(3)}_{\alpha'} 
 &   \begin{array}{c}   \alpha=\alpha'=1,2,3; \; \alpha \leftrightarrow \alpha' = 4,5  \end{array}
& k=3
\\\hline\hline
{\bf BBB} & - \rho_1 & \rho_2 - \rho_1 & \eta_{(k)} \; \varepsilon^{(k)}_{\alpha'}  & \eta_{(k)} \;  \left( \varepsilon^{(k)}_{\alpha'} - \tilde{\varepsilon}^{(k)}_{\alpha'} \right)
&   \alpha=\alpha'=1,2,3 \; \;  \alpha \leftrightarrow \alpha' = 4,5   &  k=1,2,3 
\\\hline
\end{array}
}{OR-on-cycles-Z2Z6p}{Orientifold images of bulk and exceptional three-cycles in dependence of the background lattice and choice of exotic O6-plane orbit with $\eta_{(k)} \equiv \eta_{\OR} \cdot \eta_{\OR\Z_2^{(k)}}$.}

The exceptional RR tadpole cancellation conditions resulting from the $\OR$-projection are summarised for the anticipated pairwise related  {\bf AAA} and {\bf ABB} 
lattices in table~\ref{tab:twistedRR-AAA+ABB-Z2Z6p} and for the pairing of lattices {\bf AAB} and {\bf BBB} in table~\protect\ref{tab:twistedRR-AAB+BBB-Z2Z6p} 
of appendix~\protect\ref{A:identifications_AAB+BBB}.
\mathtabfix{
\begin{array}{|c|c||c|c|}\hline
\multicolumn{4}{|c|}{\text{\bf Twisted RR tadpole cancellation conditions on $T^6/(\Z_2 \times \Z_6' \times \OR)$ with discrete torsion $(\eta=-1)$, Part I}}
\\\hline\hline
i & \alpha & {\bf AAA} & {\bf ABB}
\\\hline\hline
1 & 1,2,3 & \sum_a N_a \left[ x_{\alpha,a}^{(1)}-\eta_{(1)} \; \left(x_{\alpha,a}^{(1)} + y_{\alpha,a}^{(1)} \right) \right]=0 =  \sum_a N_a (1+\eta_{(1)}) \; y_{\alpha,a}^{(1)}
&  \sum_a N_a \left[ y_{\alpha,a}^{(1)}-\eta_{(1)} \; \left(x_{\alpha,a}^{(1)}+  y_{\alpha,a}^{(1)} \right) \right] =0 = \sum_a N_a (1+\eta_{(1)})  \;x_{\alpha,a}^{(1)} =0
\\\hline
1 & 4,5 & \sum_a N_a \left[ \left( 2 \,  x^{(1)}_{4,a} +y^{(1)}_{4,a} \right) -\eta_{(1)} \; \left( 2 \,  x^{(1)}_{5,a} +  y^{(1)}_{5,a} \right) \right]=0 = \sum_a N_a \left[ y^{(1)}_{4,a} + \eta_{(1)} \; y^{(1)}_{5,a} \right]
&\sum_a N_a  \left[ \left( x^{(1)}_{4,a}+ 2 \, y^{(1)}_{4,a} \right)- \eta_{(1)} \; \left( x^{(1)}_{5,a} + 2 \, y^{(1)}_{5,a} \right)  \right] =0 =\sum_a N_a  \left[ x^{(1)}_{4,a}+ \eta_{(1)}  \; x^{(1)}_{5,a} \right]
\\\hline\hline
2,3 & 1,5 & \text{analogous to } i=1 &  \sum_a N_a \left[ x^{(i)}_{\alpha,a} -\eta_{(i)}  \;  y^{(i)}_{\alpha,a} \right]=0
\\\cline{1-2}\cline{3-4}
& 3,4 & & \sum_a N_a \left[ x^{(i)}_{3,a} -\eta_{(i)} \;  y^{(i)}_{4,a} \right]=0 =\sum_a N_a  \left[ x^{(i)}_{4,a} -\eta_{(i)}  \; y^{(i)}_{3,a} \right]
\\\hline
& 2 & & \sum_a N_a  \left[  x^{(i)}_{2,a}  +\eta_{(i)} \; \left( x^{(i)}_{2,a} + y^{(i)}_{2,a} \right) \right]=0 =  \sum_a N_a (1 -\eta_{(i)}) \;  y^{(i)}_{2,a}
\\\hline
\end{array}
}{twistedRR-AAA+ABB-Z2Z6p}{Twisted RR tadpole cancellation conditions for the $\Z_2^{(i)}$ twisted sectors on the {\bf AAA} and {\bf ABB} background lattices.
The corresponding equations for the {\bf AAB} and {\bf BBB} backgrounds are given in table~\protect\ref{tab:twistedRR-AAB+BBB-Z2Z6p} of appendix~\protect\ref{A:identifications_AAB+BBB}.
}
This completes the list of basic ingredients  - together with the computation of full massless open string spectra briefly reviewed in 
appendix~\ref{A:intersections+betas} 
- required for globally consistent D6-brane model building on $T^6/(\Z_2 \times \Z_6' \times \OR)$ with discrete torsion,
for which  pairwise equivalence relations are rigorously proven in the following section~\ref{Ss:identifications}.

\subsection{Pairwise identification of lattice orientations: ${\bf AAA} \leftrightarrow {\bf ABB}$}\label{Ss:identifications}

In section~\ref{Ss:Geometry}, it was argued that the counting of twisted massless closed string states, the bulk RR tadpole cancellation and supersymmetry conditions 
as well as the  abundance of three-cycles of a given length indicate a pairwise symmetry among the {\bf AAA} and {\bf ABB} as well as the {\bf AAB} and {\bf BBB} lattice orientations. 
In this section,   the equivalence is proven for the first pair of lattices ${\bf AAA} \leftrightarrow {\bf ABB}$ at the level of global consistency conditions, intersection numbers, full massless spectra and string one-loop amplitudes without 
operator insertions.  The latter imply that the K\"ahler metrics at leading order and the perturbatively exact holomorphic gauge kinetic functions are also identical 
for a given pairing.
The analogous discussion for the second pair of lattices ${\bf AAB} \leftrightarrow {\bf BBB}$  is relegated to appendix~\ref{A:identifications_AAB+BBB}.

Since the discussion for the $T^6/\Z_2 \times \Z_6'$ orbifold with discrete torsion is quite involved due to the presence of three $\Z_2^{(k)}$ 
subsectors, and it is a priori not clear if a map from the {\bf AAA} to {\bf ABB} background involves a permutation of these $\Z_2^{(k)}$ sectors, it is instructive to
first consider the simpler set-up of the Type IIA string theory with D6-branes on the $T^6/(\Z_6 \times \OR)$ orbifold with $\Z_6 \subset \Z_2 \times \Z_6'$.

\subsubsection{Identifications of three-cycle lattices on {\bf AAA} and {\bf ABB} backgrounds for the $T^6/(\Z_6 \times \OR)$ orientifold}\label{Sss:T6Z6_identifications}

The unimodular lattice of three-cycles  of the $T^6/\Z_6$ orbifold on an $SU(3)^3$ root lattice
can be obtained by restricting the lattice of three-cycles of the  $T^6/\Z_2 \times \Z_6'$ orbifold
with discrete torsion to the bulk plus one of its $\Z_2^{(i)}$ contributions while changing the overall normalisation of fractional three-cycles
and their intersection numbers
accordingly.\footnote{This procedure can only be applied to the 
three-cycles. The structure of two- and four-cycles on $T^6/\Z_6$ is richer as can already be seen from 
the number of untwisted K\"ahler moduli $h_{11}^U=5$ on $T^6/\Z_6$ versus $h_{11}^U=3$ on $T^6/\Z_2 \times \Z_6'$ with/without discrete torsion.}
The bulk RR tadpole cancellation and supersymmetry conditions for each lattice orientation can then be recovered by setting
$\eta_{\OR}=\eta_{\OR\Z_2^{(i)}}=1$ and  $\eta_{\OR\Z_2^{(j)}}=\eta_{\OR\Z_2^{(k)}}=0$ with $(ijk)$ some permutation of $(123)$
in table~\ref{tab:Bulk-RR+SUSY-Z2Z6p} if $\Z_2^{(i)} \subset \Z_6$ is the surviving $\Z_2$ twisted  sector in the truncation procedure.

In~\cite{Honecker:2004kb}, a priori six different background lattices for the $T^6/(\Z_6 \times \OR)$ background of Type IIA string theory compactifications  
with D6-branes were identified. 
While for the {\bf AAA} and {\bf BBB} lattice orientations all choices of the two-torus index $i$ are equivalent, it is necessary in the truncation procedure
to consider at least two different values of $i \in \{1,2,3\}$
in order to reproduce the global consistency conditions for the remaining four background lattices. 
The result is summarised for all $T^6/(\Z_6 \times \OR)$ backgrounds in table~\ref{tab:Bulk-RR+SUSY-Z6}
together with the massless closed string spectrum.
\mathtabfix{
\begin{array}{|c||r|r|r||c|c|c|}\hline
\multicolumn{7}{|c|}{\text{\bf Bulk RR tadpole cancellation and massless closed string spectrum on $T^6/(\Z_6 \times \OR)$ with $SU(3)^3$ lattice}}
\\\hline\hline
\text{lattice} 
&  \begin{array}{c} (1): \quad \Z_2^{(1)} \subset \Z_6 \\\eta_{\OR}=\eta_{\OR\Z_2^{(1)}}=1 \\ \eta_{\OR\Z_2^{(2)}}=\eta_{\OR\Z_2^{(3)}}=0 \end{array} 
& \begin{array}{c} (2): \quad \Z_2^{(2)} \subset \Z_6  \\ \eta_{\OR}=\eta_{\OR\Z_2^{(2)}}=1 \\ \eta_{\OR\Z_2^{(1)}}=\eta_{\OR\Z_2^{(3)}}=0 \end{array}  
&  \begin{array}{c} (3) : \quad \Z_2^{(3)} \subset \Z_6 \\ \eta_{\OR}=\eta_{\OR\Z_2^{(3)}}=1 \\ \eta_{\OR\Z_2^{(1)}}=\eta_{\OR\Z_2^{(2)}}=0 \end{array}
&   \begin{array}{c} (1) \\ (h_{11}^+,h_{11}^-)   \end{array} &  \begin{array}{c} (2) \\ (h_{11}^+,h_{11}^-)  \end{array} &  \begin{array}{c} (3) \\ (h_{11}^+,h_{11}^-)  \end{array}
\\\hline\hline
{\bf AAA} & \multicolumn{3}{|c||}{\sum_a N_a \left(2 \,X_a+Y_a \right)= 16} & \muc{3}{|c|}{ (8,21) } 
\\\hline
{\bf ABB} & \sum_a N_a \left(X_a+  2\, Y_a \right)= 16 & \multicolumn{2}{|c||}{ \sum_a N_a \left(X_a+  2\, Y_a \right)=  24} & (8,21) & \muc{2}{|c|}{ (5,24) }
\\\hline\hline
{\bf AAB} &  \multicolumn{2}{|c|}{ \sum_a N_a \left( X_a+Y_a \right)=  8}  & \sum_a N_a \left( X_a+Y_a \right)= 16 & \muc{2}{|c|}{ (9,20) } & (2,27)
\\\hline
{\bf BBB} & \multicolumn{3}{|c||}{\sum_a N_a \, Y_a = 16} & \muc{3}{|c|}{ (2,27) } 
\\\hline
\end{array}
}{Bulk-RR+SUSY-Z6}{Bulk RR tadpole cancellation conditions for $T^6/(\Z_6 \times \OR)$ backgrounds for all choices of $\Z_2^{(i)} \subset \Z_6$
obtained by truncating the $T^6/(\Z_2 \times \Z_6' \times \OR)$ conditions  in table~\protect\ref{tab:Bulk-RR+SUSY-Z2Z6p}.
The supersymmetry conditions remain unchanged. In the last three columns, the Hodge number $h_{11}=h_{11}^+ + h_{11}^-$  for the counting of massless closed string 
vectors and K\"ahler moduli as first computed in~\cite{Honecker:2004kb} is reproduced with $h_{21}=h_{21}^{\Z_2^{(i)}}=5$ the number of complex structures. 
Both the global bulk consistency conditions and the massless closed string spectrum point to only four independent lattice orientations, as detailed in the text.}
Analogously to Type IIA string compactifications on $T^6/(\Z_2 \times \Z_6' \times \OR)$ with discrete torsion, the massless closed string spectrum and global bulk consistency conditions on $T^6/(\Z_6 \times \OR)$ 
suggest two pairwise relations
${\bf AAA} \leftrightarrow {\bf ABB}$ for $\Z_2^{(1)}\subset \Z_6$ and ${\bf AAB} \leftrightarrow {\bf BBB}$ for $\Z_2^{(3)} \subset \Z_6$,  leaving only four 
independent lattice backgrounds, as will now be proven for the first pair.
The analogous discussion for the second pairing can be found in appendix~\ref{A:identifications_AAB+BBB}.

The twisted RR tadpole cancellation conditions for the $T^6/(\Z_6 \times \OR)$ orientifold are recovered in a similar fashion to the bulk ones by setting 
$\eta_{(i)} \equiv 1$ for one $\Z_2^{(i)} \subset \Z_6$ in tables~\ref{tab:twistedRR-AAA+ABB-Z2Z6p} and~\ref{tab:twistedRR-AAB+BBB-Z2Z6p} of the
twisted RR tadpole  cancellation conditions on $T^6/(\Z_2 \times \Z_6' \times \OR)$ 
and truncating the remaining two sets of $\Z_2^{(j)}$ and $\Z_2^{(k)}$ twisted RR tadpole conditions for a given lattice orientation. 
The $\Z_2^{(1)}$ twisted RR tadpole cancellation conditions  for the lattices under consideration read
\begin{equation}
\begin{array}{ll}
 {\bf AAA} & {\bf ABB}\\
\sum_a N_a \; y_{\alpha,a}^{(1)}=0, &\sum_a N_a  \; \ov{x}_{\alpha,a}^{(1)} =0,
\qquad\qquad 
\text{ for }
\quad 
\alpha=1,2,3
\\
\sum_a N_a \left[ (2 \, x^{(1)}_{4,a} + y^{(1)}_{4,a})  - (2 x^{(1)}_{5,a} +  y^{(1)}_{5,a}) \right]=0,
 &\sum_a N_a  \left[ ( \ov{x}^{(1)}_{4,a}+ 2 \, \ov{y}^{(1)}_{4,a})- (  \ov{x}^{(1)}_{5,a} + 2 \, \ov{y}^{(1)}_{5,a} )  \right] =0,
\\
 \sum_a N_a \left[ y^{(1)}_{4,a} +  y^{(1)}_{5,a} \right]=0,
& \sum_a N_a  \left[ \ov{x}^{(1)}_{4,a}+ \ov{x}^{(1)}_{5,a} \right]=0,
\end{array}
\end{equation}
with the exceptional wrapping numbers $(x^{(1)}_{\alpha,a},y^{(1)}_{\alpha,a})$ for {\bf AAA} and $(\ov{x}^{(1)}_{\alpha,a},\ov{y}^{(1)}_{\alpha,a})$ for {\bf ABB}
each of the shape of one of the six pairs in table~\protect\ref{Tab:xy-entries-Z2Z6p}, see  table~\protect\ref{tab:exceptional-wrappings-Z2Z6} in 
appendix~\protect\ref{A:tables_bulk-cycles} for more details.

The first thing to notice is that the two-torus index $i=1$ will be preserved under any identification of wrapping numbers $(n^i_a,m^i_a)$ on the {\bf AAA} and 
$(\ov{n}^i_a,\ov{m}^i_a)$ on the {\bf ABB} lattice since $\Z_2^{(1)} \subset \Z_6$ for both. Secondly, preserving the length $L_a$ of a given bulk three-cycle 
requires that each one-cycle length $L_a^{(i)}$ remains unchanged, which in turn limits the transformations  of one-cycle 
wrapping numbers to rotations by $\pm \pi/3$ as in equation~(\ref{Eq:1-cycle-orbits}) combined with reflexions along the ${\cal R}$-invariant axis as given in 
equation~(\ref{Eq:OR_on_n+m}).
The unique solution\footnote{The solution is unique up to the permutation of two-torus indices on $T^2_{(2)}$ and $T^2_{(3)}$. This permutation does, however,
not provide physically distinct compactifications,
but is technically more involved due to the non-trivial exchange of the labels $\alpha =1 \ldots 5$ of exceptional three-cycles, e.g. 
$ \varepsilon^{(1)}_{1} \stackrel{{\bf AAA} \leftrightarrow {\bf ABB}}{\longleftrightarrow}  \varepsilon^{(1)}_{2}$
according to the definition of fixed point orbits in table~\ref{tab:Z2Z6p-fps+excycles}.} 
for the one-cycle wrapping numbers - up to choosing a different orbifold and orientifold image of each toroidal three-cycle - 
respecting the identification of bulk wrapping numbers in equation~(\ref{Eq:1st_hint_identifications}) 
and permuting the O6-plane orbits accordingly turns out to be 
\begin{equation}\label{Eq:identify_nm_AAA+ABB}
\left(\begin{array}{cc} n^1_a & m^1_a  \\ n^2_a  & m^2_a  \\ n^3_a  & m^3_a   \end{array}\right)_{\bf AAA}
=\left(\begin{array}{cc} \ov{n}^1_a  + \ov{m}^1_a  & - \ov{n}^1_a  \\ \ov{n}^2_a  & \ov{m}^2_a  \\ \ov{n}^3_a  & \ov{m}^3_a   \end{array}\right)_{\bf ABB}
.
\end{equation}
Using the general form of exceptional wrapping numbers in table~\ref{Tab:xy-entries-Z2Z6p} or the complete set in table~\protect\ref{tab:exceptional-wrappings-Z2Z6} of appendix~\protect\ref{A:tables_bulk-cycles}
in turn implies
\begin{equation}\label{Eq:mapping_x+y_AAA_to_ABB}
\left( x_{\alpha,a}^{(1)} \, , \,  y_{\alpha,a}^{(1)} \right)_{\bf AAA} = \left( \ov{x}_{\alpha,a}^{(1)} + \ov{y}_{\alpha,a}^{(1)} \, , \, -\ov{x}_{\alpha,a}^{(1)} \right)_{\bf ABB}
\quad  \text{ for each } \quad 
\alpha \in \{1 \ldots 5\}
,
\end{equation}
and the matching of the $\Z_2^{(1)}$ twisted RR tadpole cancellation conditions in table~\ref{tab:twistedRR-AAA+ABB-Z2Z6p}
is obviously fulfilled line by line, if the $\Z_2^{(1)}$ eigenvalues and  discrete displacement and Wilson line parameters $(\sigma^2_a;\sigma^3_a)$ 
and $(\tau^2_a;\tau^3_a)$ of each D6-brane remain unchanged under the mapping ${\bf AAA} \leftrightarrow {\bf ABB}$.
This completes the matching of global consistency conditions for the first pair.

It is easy to show that for the mappings  of bulk and exceptional  wrapping numbers in equations~(\ref{Eq:1st_hint_identifications}) and~(\ref{Eq:mapping_x+y_AAA_to_ABB}),
the intersection numbers and thus net-chiralities of complex representations,
\begin{equation}\label{Eq:matching_intersections_Z6(1)}
\chi^{(\N_a,\ov{\N}_b)} \equiv \Pi_a^{\text{frac}} \circ \Pi_b^{\text{frac}} = \frac{1}{2} \left(X_a Y_b - Y_a X_b - \sum_{\alpha=1}^5 \left[  x_{\alpha,a}^{(1)} y_{\alpha,b}^{(1)}  -y_{\alpha,a}^{(1)} x_{\alpha,b}^{(1)}  \right] \right)
\quad
\text{ on }
\quad 
T^6/(\Z_6 \times \OR)
,
\end{equation}
are preserved as well.

The matching of non-chiral matter spectra and field theory results at one-loop will be discussed in the next section~\ref{Sss:T6Z2Z6p_identifications}
in the context of the full $T^6/(\Z_2 \times \Z_6' \times \OR)$ orientifold with discrete torsion.

\subsubsection{Identification of three-cycle lattices for $T^6/(\Z_2 \times \Z_6' \times \OR)$ with $\eta=-1$
}\label{Sss:T6Z2Z6p_identifications}

The matching of the global consistency conditions and contributions to intersection numbers of the 
 bulk and $\Z_2^{(1)}$ twisted sectors of $T^6/(\Z_2 \times \Z_6' \times \OR)$ on the {\bf AAA} and {\bf ABB} lattices is identical to the matching presented in 
the previous section~\ref{Sss:T6Z6_identifications}  for $T^6/(\Z_6 \times \OR)$ with $\Z_2^{(1)} \subset \Z_6$. It is straightforward to check that the matching 
prescription~(\ref{Eq:mapping_x+y_AAA_to_ABB}) of exceptional wrapping numbers also respects the matching of $\Z_2^{(1)}$ twisted tadpoles  in 
table~\ref{tab:twistedRR-AAA+ABB-Z2Z6p} for the choice $\eta_{(1)}=-1$ of the orientifold projection in the presence of an exotic O6-plane orbit.

The matching of the $\Z_2^{(2)}$ and $\Z_2^{(3)}$ twisted RR tadpole cancellation conditions and contributions to intersection numbers
of fractional D6-branes is shown as follows.
According to the matching prescription~(\ref{Eq:identify_nm_AAA+ABB}) for toroidal one-cycle wrapping numbers, the $\Z_2$ fixed points along $T^2_{(1)}$ 
in table~\ref{Tab:Reference_Point+Signs} and corresponding indices $\alpha$ of exceptional three-cycles in table~\ref{tab:Z2Z6p-fps+excycles} are permuted as follows,
\begin{equation}
\begin{array}{ccc}
\begin{array}{c}
(\ov{n}^1_a \, , \, \ov{m}^1_a )_{\bf ABB}\to (n^1_a \, , \, m^1_a)_{\bf AAA}
\\
\text{(odd,odd)}_{\bf ABB} \to \text{(even,odd)}_{\bf AAA} \\
 \text{(even,odd)}_{\bf ABB} \to \text{(odd,even)}_{\bf AAA} \\  \text{(odd,even)}_{\bf ABB} \to \text{(odd,odd)}_{\bf AAA}
\end{array}
&
\begin{array}{c}
\text{f.p. on } T^2_{(1)} \times T^2_{(2)} 
\\
61 \to 51\\ 51 \to 41\\  41 \to 61 \\ \quad 64 \to 54
\end{array}
&
\begin{array}{c}
\ov{\alpha}=1 \to \alpha=1\\
\quad 2 \to 2\\
\quad 3 \to 4\\
\quad 4 \to 5\\
\quad 5 \to 3\\
\end{array}
\end{array}
.
\end{equation}
This provides the matching of exceptional wrapping numbers in the $\Z_2^{(3)}$ twisted sector, 
\begin{equation}
\begin{aligned}
\left( x^{(3)}_{\alpha,a} \, , \, y^{(3)}_{\alpha,a} \right)_{\bf AAA} &= \left\{\begin{array}{cc} 
\left( \ov{y}^{(3)}_{\ov{\alpha},a} \, , \, - ( \ov{x}^{(3)}_{\ov{\alpha},a} + \ov{y}^{(3)}_{\ov{\alpha},a} ) \right)_{\bf ABB}
& \alpha=(1,3,4,5); \, \ov{\alpha}=(1,5,3,4)
\\ \left( \ov{x}^{(3)}_{\ov{\alpha},a} \, , \, \ov{y}^{(3)}_{\ov{\alpha},a} \right)_{\bf ABB}  & \alpha=\ov{\alpha}=2
\end{array}\right.
,
\end{aligned}
\end{equation}
as can be verified on a case-by-case basis for each combination of displacement parameters $(\sigma^1_a;\sigma^2_a)$ 
using table~\ref{tab:exceptional-wrappings-Z2Z6} of appendix~\ref{A:tables_bulk-cycles}.
 
Combining with the  in caption of table~\ref{Tab:ClosedSpectrum_Z2Z6p} anticipated flip of the orientifold projection in the $\Z_2^{(3)}$ twisted sector, 
\begin{equation}
\left(\eta_{(3)}\right)_{\bf AAA} =   \left( - \eta_{(3)}\right)_{\bf ABB}
,
\end{equation}
one can explicitly verify that the $\Z_2^{(3)}$ twisted tadpoles in table~\ref{tab:twistedRR-AAA+ABB-Z2Z6p} are indeed mapped onto each other, provided 
that the Wilson line and displacement parameters $(\vec{\tau}_a,\vec{\sigma}_a)$ do not transform. 

It is again straightforward to check that the contribution  to the intersection number~(\ref{Eq:def_frac_cycle}) from this twisted sector
$\frac{1}{4}\Pi_a^{\Z_2^{(3)}} \circ \frac{1}{4}\Pi_b^{\Z_2^{(3)}}=-\frac{ x_{\alpha,a}^{(3)} y_{\alpha,b}^{(3)}  -y_{\alpha,a}^{(3)} x_{\alpha,b}^{(3)}}{4}  $ 
 is preserved under the mapping.\\
The matching of global consistency conditions and intersection numbers in the $\Z_2^{(2)}$ twisted sector is obtained by permuting the two-torus indices
$2 \leftrightarrow 3$.

The validity of the matching of fractional D6-brane configurations on the {\bf AAA} and {\bf ABB} lattices can be further extended to the full {\bf massless spectra},
which includes the counting of matter states in the adjoint representations, or symmetric and antisymmetric representations for $SO(2N)$ and $USp(2N)$ gauge groups,
as well as non-chiral pairs of bifundamental, symmetric and antisymmetric representations of $U(N)$. 
The total amount of massless open string states is most conveniently computed from the formulas for beta function coefficients derived 
in~\cite{Honecker:2011sm} by means of Conformal Field Theory methods for open string one-loop amplitudes
on $T^6/(\Z_2 \times \Z_{2M} \times \OR)$ (see also~\cite{Lust:2003ky,Akerblom:2007np,Blumenhagen:2007ip,Gmeiner:2009fb} 
for partial results) as briefly reviewed in appendix~\ref{A:intersections+betas}. 

For example, the field theoretical beta function coefficient of an ${\cal N}=1$ supersymmetric $SU(N_a)$ gauge group is given by
\begin{equation}\label{Eq:beta-coeffs}
\begin{aligned}
b_{SU(N)} = & \underbrace{N_a(-3 + \varphi^{\Adj_a})} + \underbrace{ \frac{N_a}{2} (\varphi^{\Sym_a} + \varphi^{\Anti_a})}
+  \underbrace{(\varphi^{\Sym_a} - \varphi^{\Anti_a})} + \underbrace{\sum_{b \neq a} \frac{N_b}{2} (\varphi^{(\N_a,\ov{\N}_b)} + 
\varphi^{(\N_a,\N_b)})}
\\
= & \hspace{3mm} \sum_{k=0}^2 b_{a(\omega^k a)}^{\cal A} 
\qquad + \quad \sum_{k=0}^2 b_{(\omega^k a)(\omega^k a)'}^{\cal A} 
\hspace{4mm} +\quad  \sum_{k=0}^2 b_{(\omega^k a)(\omega^k a)'}^{\cal M}
\quad + \quad \sum_{b\neq a} \sum_{k=0}^2 \left( b_{a(\omega^k b)}^{\cal A} +  b_{a(\omega^k b')}^{\cal A} \right)
,
\end{aligned}
\end{equation}
where $\varphi^{\Adj_a},\varphi^{\Sym_a}, \varphi^{\Anti_a},\varphi^{(\N_a,\ov{\N}_b)},\varphi^{(\N_a,\N_b)}$ denote the number
of chiral multiplets in a given representations (including its hermitian conjugate for complex representations). The decomposition into 
$b_{a(\omega^k b)}^{\cal A}$ and $b_{(\omega^k a)(\omega^k a)'}^{\cal M}$ on the second line denotes the origin from an open string amplitude
computation with Annulus or M\"obius strip topology, respectively, and endpoints of the open strings on D6-branes $a$ and $(\omega^k \, b)$
or $(\omega^k \, a)'$.
The sum runs over the three orbifold images of a given toroidal D6-brane $b$ or $a'$ on $T^6/(\Z_2 \times \Z_6' \times \OR)$.
The beta function coefficients only depend on relative intersection numbers among D6-branes and with O6-planes, e.g. for three non-vanishing angles
in all sectors $a(\omega^k \, b)_{k=0,1,2}$ one can decompose the intersection number of equation~(\ref{Eq:def_frac_cycle}) per sector,
\begin{equation}\label{Eq:ChiralMatter}
\chi^{(\N_a,\ov{\N}_b)} \equiv \Pi_a^{\text{frac}} \circ \Pi_b^{\text{frac}} = - \sum_{k=0}^2 \frac{ I_{a(\omega^k b)} + \sum_{i=1}^3 I_{a(\omega^k b)}^{\Z_2^{(i)}} }{4}
\equiv \sum_{m=0}^2 \chi^{a(\omega^m b)}
,
\end{equation}
and the total amount of matter at  $a(\omega^k \, b)_{k=0,1,2}$ intersections is counted by
\begin{equation}
\varphi^{(\N_a,\ov{\N}_b)} = \sum_{k=0}^2 \left| \chi^{a(\omega^k b)} \right| = \sum_{k=0}^2 \frac{ |I_{a(\omega^k b)} + \sum_{i=1}^3 I_{a(\omega^k b)}^{\Z_2^{(i)}} |}{4}
=\sum_{k=0}^2  \frac{2}{N_b}  \; b_{a(\omega^k b)}^{\cal A}
,
\end{equation}
where in the last equality the expression in terms of the corresponding Annulus contributions to the beta function coefficient
has been inserted.
Details for the more intricate cases with some vanishing angles are collected in appendix~\ref{A:intersections+betas}.

Since both relative intersection numbers, $I_{ab}^{(1)}$ for the toroidal and  $I_{ab}^{\Z_2^{(1)},(1)}$ for the fixed point part,
do not change under an overall rotation by $\pm \pi/3$ along $T^2_{(1)}$, each Annulus contribution to the beta function coefficient 
in the third column of table~\ref{tab:beta_coeffs_Kaehler_metrics} and thus the full, chiral plus non-chiral, adjoint and bifundamental matter 
spectrum is preserved for each sector $a(\omega^k \, b)_{k\in \{0,1,2\}}$ separately
under the map~(\ref{Eq:identify_nm_AAA+ABB}) among {\bf AAA} and {\bf ABB} lattices.

The case of intersections $a(\theta^k \, b')$ or $(\theta^k \, a)(\theta^k \, a)'$ involving some orientifold image D6-brane
is slightly more complicated due to the different orientifold projections~(\ref{Eq:OR_on_n+m}) along $T^2_{(2)} \times T^2_{(3)}$ for the {\bf AAA} and {\bf ABB}
lattices. From the toroidal one-cycle intersection numbers in table~\ref{tab:aaP-intersections} of appendix~\ref{A:intersections+betas}, one can read off 
the relations
\begin{equation}
\begin{aligned}
\left[I^{(1)}_{a(\omega^k \, b')} \right]_{\bf A} &= \left[\ov{I}^{(1)}_{\ov{a}(\omega^{k-1} \, \ov{b}')}  \right]_{\bf A}
,
\\
\left[I^{(i)}_{a(\omega^k \, b')} \right]_{\bf A} &= \left[- \ov{I}^{(i)}_{\ov{a}(\omega^{k-1} \, \ov{b}')}  \right]_{\bf B}
\qquad \text{on} \quad T^2_{(2)} \times T^2_{(3)}
,
\end{aligned}
\end{equation}
where $I^{(i)}_{a(\omega^k \, b')}$ denotes intersection numbers in terms of $(n^i_a,m^i_a)$ on the {\bf AAA} lattice and $\ov{I}^{(i)}_{\ov{a}(\omega^{k-1} \, \ov{b}')}$
in terms of $(\ov{n}^i_a,\ov{m}^i_a)$ on the {\bf ABB} lattice with the relation~(\ref{Eq:identify_nm_AAA+ABB}) among the two sets of one-cycle wrapping numbers.
The analogous relations for $I_{a(\omega^k \, b')}^{\Z_2^{(i)},(i)}$ and $\ov{I}_{\ov{a}(\omega^{k-1} \, \ov{b}')}^{\Z_2^{(i)},(i)}$  can be derived on a case-by-case
basis using table~\ref{tab:Wilson_signs} of appendix~\ref{A:intersections+betas}. 
\\
As a result, the $a(\omega^k \, b')$ sectors for $k=0,1,2$ on the {\bf AAA} lattice are permuted to $k=2,0,1$ on the {\bf ABB} lattice,
and the total amount of matter in the $(\N_a,\N_b)$ or $(\ov{\N}_a,\ov{\N}_b)$ representations is identical.

This can also be verified by inspection of the intersection angles. From the mapping~(\ref{Eq:identify_nm_AAA+ABB}) of one-cycle wrapping numbers,
the angles of D6-brane $a$ with respect to the $\OR$-invariant plane are deduced,
\begin{equation}\label{Eq:trafo-angles_AAA+ABB}
\begin{aligned}
{\bf AAA}  \quad & \leftrightarrow  \quad {\bf ABB}
\\
\pi \left(\phi^{(1)}_a, \phi^{(2)}_a, \phi^{(3)}_a \right)  \quad & =  \quad \pi  \bigl(\ov{\phi}^{(1)}_{\ov{a}},  \ov{\phi}^{(2)}_{\ov{a}}, \ov{\phi}^{(3)}_{\ov{a}} \bigr) + \pi \bigl(-\frac{1}{3},  \frac{1}{6},\frac{1}{6} \bigr)
,
\end{aligned}
\end{equation}
and relative angles among D6-branes $a$ and $(\omega^k \, b)$ are preserved,
\begin{equation}\label{Eq:ab-angles}
\begin{aligned}
\pi \bigl( \vec{\phi}_{a(\omega^k \, b)} \bigr)_{\bf AAA} \equiv &  \pi  \bigl( \vec{\phi}_{(\omega^k \, b)} -  \vec{\phi}_{a} \bigr)_{\bf AAA}  = \pi  \bigl( \vec{\phi}_{b} - \vec{\phi}_{a} \bigr)_{\bf AAA}  + k \, \pi \bigl(-\frac{2}{3},\frac{1}{3},\frac{1}{3}  \bigr)
\\
=&  \pi \bigl( \vec{\ov{\phi}}_{\ov{a}(\omega^k \, \ov{b})} \bigr)_{\bf ABB}
.
\end{aligned}
\end{equation}
For intersections of $a$ with orientifold images $(\omega^k \, b')$, one obtains instead
\begin{equation}\label{Eq:abP-angles}
\begin{aligned}
\pi \bigl( \vec{\phi}_{a(\omega^k \, b')} \bigr)_{\bf AAA} \equiv & \pi  \bigl( \vec{\phi}_{(\omega^k \, b')} -  \vec{\phi}_{a} \bigr)_{\bf AAA}  = \pi  \bigl(- \vec{\phi}_{b} - \vec{\phi}_{a} \bigr)_{\bf AAA}  + k \, \pi \bigl(-\frac{2}{3},\frac{1}{3},\frac{1}{3}  \bigr)
\\
= &  \pi \bigl( \vec{\ov{\phi}}_{\ov{a}(\omega^k \, \ov{b}')} \bigr)_{\bf ABB} - \pi \bigl(-\frac{2}{3},\frac{1}{3},\frac{1}{3}  \bigr) 
=  \pi \bigl( \vec{\ov{\phi}}_{\ov{a}(\omega^{k-1} \, \ov{b}')} \bigr)_{\bf ABB}
\end{aligned}
\end{equation}
in accord with the relations among intersection numbers on the pair of lattice orientations.

Last but not least, the net-chirality of symmetric and antisymmetric representations is given by 
\begin{equation}\label{Eq:ChiralMatter-Sym+Anti}
\begin{aligned}
\chi^{\Anti_a/\Sym_a}
& \equiv 
\frac{\Pi_a^{\text{frac}} \circ \Pi_{a'}^{\text{frac}} \pm \Pi_a^{\text{frac}} \circ \Pi_{O6}}{2}\\
 &= - \sum_{m=0}^2 \frac{ \left(I_{(\omega^m a)(\omega^m a)'} + \sum_{i=1}^3 I_{(\omega^m a)(\omega^m a)'}^{\Z_2^{(i)}} \right)
\pm \left( \sum_{n=0}^3 \eta_{\OR\Z_2^{(n)}} \, \tilde{I}_{(\omega^m a)}^{\OR\Z_2^{(n)}} \right) }{8}
\\
& \equiv \sum_{m=0}^2 \left( \chi^{(\omega^m a)(\omega^m a)'} \pm \chi^{(\omega^m a)\OR}  \right)
,
\end{aligned}
\end{equation}
and for non-vanishing angles everywhere, the total amount of (chiral plus non-chiral) symmetric and antisymmetric representations
reads
\begin{equation}
\varphi^{\Anti_a/\Sym_a}=\sum_{m=0}^2 \left| \chi^{(\omega^m a)(\omega^m a)'} \pm \chi^{(\omega^m a)\OR} \right|
.
\end{equation}
The equal amount of representations of each kind on the {\bf AAA} and {\bf ABB} lattices under the identifications of
one-cycle wrapping numbers~(\ref{Eq:identify_nm_AAA+ABB}) and permutation of exotic O6-planes~(\ref{Eq:1st_hint_identifications}) is obtained from the relations
\begin{equation}
\begin{aligned}
\left[ \tilde{I}_{(\omega^k \, a)}^{\OR,(1)} \right]_{\bf A} &=\left[  - \tilde{\ov{I}}_{(\omega^{k-1} \, \ov{a})}^{\OR,(1)}   \right]_{\bf A}
,\qquad 
\left[ \tilde{I}_{(\omega^k \, a)}^{\OR\Z_2,(1)}  \right]_{\bf A} =  \left[  - \tilde{\ov{I}}_{(\omega^{k-1} \, \ov{a})}^{\OR\Z_2,(1)}  \right]_{\bf A}
,
\\
\left[ \tilde{I}_{(\omega^k \, a)}^{\OR,(i)} \right]_{\bf A} &= \left[  \tilde{\ov{I}}_{(\omega^{k-1} \, \ov{a})}^{\OR\Z_2,(i)}  \right]_{\bf B},
\qquad 
\left[ \tilde{I}_{(\omega^k \, a)}^{\OR\Z_2,(i)}  \right]_{\bf A} = \left[  - \tilde{\ov{I}}_{(\omega^{k-1} \, \ov{a})}^{\OR,(i)}   \right]_{\bf B}
\qquad \text{on} \quad T^2_{(2)} \times T^2_{(3)}
,
\end{aligned}
\end{equation}
with explicit expressions given in table~\ref{tab:aaP-intersections} of appendix~\ref{A:intersections+betas}.
The validity of the mapping is again extended to cases with some vanishing angles in a straightforward manner using the corresponding expressions for 
beta function coefficients in table~\ref{tab:beta_coeffs_Kaehler_metrics} of appendix~\ref{A:intersections+betas}.

As briefly reviewed in appendix~\ref{A:intersections+betas}, string one-loop vacuum amplitudes on factorisable toroidal orbifolds only depend on
the {\it relative} intersection numbers and angles as well as the lengths of one-cycles. They are thus - up to the permutation of orbifold images $(\omega^k \, b')$ - 
identical on the pairwise related lattices.
Besides the total amount of matter multiplets and corresponding beta function coefficients, this implies that the perturbatively exact holomorphic gauge kinetic function
and the K\"ahler metrics at leading order displayed in tables~\ref{tab:hol-gauge-kin}
and~\ref{tab:beta_coeffs_Kaehler_metrics}, respectively, are identical on the {\bf AAA} and {\bf ABB} lattice orientations
for pairs of D6-brane configurations related by~(\ref{Eq:identify_nm_AAA+ABB}) and~(\ref{Eq:1st_hint_identifications}).

Since it has been shown in the present section for ${\bf AAA} \leftrightarrow {\bf ABB}$ and in appendix~\ref{A:identifications_AAB+BBB} for ${\bf AAB} \leftrightarrow {\bf BBB}$
that at the present level of knowledge of the matter spectra and perturbative field theory results only two lattices yield physically distinguishable results, 
the focus of the remainder of this article will be on the {\bf AAA} and {\bf BBB} lattices.


\section{D6-Brane Model Building}\label{S:Modelbuilding}

In this section, searches for $USp(2)$ gauge factors are performed, which are on the one hand used as probe branes 
for testing the K-theory constraint on a given globally defined D6-brane model~\cite{Uranga:2002vk}
and on the other hand provide potential $SU(2)_L$ or $SU(2)_R$ gauge factors.
It is furthermore shown that only a small number of `short' fractional three-cycles with special choices of displacement 
parameters $(\vec{\sigma})$ and Wilson lines $(\vec{\tau})$ provides completely rigid D6-branes without any
adjoint representation. The occurrence of D6-branes without matter in symmetric or antisymmetric representations
is equally constrained, and finally the conditions from demanding three Standard Model or GUT particle generations are given.
The content of this section is required for the model building in section~\ref{S:model_building}.

\subsection{Classification of $USp(2M)$ and $SO(2M)$ Enhancements}\label{Ss:Gaugeenhanc}

Enhancements of gauge groups from $U(M)$ to $SO(2M)$ or $USp(2M)$ occur iff a stack of D6-branes wraps an orientifold invariant 
three-cycle. In~\cite{Forste:2010gw}, it was shown for all factorisable $T^6/(\Z_2 \times \Z_{2N \in \{2,6,6'\}} \times \OR)$ orbifolds with discrete torsion
that the conditions on orientifold invariance can be concisely expressed in terms
of bulk parts parallel to some exotic or ordinary O6-plane orbit combined with untilted or tilted tori, $b_i=0,\frac{1}{2}$ respectively, and for the 
latter with certain combinations of discrete Wilson lines $(\vec{\tau})$ and displacements $(\vec{\sigma})$ as displayed in table~\ref{Tab:OR-inv-branes}.
\begin{table}
\begin{center}
\begin{equation*}
\begin{array}{|c|c|}\hline
\muc{2}{|c|}{\text{\bf $\OR$ inv. three-cycles on $T^6/(\Z_2 \times \Z_{2N} \times \OR)$ ($\eta=-1$)}}
\\\hline\hline
a\pp \text{orbit} & 
 \left(\eta_{(1)},\eta_{(2)},\eta_{(3)} \right) \stackrel{!}{=} 
\\\hline\hline
\OR & \!\!\!\left( - (-1)^{\delta^2_a + \delta^3_a} , - (-1)^{\delta^1_a + \delta^3_a} , - (-1)^{\delta^1_a + \delta^2_a} \right)\!\!\!
\\
\OR\Z_2^{(1)} &  \left( - (-1)^{\delta^2_a + \delta^3_a} , (-1)^{\delta^1_a + \delta^3_a} , (-1)^{\delta^1_a + \delta^2_a} \right)
\\
\OR\Z_2^{(2)} &  \left( (-1)^{\delta^2_a + \delta^3_a} , - (-1)^{\delta^1_a + \delta^3_a} ,  (-1)^{\delta^1_a + \delta^2_a} \right)
\\
\OR\Z_2^{(3)} &  \left(  (-1)^{\delta^2_a + \delta^3_a} ,  (-1)^{\delta^1_a + \delta^3_a} , - (-1)^{\delta^1_a + \delta^2_a} \right)
\\\hline
\end{array}
\end{equation*}
\end{center}
\caption{ 
Topological condition for a fractional D6-brane $a$ on $T^6/(\Z_2 \times \Z_{2N} \times \OR)$ with $2N \in \{2,6,6'\}$ and $\eta=-1$
to wrap an orientifold invariant three-cycle with 
$\delta^i_a \equiv 2b_i \, \sigma^i_a \tau^i_a \in \{0,1\}$~\cite{Forste:2010gw}.
For $T^6/(\Z_2 \times \Z_2 \times \OR)$  on square tori ($b_1=b_2=b_3=0$), D6-branes with arbitrary $(\vec{\sigma}_a;\vec{\tau}_a)$
parallel to the exotic O6-plane orbit support $USp(2M_a)$ gauge groups, while all other D6-branes are of $U(M_a)$ type~\cite{Blumenhagen:2005tn}.
For tilted tori ($b_1=b_2=b_3=\frac{1}{2}$), specific combinations of  $(\vec{\sigma}_a;\vec{\tau}_a)$ provide orientifold invariant D6-branes
parallel to each of the O6-plane orbits, roughly half of which support $USp(2M_a)$ and the other half $SO(2M_a)$ gauge groups as detailed in the text.
}
\label{Tab:OR-inv-branes}
\end{table}

For $T^6/(\Z_2 \times \Z_2 \times \OR)$ with discrete torsion on untilted tori ($b_1=b_2=b_3=0$) discussed in~\cite{Blumenhagen:2005tn}, the orientifold 
invariance constraints in  table~\ref{Tab:OR-inv-branes} can only be solved by fractional D6-branes parallel to the exotic O6-plane orbit, in this case 
with arbitrary discrete Wilson lines  $(\vec{\tau}_a)$ and displacements $(\vec{\sigma}_a)$. Each of the D6-brane stacks leads to the enhancement 
$U(M_a) \to USp(2M_a)$, as can be derived by considerations of Chan-Paton labels or via the beta function coefficients in table~\ref{tab:beta_coeffs_Kaehler_metrics}
 of appendix~\ref{A:intersections+betas}. The total amount of allowed $USp(2M_a)$ gauge factors is then $2^2 \cdot 2^3 \cdot 2^3=256$ with multiplicities due to 
the choices of $\Z_2$ eigenvalues, displacements and Wilson lines.

For $T^6/(\Z_2 \times \Z_6' \times \OR)$ with discrete torsion, all tori are tilted ($2b_1=2b_2=2b_3=1$), and the pattern of gauge enhancements
changes radically: there exist combinations of  $(\vec{\tau}_a)$ and  $(\vec{\sigma}_a)$ for D6-branes $a$ parallel to any exotic or ordinary O6-plane
orbit, for which the orientifold invariance constraint is solved. In~\cite{Honecker:2011sm}, it was shown that for  $(\vec{\sigma}_a)=(\vec{0})$
and the D6-branes on top of the exotic O6-plane orbit, the same gauge enhancement $U(M_a) \to USp(2M_a)$ occurs as for untilted tori.
However, e.g. for the choice $\eta_{\OR}=-1$ of exotic O6-plane, also D6-branes parallel to the ordinary $\OR\Z_2^{(1)}$-plane orbit
with $\sigma^1\tau^1 \neq \sigma^2\tau^2 = \sigma^3\tau^3$
are orientifold invariant according to the conditions in table~\ref{Tab:OR-inv-branes}, and the corresponding beta function coefficient
in table~\ref{tab:beta_coeffs_Kaehler_metrics}  of appendix~\ref{A:intersections+betas} reads
\begin{equation}\label{Eq:example-enhance}
\begin{aligned}
\frac{1}{2} \left(b^{\cal A}_{aa} + b^{\cal A}_{aa'} + b^{\cal M}_{aa'} \right)= &  -3 \, N_a -  \eta_{\OR} \, \left( (-1)^{\sigma^1_a\tau^1_a} + \eta_{(2)} \, (-1)^{\sigma^3_a \tau^3_a} + \eta_{(3)} \, (-1)^{\sigma^2_a \tau^2_a} \right)
\\
\stackrel{\eta_{\OR}=-1}{=} &   -3 \, N_a + \left( (-1)^{\sigma^1_a\tau^1_a}  - (-1)^{\sigma^3_a \tau^3_a} - (-1)^{\sigma^2_a \tau^2_a} \right)
\\
=&  -3 \, N_a  +  \left\{\begin{array}{cr}
-3  & \sigma^1_a\tau^1_a \neq \sigma^2_a\tau^2_a = \sigma^3_a\tau^3_a =1 \\
3  & 0
\end{array}\right.
\quad
\leadsto 
\left\{\begin{array}{c}
USp(2N_a) \\ SO(2N_a)
\end{array}\right.
,
\end{aligned}
\end{equation}
where the global factor 1/2 was inserted due to the different normalisation of beta function coefficients of $SO(2N_a)$ and $USp(2N_a)$ 
gauge groups compared to $U(N_a)$.
There exist $2^2 \cdot 3 \cdot 1 \cdot 1=12$ combinations of $\Z_2$ eigenvalues, displacements and Wilson lines leading to enhancements to
$USp(2N_a)$ and $4 \cdot 1 \cdot 3 \cdot 3=36$ combinations leading to $SO(2N_a)$ gauge factors instead.

The factor $(-1)^{\sigma^i_a\tau^i_a}$ was first advertised in~\cite{Forste:2010gw}, 
and it is indeed required in equation~(\ref{Eq:example-enhance})
to ensure the constant factor $\pm 3$. It originates from the momentum and winding sum in the M\"obius strip amplitude, 
which is well-known in the absence of Wilson lines $(\vec{\tau}_a)=(\vec{0})$, but - partly also due to missing 
T-dual results of the Type IIB/$\Omega$ string theory with $B$-field background 
to the Type IIA/$\OR$ string theory on tilted tori - has not yet been rigorously derived for arbitrary combinations of $(\vec{\tau}_a)$ 
and $(\vec{\sigma}_a)$ from first principles. 
Further details and examples supporting the conjecture of the factor $(-1)^{2 b_i \, \sigma^i_a\tau^i_a}$ are given in appendix~\ref{Aa:sublety_Moebius}.

For the choice $\eta_{\OR}=-1$ of exotic O6-plane and the orientifold invariant D6-brane orbit parallel to the $\OR$-invariant orbit, 
\begin{equation}
\frac{1}{2} \left(b^{\cal A}_{aa} + b^{\cal A}_{aa'} + b^{\cal M}_{aa'} \right)= - 3 N_a + \left\{\begin{array}{cr}
-3  &\sigma^i_a\tau^i_a = 0  \quad  \forall \, i \\ 3  &  \sigma^i_a\tau^i_a = 1 \quad  \forall \, i
\end{array}\right.
\quad
\leadsto 
\left\{\begin{array}{c}
USp(2N_a) \\ SO(2N_a)
\end{array}\right. .
\end{equation}
reproduces the known result for $(\vec{\sigma}_a,\vec{\tau}_a)=(\vec{0},\vec{0})$. The enhancement to $USp(2N_a)$ occurs for $2^2 \cdot 3^3=108$
choices of $\Z_2$ eigenvalues, displacements and Wilson lines, and the enhancement to $SO(2N_a)$ takes place for $2^2=4$ configurations.
The remaining combinations of $(\vec{\sigma}_a)$ and $(\vec{\tau}_a)$ with $\sigma^i_a\tau^i_a \neq\sigma^j_a\tau^j_a =\sigma^k_a\tau^k_a$
do not lead to an enhancement of the $U(N_a)$ gauge group, but to non-chiral matter in the symmetric or antisymmetric representation as shown in subsection~\ref{Ss:no-anti+sym}.

For the choice $\eta_{\OR}=-1$, which is for the {\bf AAA} lattice of particular interest due to its maximal rank 16 in equation~(\ref{Eq:max-rank}),
the results for gauge enhancements on D6-branes parallel to $\OR\Z_2^{(k)}$  with $k \in \{2,3\}$ are equivalent to the one for $\OR\Z_2^{(1)}$ discussed above.
In total, an enhancement to $USp(2N_a)$ occurs for $108 + 3 \times 12 = 144$ orientifold invariant fractional D6-branes and to $SO(2N_a)$ for 
$4 + 3 \times 36=112$ D6-branes. Similar results apply to the other choices of an exotic O6-plane orbit.

The number of D6-branes with $USp(2)$ gauge group is thus roughly reduced by a factor of two compared to $T^6/(\Z_2 \times \Z_2 \times \OR)$ 
with discrete torsion on untilted tori, facilitating the computation of K-theory constraints and constraining model building with 
\mbox{$USp(2)=SU(2)_L$} or $SU(2)_R$ gauge groups. 

Further constraints on model building from minimising the non-chiral matter content are discussed in the following sections~\ref{Ss:no-adjoints}
and~\ref{Ss:no-anti+sym}.

\subsection{Rigid D6-branes without Adjoint Matter}\label{Ss:no-adjoints}
One of the motivations to study model building on the orientifold $T^6/(\Z_2\times \Z_6'\times \Omega{\cal R})$ with discrete torsion arises from the 
fact that the fractional three-cycles in equation (\ref{Eq:def_frac_cycle}) are stuck at the $\Z_2 \times \Z_2$ fixed points. The usual 
three chiral multiplets in the adjoint representation, which are characteristic for bulk three-cycles, containing the position moduli of the D6-brane 
along the compact directions are projected out. However, this does not guarantee the complete absence of matter in the adjoint representation. 
New matter in the adjoint representation can arise at the intersections between a cycle and its orbifold images, i.e. in the $a(\omega^k \, a)_{k = 1,2}$ sectors. 
These new multiplets in the adjoint representation encode the deformation moduli of the intersection point with some {\it vev} leading to recombination of the
orbifold images. 

In a slight abuse of notation, the expressions for net-chiralities~(\ref{Eq:ChiralMatter}) per intersection sector,
 \begin{equation}
0= \chi^{\Adj_a} \equiv \Pi_a^{\text{frac}} \circ \Pi_a^{\text{frac}} = \sum_{k=0}^2 \chi^{a\, (\omega^k \, a)}
\quad \text{ with }\quad 
\chi^{a\,a}=0,  \quad
\chi^{a\, (\omega \, a)}= -\chi^{a\, (\omega^2 \, a)} 
,
\end{equation}
can be applied to the adjoint representation as well. The amount of matter in the adjoint representation at intersections of orbifold images is given by
\begin{equation}
\varphi^{\Adj_a} = |\chi^{a\, (\omega \, a)}| = |\chi^{a\, (\omega^2 \, a)}| 
,
\end{equation}
as can be seen by comparing with the expression for the contribution to the beta function coefficient in table~\ref{tab:beta_coeffs_Kaehler_metrics} 
at intersection angles $(\vec{\phi}_{a(\omega^k \, a)})= \pm \pi \, \bigl( - \frac{2}{3},\frac{1}{3},\frac{1}{3}\bigr)$.
The condition on the absence of any matter in the adjoint representation can now be written as ($m \in \{1,2\}$):
\begin{equation}\label{Eq:constrain-no-adjoints}
\begin{aligned}
I_{a(\omega^m a)} + \sum_{k=1}^3 I_{a(\omega^m a)}^{\Z_2^{(k)}} &=0 \qquad \Leftrightarrow \qquad 1 + \sum_{i<j} \frac{1}{p_i p_j} =0
\\
\text{ with } \quad 
p_i &\equiv \frac{(-1)^m I_{a(\omega^m a)}^{(i)}}{ I_{a(\omega^m a)}^{\Z_2^{(i)},(i)}} =(-1)^{\sigma^i_a \tau^i_a} 
\left[ (n^i_a)^2 + n^i_a m^i_a + (m^i_a)^2 \right]
\in \Z_{\text{odd}}
.
\end{aligned}
\end{equation}
The odd integer $p_i$ is related to the (length)${}^2$ $(L^{(i)}_a)^2$ of the one-cycle on the $i$-th torus $T^2_{(i)}$ measured in units of the 
radius $r_i$ as detailed in the caption of table~\ref{Tab:Z2Z6p-ORonBulkO6plane-bulk}. 
Clearly, the relation on the right-hand side can only be satisfied if $|p_i| = 1$ for all $i \in \{1,2,3\}$, or if at most one one-cycle is 
characterised by $|p_i| \geq 3$, while the other two one-cycles are constrained to have $p_j= -p_k = 1$ (with $(ijk)$ some permutation of $(123)$).
This implies that only the shortest cycles on the {\bf AAA} and the {\bf BBB} lattices have a chance of being free of adjoint matter
as can be read off from the toroidal wrapping numbers in tables~\ref{Tab:all-cycles-AAA} and~\ref{Tab:all-cycles-BBB}, respectively, with shortest one-cycles 
implying 
$(n^i,m^i) \in \{(\pm 1,0),(0,\pm 1),(\pm 1,\mp 1)\}$. 
Whether or not adjoint matter occurs in the $a(\, \omega^k a)_{k = 1,2}$ sectors for these shortest three-cycles still depends on the displacements $(\vec{\sigma_a})$ and the discrete 
Wilson lines $(\vec{\tau_a})$, but is independent of the choice of $\Z_2$ eigenvalues $(-1)^{\tau^{\Z_2^{(i)}}_a}$. 
Taking for instance the two shortest cycles on the {\bf AAA} lattice, at  most one chiral multiplet in the adjoint representation can arise,
\begin{equation}
\left(\varphi^{\Adj_a} \right)_{\bf AAA} = \left\{ \begin{array}{ll} 0 &\sigma^i_a \tau^i_a \neq  \sigma^j_a \tau^j_a = \sigma^k_a \tau^k_a \in \{0,1\} \\
1 &  \sigma^i_a \tau^i_a =\sigma^j_a \tau^j_a = \sigma^k_a \tau^k_a \in \{0,1\}     \end{array}  \right.
,
\end{equation}
for any permutation  $(ijk)$ of $(123)$ and any choice of $\Z_2$ eigenvalues.
For each shortest cycle on the {\bf AAA} lattice there are $(2^2\cdot 1 \cdot 3 \cdot 3) \times 3 = 36 \times 3 = 144$ combinations free of adjoint representations, and $2^2\cdot ( 3\cdot 3 \cdot 3 +1\cdot 1 \cdot 1) = 108$ combinations with one chiral multiplet in the adjoint representation. 

For the {\bf BBB} lattice, the six shortest three-cycles carry no, one or two chiral multiplets in the adjoint representation
depending on the combinations of discrete Wilson lines and displacements,
\begin{equation}
\left(\varphi^{\Adj_a}\right)_{\bf BBB}  = \left\{ \begin{array}{ll} 0 & \sigma^i_a \tau^i_a \, \text{ arbitrary}, \; \sigma^j_a \tau^j_a  \neq \sigma^k_a \tau^k_a \in \{0,1\} 
\\ 1 & \sigma^i_a \tau^i_a \neq  \sigma^j_a \tau^j_a = \sigma^k_a \tau^k_a \in \{0,1\}  
\\ 2 & \sigma^i_a \tau^i_a =\sigma^j_a \tau^j_a = \sigma^k_a \tau^k_a \in \{0,1\} 
  \end{array}  \right.
,
\end{equation}
where $i$ labels the two-torus along which the one-cycle has next-to-shortest length. For each of the shortest three-cycles $(2^2\cdot 4 \cdot 3\cdot 1)\times 2 = 96$ combinations are free of adjoint representations, $2^2 \cdot ( 3\cdot 1\cdot 1 + 1 \cdot 3 \cdot 3 ) = 48$ fractional D6-branes provide one chiral multiplet and 
$2^2 \cdot(3\cdot 3 \cdot 3 + 1 \cdot 1 \cdot 1)=112$ fractional D6-branes two chiral multiplets in the adjoint representation. 

This completes the classification of completely rigid D6-branes on $T^6/(\Z_2 \times \Z_6' \times \OR)$.

\subsection{D6-branes without Symmetric and Antisymmetric Matter}\label{Ss:no-anti+sym}

As is well known, in an orientifold background  a fractional cycle $\Pi^{\text{frac}}_a$ is accompanied by its orientifold image $\Pi^{\text{frac}}_{a'}$,  
and matter in the symmetric or antisymmetric representation arises at intersections between the orientifold image three-cycles. 
In some cases, chiral matter in symmetric and antisymmetric representations
can be used for model building, e.g. antisymmetrics of $U(3)_a$ as candidates for right-handed quarks and symmetrics of $U(1)_c$ for right-handed electrons.

In the case of the orientifold $T^6/(\Z_2\times \Z_6'\times \Omega{\cal R})$, however,  the net-chirality of supersymmetric matter in symmetric and  antisymmetric representations will always be the same. This can be seen quite straightforwardly from the expression in equation~(\ref{Eq:ChiralMatter-Sym+Anti}). As there is only one supersymmetric bulk cycle, the intersection number of a fractional cycle with the O6-plane orbits always vanishes, i.e.~$\Pi^{\text{frac}}_a \circ \Pi_{O6} = 0$, and thus 
\begin{equation}\label{Eq:CSeqCA}
\chi^{\Sym_a} = \chi^{\Anti_a} = \frac{1}{2} \, \Pi_a^{\text{frac}} \circ \Pi_{a'}^{\text{frac}} .    
\end{equation}
This relation rules antisymmetric representations as Standard Model particles out and excludes $SU(5)$ GUT model building.

Besides the chiral matter in symmetric and antisymmetric representations, the possible appearance of non-chiral matter in these sectors
should also be taken into consideration. To determine the total amount of non-chiral matter in symmetric and antisymmetric representations,
the beta function coefficients displayed in table~\ref{tab:beta_coeffs_Kaehler_metrics} of appendix~\ref{A:intersections+betas} are used. 
To investigate the constraints on fractional three-cycles free of such matter, three different cases need to be distinguished
depending on the relative angle between D6-brane $a$ and its orientifold image $a'$:
\begin{itemize}
\item[\bf (1)] {\bf For three non-vanishing angles}
\end{itemize}
This is the general situation. In analogy to the condition on no matter in the adjoint representation~(\ref{Eq:constrain-no-adjoints}), neither
symmetric nor antisymmetric representations occur in a given sector if 
\begin{equation}\label{Eq:cond-no-anti+sym-3angles}
I_{(\omega^k \, a)(\omega^k \, a)'} + \sum_{l=1}^3 I_{(\omega^k \, a)(\omega^k \, a)'}^{\Z_2^{(l)}} =0 \quad \Leftrightarrow \quad  1 + \sum_{i<j} \frac{1}{q_i q_j} =0
\quad \text{ with } \quad q_i \equiv  \frac{I^{(i)}_{(\omega^k \, a)(\omega^k \, a)'}}{I^{\Z_2^{(i)},(i)}_{(\omega^k \, a)(\omega^k \, a)'}} \in \Z
,
\end{equation}
with  the intersection numbers $I^{(i)}_{(\omega^k \, a)(\omega^k \, a)'}$ given in table~\ref{tab:aaP-intersections} and 
$I^{\Z_2^{(i)},(i)}_{(\omega^k \, a)(\omega^k \, a)'}$  in equation~(\ref{Eq:IaapZ2-values}) of appendix~\ref{A:intersections+betas}.
The integers $q_i$ can now also take even values, but the conclusion that at most one $|q_i|>1$ is identical to the constraint on the absence
of matter in the adjoint representation. This condition can at most be satisfied in all sectors $(\omega^k \, a)(\omega^k \, a)^{\prime}_{k=1,2}$ 
with non-vanishing angles
for the shortest cycles given in table~\ref{Tab:all-cycles-AAA}  for the {\bf AAA} lattice and table~\ref{Tab:all-cycles-BBB} for the {\bf BBB} lattice.

This situation occurs in particular in the $(\omega^k \, a) (\omega^k \, a)^{\prime}_{k=1,2}$ sectors for D6-branes $a$ parallel to some O6-plane $\OR$ or 
$\OR\Z_2^{(m)}$.
All one-cycle intersection numbers $I^{(i)}_{(\omega^k \, a)(\omega^k \, a)'}$ for these cases turn out to be odd, leaving only the dependence on the 
choice of exotic O6-plane $I^{\Z_2^{(i)},(i)}_{(\omega^k \, a)(\omega^k \, a)'}=-  \eta_{(i)}$. 
The amount of matter in the symmetric and antisymmetric representations is thus
independent of (combinations of) displacements $(\vec{\sigma}_a)$, discrete Wilson lines $(\vec{\tau}_a)$ and $\Z_2$ eigenvalues.

On the {\bf AAA} lattice, a stack of D6-branes parallel to the $\OR$-invariant plane yields
\begin{equation}
\sum_{k=1}^2 \left[b^{\cal A}_{(\omega^k a) (\omega^k a)'} + b^{\cal M}_{(\omega^k a) (\omega^k a)'} \right]= 
\left\{\begin{array}{cc}
2 \times \left[\frac{N_a}{2} -1 \right] & \eta_{\OR}=-1 \\
0 & \text{else}
\end{array}\right.
.
\end{equation}
The two multiplets in the antisymmetric representation for $\eta_{\OR}=-1$ have opposite chirality.

For the D6-branes parallel to the $\OR\Z_2^{(i)}$-invariant planes on the {\bf BBB} lattice, one finds the following contributions:
\begin{equation}
\sum_{k=1}^2 \left[b^{\cal A}_{(\omega^k a) (\omega^k a)'} + b^{\cal M}_{(\omega^k a) (\omega^k a)'} \right]= 
\left\{\begin{array}{cc}
2 \times \left[ \frac{N_a}{2} +1 \right] & \eta_{\OR}=-1  \\
2 \times \left[ N_a - 2 \right] & \eta_{\OR\Z_2^{(i)}}=-1
\\ 0 & \eta_{\OR\Z_2^{(j)}}=-1  \text{ or }  \eta_{\OR\Z_2^{(k)}}=-1
\end{array}\right.
.
\end{equation}
The two multiplets in the symmetric representation for $\eta_{\OR}=-1$ have opposite chirality. Also the net-chirality for the four multiplets in the 
antisymmetric representation for $\eta_{\OR \Z_2^{(i)}}=-1$ is zero, such that two multiplets have the opposite chirality with respect to the other two.

\begin{itemize}
\item[\bf (2)]{\bf For one vanishing angle $\phi^{(i)}_{aa'}=0$}
\end{itemize}
In this case, the condition to have no symmetric or antisymmetric representation reads
\begin{equation}
|I_{aa'}^{(j \cdot k)}| -  I_{aa'}^{\Z_2^{(i)},(j \cdot k)}=0 \quad \Leftrightarrow \quad  1  + \frac{1}{q_jq_k}=0
,
\end{equation}
with $q_j$ defined in equation~(\ref{Eq:cond-no-anti+sym-3angles}).
It can only be satisfied for $q_j = -q_k = 1$, in other words only if the D6-brane wraps the shortest possible one-cycles 
along $T^2_{(j)} \times T^2_{(k)}$. This is exactly the case for every sector of the one shortest three-cycle not parallel to an O6-plane
in table~\ref{Tab:all-cycles-AAA} on the {\bf AAA} lattice, and for one of the $(\omega^k a)(\omega^k a)'$ sectors for each shortest three-cycle 
in table~\ref{Tab:all-cycles-BBB} on the~{\bf BBB} lattice not parallel to some O6-plane orbit.

In contrast to the case with three non-vanishing angles, not only the choice of exotic O6-plane defines the number of multiplets in the 
symmetric or antisymmetric representation, but the combinations of displacements $(\vec{\sigma}_a)$ and discrete Wilson lines $(\vec{\tau}_a)$
determine which of these representations occur, as can be read off from the factor $(-1)^{\sigma^i_a\tau^i_a}$ in 
the contribution $b^{\cal M}_{(\omega^k a) (\omega^k a)'}$ to the beta function coefficient in table~\ref{tab:beta_coeffs_Kaehler_metrics}
of appendix~\ref{A:intersections+betas}.
 
For the shortest three-cycle (0,1;1,0;1,-1) on the {\bf AAA} lattice not parallel to the $\OR$-invariant orbit, one can easily compute that in each of the three $(\omega^k a) (\omega^k a)'_{k=0,1,2}$ sectors one of the angles 
$\phi_{(\omega^k a) (\omega^k a)'}^{(i)}$ vanishes, e.g. in the $aa'$ sector the angles are \mbox{$(\vec{\phi}_{aa'}) = \pi \, \bigl(\frac{1}{3},0,-\frac{1}{3} \bigr)$}. 
The contributions to the beta function coefficient of $SU(N_a)$ take the form 
\begin{equation}
b^{\cal A}_{(\omega^k a) (\omega^k a)'} + b^{\cal M}_{(\omega^k a) (\omega^k a)'} = \left\{ \begin{array}{cr}
  \big(1+ \eta_{(2)}\big) \big( \frac{N_a}{4} - \frac{\eta_{\Omega \cal R}}{2} (-1)^{\sigma_a^2 \tau_a^2} \big)  & k=0,  \\
  \big(1+ \eta_{(3)}\big) \big( \frac{N_a}{4} - \frac{\eta_{\Omega \cal R}}{2} (-1)^{\sigma_a^3 \tau_a^3} \big) & 1,  \\ 
  \big(1+ \eta_{(1)}\big) \big( \frac{N_a}{4} - \frac{\eta_{\Omega \cal R}}{2} (-1)^{\sigma_a^1 \tau_a^1} \big)  &2,   \end{array} \right.
\end{equation}
and consequently only if $\Omega \cal R$ is chosen as the exotic O6-plane $(\eta_{\OR}=-1=\eta_{(i)}$ for all $i \in \{1,2,3\}$), 
the intersections of (orbifold images of) D6-brane $a$  with its orientifold image $a'$
do not support any matter in the symmetric or antisymmetric representation.

On the {\bf BBB} lattice there are three shortest cycles which are not parallel to an O6-plane orbit $\OR \Z_2^{(i)}$. 
Also these three-cycles fit completely in this case, as one of the angles $\phi_{(\omega^k a) (\omega^k a)'}^{(i)}$ vanishes for each sector $(\omega^k a) (\omega^k a)'_{k=0,1,2}$. In table \ref{Tab:SymAntiShortBBB} the contributions to the beta function coefficients of $SU(N_a)$ are listed for each cycle and for each sector separately.
\begin{table}[h]
\begin{center}
\begin{tabular}{|c|c|c|c|}
\hline \multicolumn{4}{|c|}{\bf (Anti-)symmetric representations for the shortest cycles on the {\bf \color{blue}{BBB}} lattice}\\
\hline \hline
\bf  cycle &\bf sector & \bf Beta function contribution $b^{\cal A}_{(\omega^k a) (\omega^k a)'} + b^{\cal M}_{(\omega^k a) (\omega^k a)'}$&
$(\vec{\phi}_{(\omega^k a) (\omega^k a)'})$\\
\hline
(1,-1;1,0;-2,1)&$ \begin{array}{c} (k=0) \\ (k=1) \\ (k=2) \end{array}$ & $ \begin{array}{l}
  \frac{N_a}{4}\left[3-\eta_{(1)}\right]- \frac{\eta_{\OR}}{2} (-)^{\sigma^1_a \tau^1_a} \left[\eta_{(2)}+3\eta_{(3)}\right]   \\
   \big(1+ \eta_{(3)}\big) \big( \frac{N_a}{4} - \frac{\eta_{\Omega \cal R}}{2} (-1)^{\sigma_a^3 \tau_a^3} \big)  \\ 
 \frac{N_a}{4}\left[3-\eta_{(2)}\right]- \frac{\eta_{\OR}}{2} (-)^{\sigma^2_a \tau^2_a} \left[\eta_{(1)}+3\eta_{(3)}\right]    \end{array} $ 
&$\begin{array}{c} \pi (0,\frac{1}{3},-\frac{1}{3}) \\ \pi (\frac{1}{3},-\frac{1}{3},0) \\ \pi (-\frac{1}{3},0,\frac{1}{3}) \end{array}$  \\
\hline
(1,-1;2,-1;-1,0)&$ \begin{array}{c} (k=0) \\ (k=1) \\ (k=2) \end{array}$  &  $ \begin{array}{l}
 \frac{N_a}{4}\left[3-\eta_{(1)}\right]- \frac{\eta_{\OR}}{2} (-)^{\sigma^1_a \tau^1_a} \left[3\eta_{(2)}+\eta_{(3)}\right]    \\
 \big(1+ \eta_{(2)}\big) \big( \frac{N_a}{4} - \frac{\eta_{\Omega \cal R}}{2} (-1)^{\sigma_a^2 \tau_a^2} \big)  \\ 
\frac{N_a}{4}\left[3-\eta_{(3)}\right]- \frac{\eta_{\OR}}{2} (-)^{\sigma^2_a \tau^2_a} \left[\eta_{(1)}+3\eta_{(2)}\right]    \end{array} $ 
&$\begin{array}{c} \pi (0,-\frac{1}{3},\frac{1}{3}) \\ \pi (\frac{1}{3},0 ,-\frac{1}{3}) \\ \pi (-\frac{1}{3},\frac{1}{3},0) \end{array}$ \\
\hline
(1,1;1,0;0,1)&$ \begin{array}{c} (k=0) \\ (k=1) \\ (k=2) \end{array}$ & $\begin{array}{l}
 \big(1+ \eta_{(1)}\big) \big( \frac{N_a}{4} - \frac{\eta_{\Omega \cal R}}{2} (-1)^{\sigma_a^1 \tau_a^1} \big)  \\
 \frac{N_a}{4}\left[3-\eta_{(3)}\right]- \frac{\eta_{\OR}}{2} (-)^{\sigma^3_a \tau^3_a} \left[3\eta_{(1)}+\eta_{(2)}\right]      \\ 
 \frac{N_a}{4}\left[3-\eta_{(2)}\right]- \frac{\eta_{\OR}}{2} (-)^{\sigma^2_a \tau^2_a} \left[3\eta_{(1)}+3\eta_{(3)}\right]   \end{array} $ 
&$\begin{array}{c} \pi (0,\frac{1}{3},-\frac{1}{3}) \\ \pi (\frac{1}{3},-\frac{1}{3},0) \\ \pi (-\frac{1}{3},0,\frac{1}{3}) \end{array}$\\
 \hline
\end{tabular}
\caption{Overview of the Annulus and M\"obius strip contributions to the beta function coefficient per sector for the three shortest cycles on the {\bf BBB} 
lattice not parallel to an $\OR\Z_2^{(i)}$ invariant O6-plane. The total amount of matter in  symmetric and antisymmetric representations
depends on the choice of the exotic O6-plane orbit and the combinations of discrete Wilson lines and displacements. \label{Tab:SymAntiShortBBB}}
\end{center}
\end{table}
Table \ref{Tab:SymAntiShortBBB} shows that  for each shortest three-cycle there exists one sector where the bulk three-cycle $(\omega^k a)$ and its orientifold image 
$(\omega^k a)'$ are parallel to two $\OR \Z_2^{(m), m \in\{0,1,2,3\}}$ invariant planes along one torus $T^2_{(i)}$. Equivalently, the angles 
$(\vec{\phi}_{(\omega^k a) (\omega^k a)'})$ for this sector are of the form $\pi (0_{i}, \frac{1}{3}_{j}, -\frac{1}{3}_{k})$. Choosing the exotic O6-plane to be $\OR$ 
(i.e. $\eta_{\OR} = -1$) or $\OR  \Z_2^{(i)}$ (i.e. $\eta_{\OR  \Z_2^{(i)}} = -1$), this sector will not yield matter in the symmetric or antisymmetric representation. 
The other two sectors on the other hand will always give rise to matter in the symmetric and/or antisymmetric representation at the intersections between the bulk 
cycle $(\omega^k a)$ and its orientifold image $(\omega^k a)'$. The total amount of matter depends on the choice of the exotic O6-plane, while the actual 
representation - symmetric or antisymmetric - is determined by the combinations of discrete Wilson lines and displacements. Hence, the three shortest three-cycles 
not parallel to an O6-plane $\OR  \Z_2^{(i)}$ do not represent suitable candidates for the `QCD stack' of the strong interactions. Furthermore, as also the 
three-cycles  parallel to some O6-plane always yield some non-chiral matter in the symmetric or antisymmetric representation, there does not exist a natural 
candidate for the `QCD stack' on the {\bf BBB} lattice.

\begin{itemize}
\item[\bf (3)]{\bf  For three vanishing angles} 
\end{itemize}

The condition for the absence of matter in the symmetric or antisymmetric representation for vanishing angles reads
\begin{equation}
\sum_{i=1}^3 I_{aa'}^{\Z_2^{(i)},(j \cdot k)} =0 
.
\end{equation}
This situation occurs exactly if the stack of D6-branes $a$ is parallel to some  (ordinary or exotic) O6-plane orbit, provided the combination of displacements 
$(\vec{\sigma}_a)$ and discrete Wilson lines $(\vec{\tau}_a)$ does not satisfy the 
topological condition for enhancement of the gauge group in table \ref{Tab:OR-inv-branes}.

In case of a stack of D6-branes parallel to the exotic $\Omega {\cal R}$-plane ($\eta_{\Omega \cal R} = -1$), the contribution to the beta function coefficient 
from the $aa'$ sector encoding the amount of symmetric and antisymmetric matter depending on $(\vec{\sigma}_a)$  and $(\vec{\tau}_a)$ is given by:
\begin{equation}\label{Eq:betas_pp_OR_exotic}
b^{\cal A}_{aa'} + b^{\cal M}_{aa'} = N_a +  \left\{ \begin{array}{lr} -2&  (\sigma^i_a \tau^i_a = 1, \sigma^j_a \tau^j_a = \sigma^k_a \tau^k_a =0)  \rightarrow   
\Anti_a + \ov \Anti_a \text { of } U(N_a)
   \\ +2 &  (\sigma^i_a \tau^i_a = 0, \sigma^j_a \tau^j_a = \sigma^k_a \tau^k_a = 1) \rightarrow \Sym_a + \ov\Sym_a  \text { of } U(N_a)  \end{array} \right. .
\end{equation} 
If the stack of D6-branes is parallel to an ordinary $\OR$-plane $(\eta_{\OR}=+1)$, the contribution to the beta function coefficient reads
\begin{equation}\label{Eq:betas_pp_OR_ordinary}
b^{\cal A}_{aa'} + b^{\cal M}_{aa'} = N_a  + \left\{\begin{array}{ccc} 
2 & \left( \begin{array}{c} 0=\sigma^i_a\tau^i_a = \sigma^j_a\tau^j_a \neq \sigma^k_a\tau^k_a 
\\  1=\sigma^1_a\tau^1_a = \sigma^2_a\tau^2_a = \sigma^3_a\tau^3_a \end{array} \right) & 
\rightarrow \Sym_a + \ov\Sym_a   \text { of } U(N_a)
\\-2 &\left(\begin{array}{c}  1=\sigma^i_a\tau^i_a = \sigma^j_a\tau^j_a \neq \sigma^k_a\tau^k_a
\\  0=\sigma^1_a\tau^1_a = \sigma^2_a\tau^2_a = \sigma^3_a\tau^3_a  \end{array}\right)
& \rightarrow   \Anti_a + \ov \Anti_a   \text { of } U(N_a) 
\end{array}\right.
.
\end{equation} 

Analogous results arise for D6-branes parallel to some $\OR\Z_2^{(m)}$-invariant plane.

The results in this sector with three vanishing angles is independent of the choice of background lattice orientation.
Since the contributions of this sector are necessarily combined with those from three non-vanishing angles in the 
$(\omega^k \, a)(\omega^k \, a)^{\prime}_{k \in\{1,2\}}$ sectors, stacks of D6-branes parallel to some O6-plane orbit always 
provide some amount of non-chiral matter in the symmetric or antisymmetric representation of $SU(N_a)$.

The shortest three-cycle on the {\bf AAA} lattice, which is not parallel but at angles $\pi (\frac{1}{3}_i,-\frac{1}{3}_j,0_k )$ to the $\OR$-plane orbit, 
provides thus the only candidate for wrapping completely rigid D6-branes without any matter in the symmetric or antisymmetric representation. 
This is the unique natural candidate for the `QCD stack' with gauge group $SU(3)$ or $SU(4)$ in section~\ref{S:model_building}.

\subsection{Three Quark Generations}\label{Ss:3-generations}

In the previous sections, it has been argued that rigid D6-branes without adjoint matter can be found on the {\bf AAA} lattice, if the two so-called `shortest' 
three-cycles in table~\ref{Tab:all-cycles-AAA} 
with wrapping numbers $(n^1_a,m^1_a;n^2_a,m^2_a;n^3_a,m^3_a)$=(1,0;1,0;1,0) and (0,1;1,0;1,-1) form the bulk part of the fractional three-cycles. 

For the first three-cycle, which is parallel to the $\OR$ invariant orbit, however, it was also observed 
in section~\ref{Ss:Gaugeenhanc} that gauge enhancements to  $USp(2N_a)$ or $SO(2N_a)$ occur for special choices of  displacements 
$(\vec{\sigma}_a)$ and discrete Wilson lines $(\vec{\tau}_a)$, which according to the analysis in section~\ref{Ss:no-adjoints}
are accompanied by one chiral multiplet in the antisymmetric representation for the choice of exotic O6-plane $\eta_{\OR}=-1$ with maximal allowed total rank 16
at intersections of orbifold/orientifold images. All other choices of an exotic O6-plane $\eta_{\OR\Z_2^{(k)}}=-1$ provide $SO(2N_a)$ or $USp(2N_a)$ gauge factors
without any matter in the symmetric or antisymmetric representation.
For the same bulk three-cycle parallel to the $\OR$-invariant orbit with $U(N_a)$ gauge group, equations~(\ref{Eq:betas_pp_OR_exotic}) 
and~(\ref{Eq:betas_pp_OR_ordinary}) show that any choice of exotic O6-plane provides at least one non-chiral pair of multiplets in the symmetric or 
antisymmetric representation.
The three-cycle with wrapping numbers (1,0;1,0;1,0) is thus not the most favourable for engineering the strong interactions without massless non-chiral
 charged exotic matter, but it is a good candidate for the weak $USp(2)$ or $U(2)$ gauge group, provided that the constraint on the spectrum is loosened by 
allowing for matter in the antisymmetric representation, which is vector-like with respect to the Standard Model gauge group. 
Similar considerations apply to the additionally required $U(1)$ gauge factors.

The second shortest three-cycle with wrapping numbers (0,1;1,0;1,-1) and angles $(\vec{\phi}_a)=\pi \, (\frac{1}{3},0,-\frac{1}{3})$ on the {\bf AAA} lattice
is for specific combinations of displacements $(\vec{\sigma}_a)$ and discrete Wilson lines $(\vec{\tau}_a)$ the only possible choice of a bulk three-cycle leading 
to an $U(N_a)$ gauge group without any matter in the adjoint, symmetric or antisymmetric representations. It is thus the natural choice for the `QCD stack'
with $SU(3)$ or $SU(4)$ gauge group.

In table~\ref{Tab:U(3)searchforU(2)}, a systematic scan of net-chiralities among two shortest D6-branes is presented, where for the `QCD stack', also
the choice $(1,0;1,0;1,0)$ with two non-chiral matter pairs in the antisymmetric representation, or one pair of non-chiral matter in the antisymmetric and 
one pair in the symmetric representation is considered.
Zero, one and two generations turn out to arise for various choices of discrete parameters, while three particle generations with $\chi^{ab}+\chi^{ab'}=3$ are only possible for 27 combinations of discrete parameters
on the (0,1;1,0;1,-1)  cycle for $a$ and $b$, both of them completely rigid without any matter in the adjoint representation. 
The complete list of the corresponding discrete parameters is given in tables~\ref{Tab:RigFreeSet1} to~\ref{Tab:RigFreeSet3} 
of appendix \ref{A:RigidD63Gen}. Negative values for net-chiralities  $\chi^{ab}+\chi^{ab'}=-3$ are obtained in an analogous way
for another 27 combinations.
\begin{table}[h]
\begin{center}
\begin{tabular}{|c||c|c||c|c|}
\hline
\multicolumn{5}{|c|}{\bf Net-chirality ($\chi^{ab},\chi^{ab'}$) between D6-branes $a$  and $b$ on the {\bf AAA} lattice}\\
\hline \hline \bf  $D6_b$ stack & \multicolumn{2}{|c||}{\bf $D6_a$ on (0,1;1,0;1,-1) } & \multicolumn{2}{|c|}{\bf $D6_a$ on (1,0;1,0;1,0)}\\
\hline \hline
(1,0;1,0;1,0) with adjoints& \multicolumn{2}{c||}{(0,0)} & \multicolumn{2}{c|}{(0,0)} \\
& (1,1) & (-1,-1)& (1,1) & (-1,-1)\\
\hline \hline(1,0;1,0;1,0) without adjoints& \multicolumn{2}{c||}{ (0,0), (1, -1), (-1,1),} &  \multicolumn{2}{c|}{(0,0)} \\
&\multicolumn{2}{c||}{(2, -2), (-2,2)} && \\
& (1,0), (0,1), & (-1,0), (0,-1),& (1,0), (0,1)&(-1,0), (0,-1)\\
&(-1,2), (2,-1)&   (1,-2), (-2,1)&&\\
\hline\hline(0,1;1,0;1,-1) with adjoints &  \multicolumn{2}{c||}{ (0,0), (1, -1), (-1,1)}& \multicolumn{2}{c|}{(0,0)} \\
& (1,0), (0,1) & (-1,0), (0,-1) & (1,0), (0,1) & (-1,0), (0,-1)\\ 
& (1,1) & (-1,-1)& (1,1) & (-1,-1)\\
\hline\hline (0,1;1,0;1,-1) without adjoints& \multicolumn{2}{c||}{ (0,0), (1, -1), (-1,1)}& \multicolumn{2}{c|}{(0,0)} \\
& (1,0), (0,1),& (-1,0), (0,-1), & (1,0), (0,1)& (-1,0), (0,-1) \\
&  (-1,2)  & (1,-2)  &&\\
&(1,1), (0,2) & (-1,-1), (0,-2) & (1,1) & (-1,-1)\\
& \color{darkgreen}(1,2) & \color{darkgreen}(-1,-2) &\color{darkgreen}(1,2), (2,1) &\color{darkgreen} (-1,-2), (-2,-1)\\
& &&(2,2)&(-2,-2)\\
\hline
\end{tabular}
\caption{Overview of the net-chirality ($\chi^{ab},\chi^{ab'}$) between a completely rigid D6-brane stack $a$ without adjoint 
representations and a D6-brane stack $b$ on the {\bf AAA} lattice with the choice of exotic O6-plane $\eta_{\OR}=-1$,  
both wrapped on fractional three-cycles with the bulk wrapping numbers 
$(n^i,m^i)_{i \in \{1,2,3\}}$ given in the top row and left column. The combinations of displacement and Wilson line parameters of the D6-brane stack 
$a$ in the right column imply the presence of two non-chiral matter pairs in the antisymmetric representation, or of one non-chiral matter pair in the 
antisymmetric and another one in the symmetric representation. 
}
\label{Tab:U(3)searchforU(2)}
\end{center}
\end{table}


For a `QCD stack' on a rigid fractional three-cycle with toroidal wrapping numbers (1,0;1,0;1,0) and some non-chiral pairs of symmetric or antisymmetric 
representations, three generations of quarks can be found, provided that the $SU(2)_L$ stack wraps a rigid fractional cycle with toroidal wrapping numbers 
(0,1;1,0;1,-1), as can be seen from right column of table~\ref{Tab:U(3)searchforU(2)}. The rigidness of the fractional three-cycles suitable for the 
`QCD stack' and the requirement to obtain three generations of quarks also enforce the $SU(2)_L$ stack to wrap a rigid fractional three-cycle.
This argument extends to left-right symmetric and Pati-Salam models discussed in sections~\ref{Ss:LR-symmetric} and~\ref{Ss:PS-model}, while for MSSM like spectra in 
section~\ref{Ss:MSSMs}, D6-branes $c$ with intersection numbers $\chi^{ac}+ \chi^{ac'}= \pm 6$ have to be taken into account as well.

\section{From D6-Branes to Particle Physics}\label{S:model_building}

In this section, explicit intersecting D-brane models on the {\bf AAA} lattice orientation of the orientifold 
$T^6/(\Z_2 \times \Z_6' \times \OR)$ are constructed. Recall from the discussion in subsection \ref{Sss:Geo_bulk} that large hidden gauge groups are 
only be possible on the {\bf AAA} lattice and only for the situation where the $\OR$-plane is the exotic O6-plane (i.e.~$\eta_{\OR} = -1$). 
In the previous section, it was shown that completely rigid fractional three-cycles on this orientifold background exist, and the ansatz of  requiring
a completely rigid `QCD stack' for the strong interactions implies that also the $SU(2)_L$ stack is wrapped on a rigid three-cycle. 
In addition, the constraint that the fractional three-cycle for the `QCD stack' is free of 
symmetric and antisymmetric representations is imposed in order to avoid strongly coupled
exotic matter, which according to section~\ref{Ss:no-anti+sym} implies that the orbit for the `QCD stack' is uniquely represented by the 
toroidal wrapping numbers (0,1;1,0;1,-1). The requirement of three quark generations 
leads to the $SU(2)_L$  stack being wrapped on the same bulk cycle (0,1;1,0;1,-1) with specific choices of $(\vec{\tau}_b)$ and $(\vec{\sigma}_b)$ 
for rigidity. This equally implies that the models will not contain chiral matter in the symmetric 
nor in the antisymmetric representation of $SU(2)_L$ as discussed in detail in section~\ref{Ss:3-generations}.   

Furthermore, as the tree-level value of the square of the gauge coupling is inversely proportional to the length of the three-cycle wrapped by the 
D6-brane stack, see e.g.~\cite{Blumenhagen:2006ci},
\begin{equation}\label{Eq:tree-gauge-coupling}
\frac{4 \pi}{g_{SU(N_a),{\rm tree}}^2} =  2\pi \Re \left({\rm f}_{SU(N_a)}^{\text{tree}} \right)
= \frac{1}{4} \, \frac{1}{g_{\text{string}}} \frac{\prod_{i=1}^3 L_a^{(i)}}{\ell_s^3}
,
\end{equation}
with $L_a^{(i)}$ the length of the toroidal one-cycle and the factor $1/4$ for fractional D6-branes as defined in 
equation~(\ref{Eq:def_frac_cycle})
 and $\ell_s \equiv 2\pi \sqrt{\alpha'}$ the string length,
the tree-level values of the gauge couplings for the `QCD stack' $a$ and the $SU(2)_L$ stack $b$ are equal to each other with 
$\prod_{i=1}^3  L_a^{(i)}=\prod_{i=1}^3  L_b^{(i)}= r_1r_2r_3$. 
The field theoretic one-loop running of the gauge couplings and threshold corrections, however, are  determined by the beta function coefficients, which depend 
on the amount and the representations of the massless chiral and 
non-chiral matter states charged under the respective gauge group. The beta function coefficient itself is thus model-dependent and one has to check explicitly 
on a case-by-case basis whether the beta function coefficients provide the correct phenomenology such as asymptotic freedom for the `QCD stack'
at low energies. 

The first two models that were constructed correspond to a local MSSM model and a local left-right symmetric model. 
Both models are built on four stacks of D6-brane, and in both cases the twisted RR tadpoles cannot be cancelled
  by adding appropriate supersymmetric `hidden' D6-branes. The third type of models discussed here comprises
globally consistent Pati-Salam models constructed from five or six stacks of D6-branes. 
  For such a model, the perturbative Yukawa and some other three-point couplings are given. 
The corresponding discussion of the one-loop behaviour of gauge couplings is relegated to section~\ref{S:U1s}.
  In appendix~\ref{A:RigidD63Gen} a six-stack Pati-Salam model is built starting from a rigid `QCD stack', 
yet loosening the constraint of disallowing matter in symmetric or antisymmetric representations.

\subsection{Local MSSM Models}\label{Ss:MSSMs}

To construct an MSSM-like model, at least two appropriate D6-branes $c$, $d$ with gauge group $U(1)_a \times U(1)_b$ need to be added  to the QCD and $SU(2)_L$  
stacks $a$ and $b$  in order to recover the correct amount of right-handed quarks and both left- and right-handed 
leptons.\footnote{A $USp(2)_b$ gauge factor for the weak interactions is excluded since intersections with a completely rigid QCD 
stack can only provide zero or two particle generations.} 
The present search for MSSM-like spectra is confined to $U(1)_c \times U(1)_d$ gauge groups on shortest or next-to-shortest three-cycles, allowing for matter in 
the `adjoint' representation.
This amounts to allowing singlets under the gauge group, which according to  section~\ref{Ss:no-adjoints} will not be abundant. 
Under these assumptions one can show that a single $U(1)$, or a combination of two $U(1)$s on shortest cycles, or on one shortest and one next-to-shortest cycle do 
not  yield the required chiral particle spectrum of the MSSM. A `local' four-stack model with a phenomenologically attractive MSSM-like spectrum can, however, be 
constructed when both D6-branes with $U(1)_c \times U(1)_d$ gauge groups are wrapped on a next-to-shortest cycle.\footnote{A particular set-up for a four-stack model 
consists in interpreting one $U(1)_c$ as a `right' stack and the other $U(1)_d$ as a pure `leptonic' stack. In this construction the right-handed up-type quarks 
$\bar u_R$ arise from the $ac$ sector ($\chi^{ac} = -3$), while the other
 right-handed down-type quarks $\bar d_R$ arise from the $ac'$ sector ($\chi^{ac'}= -3$). One can show that none of the shortest or next-to-shortest cycles allow for 
these net-chiralities.}  The D6-brane configuration of an explicit `local' MSSM example is presented in table~\ref{Tab:MSSM-1}.    
\begin{table}[h]
\begin{center}
\begin{tabular}{|c||c|c||c|c|c||c|}\hline
\multicolumn{7}{|c|}{\bf D6-brane configuration of a `local' MSSM on $T^6/(\Z_2 \times \Z_6' \times \OR)$ with $\eta=-1$}
\\\hline \hline
& \bf wrapping numbers &\bf $\frac{\rm Angle}{\pi}$ & \bf $ \Z_2$ eigenvalues & ($\vec{\tau}$) & ($\vec{\sigma}$)&\bf gauge group\\
\hline \hline
 $a$&(0,1;1,0;1,-1)&$(\frac{1}{3},0,-\frac{1}{3})$&$(+++)$&$(0,0,1)$ & $(0,0,1)$& $U(3)$\\
 $b$&(0,1;1,0;1,-1)& $(\frac{1}{3},0,-\frac{1}{3})$&$(+--)$&$(0,1,1)$ & $(0,1,1)$&$U(2)_L$\\
 $c$&(-1,2;2,-1;1,-1)&$(\frac{1}{2},-\frac{1}{6},-\frac{1}{3})$&$(-+-)$&$(1,0,0)$ & $(1,0,0)$&$U(1)_c$\\
$d$ & (-1,2;2,-1;1,-1) &$(\frac{1}{2},-\frac{1}{6},-\frac{1}{3})$& $(--+)$ &$(0,0,1)$&$(1,0,1)$&$U(1)_d$\\
 \hline
\end{tabular}
\caption{D6-brane configuration with four stacks of D6-branes yielding a `local' MSSM spectrum with gauge group 
$SU(3)_a \times SU(2)_b \times U(1)_a \times U(1)_b\times U(1)_c \times U(1)_d$. The masses of the Abelian gauge
factors can only be determined in a global completion. 
The angles are measured with respect to the $\OR$-invariant plane.
\label{Tab:MSSM-1}}
\end{center}
\end{table}

The chiral spectrum for this D6-brane set-up can be found in table~\ref{Tab:CSMSSM-1}, while the additional massless non-chiral matter multiplets are listed in 
table~\ref{Tab:NCSMSSM-1}. The combined RR charges of  these four stacks already take a huge chunk out of the bulk RR charges generated by the O6-planes
in table~\ref{tab:Bulk-RR+SUSY-Z2Z6p},
\begin{equation}\label{Eq:max-hidden-rank}
16 - X_a N_a - X_b N_b - X_c N_c - X_d N_d = 4,
\end{equation}
which does not leave much room  for additional hidden gauge groups. 
Besides the bulk RR tadpoles, also the twisted RR tadpoles in table~\ref{tab:twistedRR-AAA+ABB-Z2Z6p} have to cancel in a globally defined D6-brane configuration, which 
can be written as follows with $i \in \{1,2,3\}$,
\begin{equation}
\sum_{x \in \{a,b,c,d\}} N_x \left( \Pi^{\Z_2^{(i)}}_x + \Pi^{\Z_2^{(i)}}_{x'}  \right) = \left\{ \begin{array}{l} 
3 \, \varepsilon^{(1)}_1 + 3 \, \varepsilon^{(1)}_2 + 4 \, \varepsilon^{(1)}_3 -  \varepsilon^{(1)}_4 + \tilde\varepsilon^{(1)}_4 - \tilde\varepsilon^{(1)}_5 \\
3 \, \varepsilon^{(2)}_1 + 3 \, \varepsilon^{(2)}_2 + 4 \, \varepsilon^{(2)}_3 - 3 \, \varepsilon^{(2)}_4 + 2 \,  \varepsilon^{(2)}_5  + 5 \, \tilde\varepsilon^{(2)}_4 
- 5 \, \tilde\varepsilon^{(2)}_5 \\
- \varepsilon^{(3)}_1 - 3 \, \varepsilon^{(3)}_2 -2 \, \varepsilon^{(3)}_3 +  \varepsilon^{(3)}_4 - \tilde\varepsilon^{(3)}_4 + \tilde\varepsilon^{(3)}_5 
\end{array}  \right. ,
\end{equation}
by summing up the twisted sector contributions of the four D6-branes presented in table~\ref{Tab:MSSM-1}. 
Since the maximal supersymmetric hidden rank in equation~(\ref{Eq:max-hidden-rank}) is four, or two if one out of two hidden D6-branes is next-to-shortest, the 
prefactors 3,4,5 for all five indices $\alpha \in \{1 \ldots 5\}$ of the $\Z_2^{(2)}$ twisted sector show that some of the twisted RR tadpoles cannot be cancelled 
for this example, if the constraints on exceptional wrapping numbers in table~\ref{Tab:xy-entries-Z2Z6p} are taken into account. Without a global completion, it 
cannot be argued whether some or all of the Abelian symmetries remain massless. 
This affects in particular the linear combination,
\begin{equation}
Q_Y = \frac 1 6 Q_a + \frac 1 2 Q_c + \frac 1 2 Q_d,
\end{equation}
which provides the hypercharge $U(1)_Y$ and would in a globally defined MSSM-like model have to remain massless. 

\begin{table}[h]
\begin{center}
\begin{tabular}{|c||c|c||c|c|}
\hline \multicolumn{5}{|c|}{\bf Chiral spectrum of a `local' four-stack MSSM on the AAA lattice}\\
\hline \hline
Matter & Sector & $U(3)_a\times U(2)_b \times U(1)_c \times U(1)_d$& $(Q_a, Q_b, Q_c, Q_d)$   &$Q_Y$\\
\hline \hline
$Q_L$&$ab$&$({\bf 3}, \bar {\bf 2})_{(0,0)} $&(1,-1,0,0)&1/6\\
$Q_L$&$ab'$&$2 \times ({\bf 3}, {\bf 2})_{(0,0)} $&(1,1,0,0)& 1/6\\
$\bar d_R$& $ac$ &$2 \times (\bar{\bf 3}, \1)_{(1,0)}$ &(-1,0,1,0)  & 1/3\\
$\bar u_R$ & $ac'$& $ (\bar{\bf 3}, \1)_{(-1,0)}$& (-1,0,-1,0) & -2/3 \\
$\bar d_R$& $ad$ &$(\bar{\bf 3}, \1)_{(0,1)} $ & (-1,0,0,1) & 1/3\\
$\bar u_R$ & $ad'$ & $2 \times (\bar{\bf 3}, \1)_{(0,-1)}$&(-1,0,0,-1)&-2/3\\
$L$ & $bc$ &$(\1, {\bf 2})_{(-1,0)} $ &(0,1,-1,0) & -1/2\\
$L$&$bd$ &$(\1, {\bf 2})_{(0,-1)}$  &(0,1,0,-1)& -1/2 \\
$L$ & $bd'$ &$(\1, \bar{\bf 2})_{(0,-1)}$ &(0,-1,0,-1)& -1/2 \\
$e_R$& $cc'$ &$2 \times (\1,\1)_{(2_\Sym,0)}$ & (0,0,2,0) & 1\\
$\nu_R$ & $cd$ &  $(\1,\1)_{(-1,1)}$ &(0,0,-1,1)& 0\\
$e_R$ &$cd'$ &$(\1,\1)_{(1,1)}$ &(0,0,1,1)& 1 \\
\hline
\end{tabular}
\caption{Chiral spectrum for the D6-brane configuration in table \ref{Tab:MSSM-1} consisting of three Standard Model quark and lepton,
but only one right-handed neutrino generation. \label{Tab:CSMSSM-1}}
\end{center}
\end{table}
The chiral spectrum in table~\ref{Tab:CSMSSM-1} contains besides three generations of left- and right-handed quarks and left-handed leptons in bifundamental
representations also three right-handed electrons $e_R$, but only one neutrino $\nu_R$. Two of the $e_R$ arise as symmetric representations of $U(1)_c$ and 
do thus not have perturbative Yukawa couplings. One might speculate that the role of additional right-handed neutrinos is assumed by some non-chiral matter 
fields in table~\ref{Tab:NCSMSSM-1} or by some closed string modulus multiplets in table~\ref{Tab:ClosedSpectrum_Z2Z6p}.
\begin{table}[h]
\begin{center}
\begin{tabular}{|c||c|c||c|c|}
\hline \multicolumn{5}{|c|}{\bf Non-chiral spectrum of a local four-stack MSSM on the AAA lattice}\\
\hline \hline
Matter & Sector & $U(3)_a\times U(2)_b \times U(1)_c \times U(1)_d$& $(Q_a, Q_b, Q_c, Q_d)$ & $Q_Y$\\
\hline \hline
&$cc$&$2 \times (\1,\1)_{(0,0)}$ & (0,0,0,0) &0\\
& $dd$& $(\1,\1)_{(0,0)}$ & (0,0,0,0) &0\\
&$dd'$ &$2 \times [ (\1,\1)_{(0,2_\Sym)} + h.c.]$  & (0,0,0,$\pm$2) &$\pm$1\\
&$ad$ & $({\bf 3},\1)_{(0,-1)} + h.c.$ & ($\pm$1,0,0,$\mp$1) & $\mp$ 1/3 \\
&$cd$& $(\1,\1)_{(1,-1)} + h.c.$&(0,0,$\pm$1,$\mp$1)&0\\
$H_u + H_d$&$bd'$& $(\1,{\bf 2})_{(0,1)} + (\1,{\bf \bar 2})_{(0,-1)}  $ & (0,$\pm$1,0,$\pm$1)&$\pm$1/2\\
\hline
\end{tabular}
\caption{Non-chiral spectrum for the D6-brane configuration in table \ref{Tab:MSSM-1}. \label{Tab:NCSMSSM-1}}
\end{center}
\end{table}

The Higgs sector is the same as in the MSSM, consisting of two Higgs multiplets ($H_u$, $H_d$) arising from the $bd'$ sector of the non-chiral matter part in 
table~\ref{Tab:NCSMSSM-1}. This implies that perturbative Yukawa-interactions exist for right-handed quarks and left-handed leptons if they arise as 
bifundamental matter states at the intersections with the $d$-stack. Hence, two up-type $\bar u_R$ and  one down-type $\bar d_R$ generation as well as one 
leptonic $e_R$ generation satisfy charge selection rules on perturbative three-point couplings involving the Higgs fields
(with $v_i \equiv \frac{\sqrt{3}}{2} \frac{r_i^2}{\alpha'}$ the two-torus volume on $T^2_{(i)}$ in units of $\alpha'$),
\begin{equation}
\begin{array}{lll}
i=2,3: &(\bar d_R)_{ad} H_d (Q^{(i)}_L)_{ab'} &\sim \left( 0, {\cal O}(e^{-v_1/4})   \right),\\
& (\bar u^{(i)}_R)_{ad'}  H_u (Q_L)_{ab} &\sim (0, 0) , \\
& (e_R)_{cd'}\, H_d\, L_{bc} &\sim 0,
\end{array} \label{Eq:YukMSSM}
\end{equation} 
where, however, it turns out that the condition on closed triangles in the presence of various orbifold images first advertised 
in~\cite{Honecker:2012jd,Honecker:2012fn} can only be satisfied for one down-type quark Yukawa coupling.
The remaining particle generations have to receive masses either through higher order or non-perturbative couplings. 
This naturally produces a hierarchy of one very massive against two light particle generations, which however, implies a much heavier 
bottom than top quark in contrast to experimental data. 

To derive the perturbative three-point couplings on $T^6/\Z_{2N}$ or $T^6/\Z_2 \times \Z_{2M}$ orbifolds such as those in equation~(\ref{Eq:YukMSSM}), one
 must first  determine the localisation of each matter state in a given sector $x(\omega^k y)$, given that the charge selection rule does not guarantee the existence of a closed triangle on orbifolds with D6-brane images $(\omega^k x)$ of a toroidal 
D6-brane $x$, as shown in~\cite{Honecker:2012jd,Honecker:2012fn}. If a closed triangle $[x,y,z]=[x,y][y,z][z,x]$ can be constructed with the three intersecting branes $x$, $y$ and $z$, the matter states $\phi^i_{xy}$, $\phi^j_{yz}$ and $\phi^k_{zx}$ arise at the respective intersection points, which serve as apexes for the closed triangle. The closed triangle then corresponds to a three-point coupling of the form $W_{ijk} \phi^i_{xy} \phi^j_{yz} \phi^k_{zx}$.
If such a closed sequence $[x,y,z]$  exists, the coefficient $W_{ijk}$ of the three-point coupling is related to the area of the triangle enclosed by the three intersecting cycles $x$, $y$ and $z$,
\begin{equation}
W_{ijk} \sim e^{- {\cal A}_{ijk}/(2 \pi \alpha')}.
\end{equation}
If the three cycles intersect in a single point, the area ${\cal A}_{ijk}$ is zero and the coupling is not suppressed but of order ${\cal O}(1)$. If there does not exist a closed sequence, the coefficient $W_{ijk}$ vanishes. 

To verify the vanishing of the Yukawa couplings in equation~(\ref{Eq:YukMSSM}), which are allowed by charge selection rules, it is thus necessary to 
take account of the localisation of each matter and Higgs state per intersection sector as given in tables~\ref{Tab:NetChiralMSSM1} and~\ref{Tab:NetChiralMSSM2}. 
For the Yukawa-couplings in equation~(\ref{Eq:YukMSSM}) there is only one set of closed triangles formed by the sequence $[a,(\omega^2 d),(\omega^2 b)']$. 
The triangle $[5,4,1]$ on torus $T^2_{(1)}$ has the $\Z_2$ fixed points 5, 4 and 1 as apices, whereas the three cycles intersect in one single point on $T^2_{(2)}\times T^2_{(3)}$, i.e.~the $\Z_2$ fixed point $e_{44}$. This is confirmed by the net-chiralities  for the local four-stack model in tables~\ref{Tab:NetChiralMSSM1} and~\ref{Tab:NetChiralMSSM2}, reflecting the localisation of the matter states at the corresponding intersection points. The enclosed area for this triangle is given by $\frac{v_1}{4} $, and the Yukawa-coupling corresponds to $(\bar d_R)_{ad} H_d (Q^{(3)}_L)_{ab'}$. For the other Yukawa-couplings in equation (\ref{Eq:YukMSSM}) it is not possible to write down closed sequences of three-cycles, implying that the perturbative Yukawa-coupling constants are zero.

Since the `local' MSSM-model in this section does not possess a global completion, phenomenological details will not be discussed further.

\begin{table}[h]
\begin{center}
\begin{tabular}{|c||c|c|c|c|}
\hline \multicolumn{5}{|c|}{\bf Counting of states  for a `local' four-stack MSSM, part I} \\
\hline \hline $(\chi^{x y}, \chi^{x (\omega y)},\chi^{x (\omega^2 y)})$  & $y=a$ & $y=b$ & $y=c$ & $y=d$\\
\hline\hline $x=a$& (0,0,0) & (0,1,0) & (0,0,-2) &  (0,1,-2)\\
$x=b$& & (0,0,0) & (0,1,0)) & (1,0,0) \\
$x=c$& & & (0,2,-2) & (0,1,-2)\\
$x=d$ &&&& (0,1,-1)\\
\hline
\end{tabular}
\caption{Counting $\chi^{x (\omega^k y)}$ of chiral states per sector $x(\omega^k y)$ for the D6-brane configuration in table \ref{Tab:MSSM-1}. For each sector in this table, the total amount of matter is given by $\varphi^{x (\omega^k y)}=|\chi^{x (\omega^k y)}|$. \label{Tab:NetChiralMSSM1}}
\end{center}
\end{table}

\begin{table}[h]
\begin{center}
\begin{tabular}{|c||c|c|c|c|}
  \hline \multicolumn{5}{|c|}{\bf Counting of states for a `local' four-stack MSSM, part II} \\
\hline \hline $(\chi^{x y'}, \chi^{x (\omega y)'},\chi^{x (\omega^2 y)'})$ & $y=a$ & $y=b$ & $y=c$ & $y=d$\\
\hline\hline $x=a$& (0,0,0) & (0,1,1) & (0,-1,0) &  (0,0,-2)\\
$x=b$& & (0,0,0) & (0,0,0)) & (0,-1,$|2|$) \\
$x=c$& & & (2,2,$|4|$) & (0,1,0)\\
$x=d$ &&&& (-2,2,$|4|$)\\
\hline
\end{tabular}
\caption{Counting of massless states per sector for the D6-brane configuration in table \ref{Tab:MSSM-1}. If  the net-chirality $\chi^{x(\omega^k y)'} =0$
vanished but the total amount of matter $\varphi^{x(\omega^k y)'} \neq 0$ is non-zero, the net-chirality is replaced by the absolute value of the total 
amount of matter $|\varphi^{x(\omega^k y)'}|$. The net-chirality $\chi^{x(\omega^k x)'}$ counts the amount of symmetric + antisymmetric matter as introduced 
in equation~(\ref{Eq:ChiralMatter-Sym+Anti}). \label{Tab:NetChiralMSSM2}}
\end{center}
\end{table}

\subsection{Constraints on Left-Right Symmetric Models}\label{Ss:LR-symmetric}

A second type of interesting models on the $T^6/(\Z_2 \times \Z_6' \times \OR)$ orientifold consists of left-right symmetric models with gauge groups 
$U(3)_a\times U(2)_b \times U(2)_c \times U(1)_d$ or $U(3)_a\times U(2)_b \times SO(2)_c \times U(1)_d$, where again $USp(2)_b$ or $USp(2)_c$ gauge factors 
are excluded by not providing three generations at intersections with a completely rigid $U(3)_a$ stack. 
For the second combination of gauge groups, the  enhancements $U(1)_c \to SO(2)_c$ for special combinations of discrete Wilson lines and displacements  
($\vec{\tau}_c, \vec{\sigma}_c$) were discussed in 
section~\ref{Ss:Gaugeenhanc}. As for the enhancements to $USp(2)_{b}$ or $USp(2)_{c}$, a detailed analysis provided only zero or two
particle generations, whereas three generations of right-handed  quarks  $(\bar u_R, \bar d_R)$ cannot be localised at $U(3)_a \times SO(2)_c$ intersections with 
stack $a$ completely rigid.

For the only viable ansatz with gauge group  $U(3)_a\times U(2)_b \times U(2)_c \times U(1)_d$, the existence of three quark generations was discussed in section~\ref{Ss:3-generations} with the constraint
$\chi^{ab} + \chi^{ab'} = - \chi^{ac}-\chi^{ac'} = \pm 3$ of opposite chiralities for the intersections of the `QCD stack' $a$ with the `left' and `right stacks'
 $b$ and $c$. As in section~\ref{Ss:MSSMs} for the local MSSM model,
the existence of three lepton generations enforces the stack $d$ with $U(1)_d$ gauge group to wrap a next-to-shortest three-cycle. Loosening the constraint on rigidness of the $U(2)_c \equiv U(2)_R$ stack as well, local left-right
symmetric models with adjoint matter of the gauge factors $U(2)_c \times U(1)_d$ can be constructed.
In table~\ref{Tab:LRSM1}, an explicit example of such a left-right symmetric model constructed on four stacks of D6-branes is given.
\begin{table}[h]
\begin{center}
\begin{tabular}{|c||c|c||c|c|c||c|}\hline 
\muc{7}{|c|}{\bf D6-brane configuration of a `local' left-right symmetric model}
\\\hline \hline
&\bf wrapping numbers & $\frac{\rm Angle}{\pi}$ &\bf $\Z_2$ eigenvalues & ($\vec \tau$) & ($\vec \sigma$)&\bf gauge group\\
\hline \hline
 $a$&(0,1;1,0;1,-1)&$(\frac{1}{3},0,-\frac{1}{3})$&$(+++)$&$(0,1,0)$ & $(0,1,0)$& $U(3)$\\
 $b$&(0,1;1,0;1,-1)&$(\frac{1}{3},0,-\frac{1}{3})$&$(--+)$&$(1,1,0)$ & $(1,1,0)$&$U(2)_L$\\
 $c$&(1,0;2,-1;1,1)&$(0,-\frac{1}{6},\frac{1}{6})$&$(--+)$&$(0,0,0)$ & $(0,0,0)$&$U(2)_R$\\
$d$ & (-1,2;2,-1;1,-1) &$(\frac{1}{2},-\frac{1}{6},-\frac{1}{3})$& $(+++)$ &$(1,0,0)$&$(1,0,0)$&$U(1)_d$\\
 \hline
\end{tabular}
\caption{Local D6-brane configuration with four stacks of D6-brane yielding a local left-right symmetric model with gauge group  $SU(3)_a\times SU(2)_b\times SU(2)_c \times U(1)_a \times U(1)_b\times U(1)_c\times U(1)_d$. \label{Tab:LRSM1}}
\end{center}
\end{table}

The chiral spectrum for this particular `local' D6-brane set-up is given in table~\ref{Tab:CSLRSM1}, 
and the non-chiral matter states are gathered in table~\ref{Tab:NCSLRSM1}. 
\begin{table}[h]
\begin{center}
\begin{tabular}{|c||c|c|c|c|}
\hline \multicolumn{5}{|c|}{\bf Chiral spectrum of a `local' four-stack left-right symmetric model}\\
\hline \hline
Matter & Sector & $U(3)\times U(2)_L \times U(2)_R \times U(1)_d$& $(Q_a, Q_b, Q_c, Q_d)$ & $Q_{B-L}$ \\
\hline
$Q_L$&$ab$ &  $({\bf 3}, {\bf\bar 2}, \1)_0$ & (1,-1,0,0) & 1/3\\
$Q_L$&$ab'$ & $2 \times ({\bf 3}, {\bf 2}, \1)_0$  & (1,1,0,0) & 1/3\\
$\left(\bar u_R, \bar d_R\right)$&$ac$ &  $2 \times ({\bf\bar 3}, \1,{\bf 2})_0$ & (-1,0,1,0) &-1/3\\
$\left(\bar u_R, \bar d_R\right)$&$ac'$ &  $({\bf\bar 3}, \1,{\bf\bar 2})_0$ & (-1,0,-1,0) &-1/3\\
$\left(H_d, H_u\right)$&$bc'$& $(\1, {\bf \bar 2}, {\bf \bar 2})_0$& (0,-1,-1,0)&0\\
$L$&$bd'$&  $3 \times (\1, {\bf  2}, \1)_1 $& (0,0,1,0,1)&-1\\
$(\bar \nu_R, \bar e_R)$& $cd$ & $3 \times (\1, \1, {\bf 2})_{-1} $&(0,0,1,-1)& 1\\
&$dd'$& $2 \times (\1,\1,\1)_{2_\Sym} $& (0,0,0,2)&-2\\
\hline
\end{tabular}
\caption{Chiral spectrum of the `local' left-right symmetric D6-brane configuration in table~\ref{Tab:LRSM1}. \label{Tab:CSLRSM1}}
\end{center}
\end{table}
The chiral matter spectrum contains besides exactly three quark and lepton generations and one Higgs family one kind of chiral exotic particles in the 
symmetric representation of  $U(1)_d$ and thus charged under the gauged $B-L$ symmetry,
\begin{eqnarray}
Q_{B-L}=\frac 1 3 Q_a - Q_d,
\end{eqnarray}
provided this linear combination remains massless in a global completion of the `local' D6-brane configuration.

The non-chiral spectrum in table~\ref{Tab:NCSLRSM1} does by construction not contain matter in the adjoint, symmetric or antisymmetric representation of $SU(3)_a$, 
but it contains matter in bifundamental representations under each non-Abelian gauge factor.
\begin{table}[h]
\begin{center}
\begin{tabular}{|c|c||c|c|}
\hline \multicolumn{4}{|c|}{\bf Non-chiral spectrum of a `local' four-stack left-right symmetric model}\\
\hline \hline sector  & $U(3)\times U(2)_L \times U(2)_R \times U(1)_d$ &sector  & $U(3)\times U(2)_L \times U(2)_R \times U(1)_d$\\
\hline
$cc$ & $4 \times (\1,\1,{\bf 4_\Adj})_0$& $ac'$ &  $({\bf 3}, \1,{\bf 2})_0 + h.c.$\\
$cc'$ &  $4 \times [(\1,\1, {\bf 1_\Anti})_0 + h.c.]$&$ad$ & $({\bf 3}, \1,\1)_{-1} + h.c.$ \\
$dd$&  $2 \times (\1,\1,\1)_{\bf 0_\Adj} $&$bd$ & $(\1, {\bf 2},\1 )_{-1} + h.c. $\\
$ac$ & $({\bf 3}, \1,{\bf \bar 2})_0 + h.c.$ & $cd'$ &$2 \times [(\1, \1, {\bf 2} )_{1} + h.c.]  $\\
 \hline
\end{tabular}
\caption{Non-chiral spectrum of the `local' left-right symmetric D6-brane configuration in table~\ref{Tab:LRSM1}. \label{Tab:NCSLRSM1}}
\end{center}
\end{table}

While the minimal Higgs content of one doublet  $(H_d, H_u)$ is an attractive feature of the model, perturbative charge selection rules  
imposed by the (anomalous) $U(1)_b \times U(1)_c$ symmetries only allow two quark  but all three lepton generations to couple to it,
\begin{equation}\label{Eq:YukLRSM}
\begin{array}{llcc}
(i,j = 1,2,3): &L^{(i)} \left(H_d, H_u\right) (\bar \nu^{(j)}_R, \bar e^{(j)}_R) & \sim &\left( \begin{array}{ccc} 0 & 0 & 0 \\ 0 & 0 & 0 \\ 0 & 0 & 
{\cal O}(e^{-v_2/24})    \end{array}  \right) ;\\ 
(i,j=2,3):&  {(Q_L^{(i)})}_{ab'} \left(H_d, H_u\right) \left(\bar u^{(j)}_R, \bar d^{(j)}_R\right)_{ac} & \sim & \left( \begin{array}{cc} 0 & 0 \\ 0 & {\cal O}(1) 
\end{array} \right).
\end{array}
\end{equation}
Similarly to the `local' MSSM model, only two  closed sequences, namely $[b, d', (\omega c)']$ and $[a,b',(\omega c)]$, and thus only two
perturbative Yukawa-couplings exist, as can be read off from the distribution of matter states per sector in tables~\ref{Tab:LR-distribution_1} 
and~\ref{Tab:LR-distribution_2}.
 \begin{table}[h]
\begin{center}
\begin{tabular}{|c||c|c|c|c|}
\hline \multicolumn{5}{|c|}{\bf Counting of states for a 4-stack LR sym.~model, part I} \\
\hline \hline $(\chi^{x y}, \chi^{x (\omega y)},\chi^{x (\omega^2 y)})$ & $y=a$ & $y=b$ & $y=c$ & $y=d$\\
\hline\hline $x=a$& (0,0,0) & (0,1,0) & (-2,-1,1) &  (0,0,0)\\
$x=b$& & (0,0,0) & (0,0,0) & (1,0,-1) \\
$x=c$& & & (0,4,-4) & (2,0,1)\\
$x=d$ &&&& (0,2,-2)\\
\hline
\end{tabular}
\caption{Counting of states per sector for the `local' left-right symmetric D6-brane configuration in table~\protect\ref{Tab:LRSM1}. For each sector in this table, 
the total amount of matter is given by $\varphi^{x (\omega^k y)}=|\chi^{x (\omega^k y)}|$.}
\label{Tab:LR-distribution_1}
\end{center}
\end{table}
\begin{table}[h]
\begin{center}
\begin{tabular}{|c||c|c|c|c|}
\hline \multicolumn{5}{|c|}{\bf Counting of states for a 4-stack LR sym.~model, part II} \\
\hline \hline $(\chi^{x y'}, \chi^{x (\omega y)'},\chi^{x (\omega^2 y)'})$ & $y=a$ & $y=b$ & $y=c$ & $y=d$\\
\hline\hline $x=a$& (0,0,0) & (1,0,1) & (1,0,-2) &  (0,0,$|2|$)\\
$x=b$& & (0,0,0) & (0,-1,0)) & (1,0,2) \\
$x=c$& & & ($|4|$,-2,2) & (1,1,-2)\\
$x=d$ &&&& (2,2,$|4|$)\\
\hline
\end{tabular}
\caption{Counting of states per sector for the `local' left-right symmetric D6-brane configuration in table~\protect\ref{Tab:LRSM1}. If  the net-chirality 
$\chi^{x(\omega^k y)'} =0$ vanished but the total amount of matter $\varphi^{x(\omega^k y)'} \neq 0$ is non-zero, the net-chirality is replaced by the absolute 
value of the total amount of matter $|\varphi^{x(\omega^k y)'}|$. The net-chirality $\chi^{x(\omega^k x)'}$ counts the symmetric + antisymmetric matter as introduced 
in equation~(\protect\ref{Eq:ChiralMatter-Sym+Anti}).}
\label{Tab:LR-distribution_2}
\end{center}
\end{table}
The first sequence corresponds to the coupling $L^{(3)} \left(H_d, H_u\right) (\bar \nu^{(3)}_R, \bar e^{(3)}_R)$. The set of triangles set up by the three cycles is 
given by $\{ 6, [5,6,2], 1\}$ and the enclosed area is given by $v_2/24$. 
The other sequence $[a,b',(\omega c)]$ is related to the coupling ${(Q_L^{(3)})}\left(H_d, H_u\right) \left(\bar u^{(3)}_R, \bar d^{(3)}_R\right)$. For this sequence 
the three cycles intersect in one single point on all three tori, implying that the enclosed area is zero and that the Yukawa-coupling is not area-suppressed. 
The other couplings in equation~(\ref{Eq:YukLRSM}) cannot be realized by a closed sequence of three cycles, implying  vanishing perturbative Yuakawa-couplings.
The couplings in equation~(\ref{Eq:YukLRSM}) thus correspond to one massive quark and one massive but lighter lepton generation, which 
can be interpreted as top- and bottom quark and tau and tau neutrino.

As for the `local' MSSM model, it unfortunately turns out to be impossible to cancel the twisted RR tadpoles with additional supersymmetric `hidden' D6-branes.
This can be seen as follows: first of all, the D6-brane configuration given in table~\ref{Tab:LRSM1} already cancels most of the bulk RR charges coming from the 
O6-planes  in table~\ref{tab:Bulk-RR+SUSY-Z2Z6p},
\begin{equation}
N_a X_a + N_b X_b + N_c X_c + X_d N_d  - 16 = -2.
\end{equation} 
Hence,  the `hidden' gauge groups can only wrap on the shortest three-cycles with total rank two. But at the same time, the twisted RR tadpoles in 
table~\ref{tab:twistedRR-AAA+ABB-Z2Z6p} need to
 cancel. Summing up the twisted RR tadpole contributions for the local four-stack model one finds ($i \in \{1,2,3\}$)
\begin{equation}
\sum_{x \in \{a, b, c, d \}} N_x \left( \Pi^{\Z_2^{(i)}}_x + \Pi^{\Z_2^{(i)}}_{x'}  \right) = \left\{ \begin{array}{l} 
-4 \, \varepsilon^{(1)}_1 + 5 \, \varepsilon^{(1)}_2 - \varepsilon^{(1)}_3 -3  \, \varepsilon^{(1)}_5 -3 \, \tilde\varepsilon^{(1)}_4 +3 \, \tilde\varepsilon^{(1)}_5 \\
 3 \, \varepsilon^{(2)}_2 + 5 \, \varepsilon^{(2)}_3 - 4 \, \varepsilon^{(2)}_4 -  \varepsilon^{(2)}_5  + 3 \, \tilde\varepsilon^{(2)}_4 - 3 \, \tilde\varepsilon^{(2)}_5 \\
9 \, \varepsilon^{(3)}_1 + 3 \, \varepsilon^{(3)}_2 +8 \, \varepsilon^{(3)}_3 -  \varepsilon^{(3)}_4 + 2 \, \varepsilon^{(3)}_5 + 3 \, \tilde\varepsilon^{(3)}_4 -3 \, \tilde\varepsilon^{(3)}_5 
\end{array}  \right. ,
\end{equation} 
and as for the `local' MSSM model, the large prefactors of several $\varepsilon^{(i)}_{\alpha}$, $\tilde{\varepsilon}^{(i)}_{\alpha}$ 
make it impossible to find a global supersymmetric completion.

\subsection{A Global Pati-Salam Model}\label{Ss:PS-model}

Given that the `local' MSSM-like and the left-right-symmetric models on rigid D6-branes on  $T^6/(\Z_2 \times \Z_6' \times \OR)$
cannot be completed to globally consistent supersymmetric models, 
it remains to be seen if globally defined supersymmetric Grand Unified Theories can be found. These are on the one hand even more restrictive, but require less stacks 
of D6-branes to construct the much simpler chiral spectrum. The most simple choice of an $SU(5)$ GUT in perturbative \mbox{Type II} string theory
is excluded on the $T^6/(\Z_2 \times \Z_6' \times \OR)$  orientifold due to the relation~(\ref{Eq:CSeqCA}), which states that the amount of chiral matter in the 
symmetric ${\bf 15}$  and antisymmetric ${\bf 10}$ representations of $SU(5)$ is  for supersymmetric D6-branes identical due to the 
vanishing intersection number with the O6-planes.
GUTs based on the gauge groups $SO(10)$ or enhancements thereof like $E_8$ are as usual not accessible for model building in the perturbative regime of Type II 
string theories.

The above considerations only leave over the option of supersymmetric Pati-Salam models, which have already proven favourable for model building on 
other Type IIA orientifolds such as on $T^6/(\Z_2 \times \Z_2 \times \OR)$~\cite{Cvetic:2011vz} and
$T^6/(\Z_2 \times \Z_4 \times \OR)$ without discrete torsion~\cite{Honecker:2003vq} and on $T^6/(\Z_6' \times \OR)$~\cite{Gmeiner:2007zz}. 

\subsubsection{D6-Brane Configuration And Matter Spectrum}\label{Sss:PS-Spectrum}

In contrast to the `local' MSSM and left-right-symmetric models, for which no global supersymmetric completion exists, 
it turns out to be straightforward to construct globally consistent Pati-Salam models on the $T^6/(\Z_2 \times \Z_6' \times \OR)$  orientifold. 
The starting assumptions are similar to the ones in the previous subsections: the $SU(4)$ and $SU(2)_L$ stacks wrap completely rigid fractional three-cycles
on the {\bf AAA} lattice, which are in addition free of matter in symmetric and antisymmetric representations. Also for the $SU(2)_R$ stack, 
a rigid fractional three-cycle on the orbit of the toroidal wrapping numbers (0,1;1,0;1,-1) can be chosen as the exhaustive lists in 
tables~\ref{Tab:RigFreeSet1},~\ref{Tab:RigFreeSet2} and~\ref{Tab:RigFreeSet3} in appendix~\ref{A:RigidD63Gen} show.
For each of the 27 three-stack configurations, the bulk RR tadpole cancellation condition requires additional supersymmetric `hidden' gauge groups with
$\sum_{x \in \text{ hidden}} N_x X_x =8$. Furthermore, the chiral spectrum charged under the gauge groups $U(4)\times U(2)_L \times U(2)_R$ given in table~\ref{Tab:CHSpPSallsets} is exactly the same for each of the 27 `local' three-stack configurations.
\begin{SCtable}
\begin{tabular}{|c|c|}
\hline \multicolumn{2}{|c|}{\bf Chiral spectrum of `local' Pati-Salam models} \\
\hline \hline\bf sector & $U(4)\times U(2)_L \times U(2)_R$\\
\hline \hline $ab$& $({\bf 4}, {\bf\bar 2}, \1)$\\
$ab'$ & $2 \times ({\bf4}, {\bf 2}, \1)$\\
$ac$ & $({\bf\bar 4}, \1,{\bf 2})$\\
$ac'$ &  $2 \times ({\bf\bar 4}, \1,{\bf\bar 2})$\\
$bc$&$(\1, {\bf2}, {\bf\bar 2})$\\
\hline
\end{tabular}
\caption{Common chiral spectrum for all `local' three-stack Pati-Salam configurations given in tables~\ref{Tab:RigFreeSet1},~\ref{Tab:RigFreeSet2} 
and~\ref{Tab:RigFreeSet3}.
The $U(4)^3$ anomaly cancels by construction, while the `generalised anomalies' of $U(2)_L^3$ and  $U(2)_R^3$ derived from the RR tadpole cancellation conditions
 require additional charged matter in a global completion.  
\label{Tab:CHSpPSallsets}}
\end{SCtable}

For all these `local' three-stack configurations, the non-Abelian $U(4)_a^3$ gauge anomaly cancels by construction while the non-Abelian $U(2)_b^3$ and $U(2)_c^3$ 
`generalised gauge anomalies' do not. The `generalised gauge anomaly' cancellation condition for any $U(N_x)$ on globally defined models,
\begin{equation}  
\begin{aligned}
\Pi_x \circ & \underbrace{\left[\sum_a N_a \left( \Pi_a^{\text{frac}} +  \Pi_{a'}^{\text{frac}} \right) - \Pi_{O6} \right]} =0
,
\\
& \qquad \stackrel{\text{RR tadpole cancellation}}{=} 0
\end{aligned}
\end{equation}
enforces in addition the existence of matter with net-chirality $6 \times (\1,\ov{\2},\1) + 6 \times (\1,\1,\2)$ under $U(2)_b \times U(2)_c$. 
Its origin from bifundamental representations involving some `hidden' gauge charges rules out a single `hidden' $U(8)$ factor, but is consistent with `hidden' 
gauge factors like $U(6)\times U(2)$, $U(3)^2\times U(2)$, $U(2)^3$, $U(2)\times U(2)$ etc., which might provide global completions.
A thorough scan, however, shows that, while the bulk RR tadpole cancellation condition and `generalised anomaly constraints' are easily satisfied, the cancellation 
of twisted  RR charges severely constricts the `hidden' sector.
The largest `hidden' gauge group that was found in an extensive search consists of two `hidden' $U(2)$ gauge groups, one of which is wrapped on a next-to-shortest 
three-cycle not parallel to any O6-plane and the other on a shortest orbit parallel to the exotic O6-plane orbit. For each of the 27 `local' models in 
tables~\ref{Tab:RigFreeSet1}, \ref{Tab:RigFreeSet2} and~\ref{Tab:RigFreeSet3}, a globally consistent Pati-Salam model with vanishing RR tadpoles was constructed. 
This allows for a division of the 27 models into three subgroups, depending on which next-to-shortest orbit is wrapped by one of the `hidden' $U(2)$ gauge groups, 
as listed in table~\ref{Tab:SummPSU2U2}.  
\begin{SCtable}
\begin{tabular}{|c|c|c|}
\hline \multicolumn{3}{|c|}{\bf Classification of Pati-Salam models with `hidden' $U(2)\times U(2)$}\\
\hline
\hline \bf (D6$_a$, D6$_b$, D6$_c$) stack  & \bf $U(2)_d$ stack & \bf $U(2)_e$ stack \\
\hline \hline {\bf set 1:} 9 combinations in table \ref{Tab:RigFreeSet1} & (-1,2;2,-1;1,-1) & (1,0;1,0;1,0) \\
{\bf set 2:} 9 combinations in table \ref{Tab:RigFreeSet2}  & (2,-1;1,0;1,1) & (1,0;1,0;1,0)   \\
{\bf set 3:} 9 combinations table in \ref{Tab:RigFreeSet3}  & (1,0;2,-1;1,1) & (1,0;1,0;1,0) \\
\hline
\end{tabular}
\caption{Classification of the globally defined Pati-Salam models according to the three-cycles wrapped by the `hidden' gauge group $U(2)^2$.\label{Tab:SummPSU2U2}}
\end{SCtable}
Within each set of 9 models in table~\ref{Tab:SummPSU2U2},  the chiral spectrum is identical, and the differences among the three different sets are minimal. 
Besides this systematically investigated set of models, there exist models with different `hidden' gauge groups such as the example with 
$U(4)_d \times U(2)_e \times U(2)_f$ given in appendix~\ref{A:6stack_PS}, by loosening some of the constraints put on the `QCD stack'.

The focus in the remainder of this section is on one particular globally defined Pati-Salam model with `hidden' gauge group $U(2)_d \times U(2)_e$, which is based on
combination 4 of set 1 in table~\ref{Tab:RigFreeSet1}. The full D6-brane configuration is given in table~\ref{Tab:decentPatiSalam1}, and the chiral and non-chiral
spectra are given in tables~\ref{Tab:CSdecentPatiSalam1} and~\ref{Tab:NCSdecentPatiSalam1}, respectively.
\begin{table}[h]
\begin{center}
\begin{tabular}{|c||c|c||c|c|c||c|}\hline 
\muc{7}{|c|}{\bf D6-brane configuration of a global Pati-Salam model}
\\\hline \hline
&\bf wrapping numbers &$\frac{\rm Angle}{\pi}$&\bf $\Z_2^{(i)}$ eigenvalues  & ($\vec \tau$) & ($\vec \sigma$)&\bf gauge group\\
\hline \hline
 $a$&(0,1;1,0,1,-1)&$(\frac{1}{3},0,-\frac{1}{3})$&$(+++)$&$(0,0,1)$ & $(1,1,1)$& $U(4)$\\
 $b$&(0,1;1,0,1,-1)&$(\frac{1}{3},0,-\frac{1}{3})$&$(--+)$&$(0,1,1)$ & $(1,1,1)$&$U(2)_L$\\
 $c$&(0,1;1,0,1,-1)&$(\frac{1}{3},0,-\frac{1}{3})$&$(-+-)$&$(1,0,1)$ & $(1,1,1)$&$U(2)_R$\\
 \hline $d$ & (-1,2;2,-1;1,-1) &$(\frac{1}{2},-\frac{1}{6},-\frac{1}{3})$ & $(--+)$ &$(0,0,1)$&$(1,1,1)$&$U(2)_d$\\
 $e$ & (1,0;1,0;1,0)& $(0,0,0)$  &$(+--)$&$(1,1,1)$& $(1,1,0)$&$U(2)_e$\\
 \hline
\end{tabular}
\caption{D6-brane configuration with five stacks of D6-branes yielding a globally defined Pati-Salam model with gauge group 
$SU(4)_a\times SU(2)_b\times SU(2)_c \times SU(2)_d \times SU(2)_e\times U(1)^5_{\text{massive}}$.\label{Tab:decentPatiSalam1}}
\end{center}
\end{table}
The stack $d$ is not wrapped on a shortest three-cycle and therefore carries some matter in the adjoint representation, whereas all other stacks $a$, $b$, $c$ 
and $e$ are wrapped on completely rigid three-cycles. Both `hidden' gauge groups carry some amount of non-chiral matter in the symmetric and antisymmetric 
representation.
\begin{table}[h]
\begin{center}
\begin{tabular}{|c||c|c|c|}
\hline \multicolumn{4}{|c|}{\bf Chiral spectrum of a global five-stack Pati-Salam model}\\
\hline \hline
Matter & Sector & $U(4)\times U(2)_L \times U(2)_R \times U(2)_d \times U(2)_e$& $(Q_a, Q_b, Q_c, Q_d, Q_e)$ \\
\hline $\left(Q_L , L\right)  $&$ab$&$({\bf 4}, {\bf\bar 2}, \1,\1,\1)$ &(1,-1,0,0,0)\\
$\left( Q_L, L \right)$&$ab'$&  $2 \times ({\bf4}, {\bf 2}, \1,\1,\1)$&(1,1,0,0,0)\\
$\left( \bar u_R, \bar d_R, \bar\nu_R, \bar e_R \right)  $ & $ac$ & $({\bf\bar 4}, \1,{\bf 2},\1,\1)$ &(-1,0,1,0,0) \\
$\left( \bar u_R, \bar d_R, \bar\nu_R, \bar e_R \right)  $& $ac'$ &  $2 \times ({\bf\bar 4}, \1,{\bf\bar 2},\1,\1)$&(-1,0,-1,0,0) \\
$\left(H_d, H_u\right)$&$bc$&$(\1, {\bf2}, {\bf\bar 2},\1,\1)$&(0,1,-1,0,0) \\
& $bd$ & $(\1, {\bf 2}, \1, {\bf \bar 2},\1 ) $&(0,1,0,-1,0)\\
&$bd'$&  $3 \times (\1, {\bf \bar  2}, \1, {\bf \bar 2},\1 ) $ &(0,-1,0,-1,0)\\
&$be'$&$(\1, {\bf \bar  2}, \1, \1, {\bf \bar 2}) $&(0,-1,0,0,-1)\\
&$cd$& $(\1, \1, {\bf \bar 2}, {\bf 2},\1 ) $   &(0,0,-1,1,0) \\
&$cd'$&  $3 \times (\1, \1, {\bf  2}, {\bf 2},\1 ) $ & (0,0,1,1,0)\\
&$ce'$&  $(\1, \1, {\bf 2}, \1, {\bf 2}) $ & (0,0,1,0,1)\\
\hline
\end{tabular}
\caption{Chiral spectrum for the D6-brane configuration in table~\ref{Tab:decentPatiSalam1}. 
All states from the Higgs pair downwards are chiral with respect to the anomalous $U(1) \subset U(2)$ factors,
but non-chiral with respect to the Pati-Salam group.\label{Tab:CSdecentPatiSalam1}}
\end{center}
\end{table}

 The Higgs-sector of the model is minimal with only one ($H_d$, $H_u$) doublet coming from the $b(\omega c)$ sector. 
Since the three left-handed quark generations split into two with charge $+1$ under $U(1)_b \subset U(2)_L$ and one with charge $-1$
and similarly for the right-handed quarks under $U(1)_c \subset U(2)_R$, charge selection rules only allow for one 
 Yukawa-interaction of the Higgs-doublet from the $bc$ sector with the left-handed quark in the $ab$ sector and  right-handed quark in the $ac$ sector,
\begin{equation}\label{Eq:PS-Yukawa}
\left( \bar u_R, \bar d_R, \bar\nu_R, \bar e_R \right)_{ac}  \left(H_d, H_u\right) \left(Q_L , L\right)_{ab}.
\end{equation} 
The remaining two generations do not possess perturbative Yukawa couplings by charge selection rules.
The perturbative three-point couplings are discussed further in section~\ref{Sss:PS-Yukawas}.

The chiral spectrum in table~\ref{Tab:CSdecentPatiSalam1} contains besides the Standard Model particles and minimal Higgs sector a number of states in  
bifundamental representations involving  one charge under $U(2)_L\times U(2)_R$ and the other under one of the `hidden' gauge groups. 
Since all $U(1)$ factors in this model are anomalous and massive as further discussed in section~\ref{S:U1s}, these states are non-chiral with respect to 
the non-Abelian gauge group and in particular the observable Pati-Salam group $SU(4)_a \times SU(2)_L \times SU(2)_R$. Last but not least, the non-chiral spectrum 
in table~\ref{Tab:NCSdecentPatiSalam1} contains one pair of multiplets
charged in the fundamental and antifundamental representation of $SU(4)$. This constitutes the minimal amount of exotic matter charged under the strong interactions, 
which has been found in any globally consistent Pati-Salam model on $T^6/(\Z_2 \times \Z_6' \times \OR)$ with discrete torsion, as can be seen by comparison with 
the six-stack example in table~\ref{Tab:OtherDecentPatiSalamNC} of appendix~\ref{A:6stack_PS}. 
\begin{table}[h]
\begin{center}
\begin{tabular}{|c|c|c|}
\hline \multicolumn{3}{|c|}{\bf Non-chiral spectrum  of a global five-stack Pati-Salam model}\\
\hline \hline sector  & $U(4)\times U(2)_L \times U(2)_R \times U(2)_d \times U(2)_e$& $(Q_a, Q_b, Q_c, Q_d, Q_e)$ \\
\hline
$dd$ & $(\1,\1,\1,{\bf 4}_\Adj,1)$ & $(0,0,0,0,0)$ \\
$dd'$ &  $2 \times [(\1,\1,\1, {\bf 3_{\Sym}},1) + h.c.]$ & $(0,0,0,\pm 2,0)$ \\
 &   $2\times [(\1,\1,\1, {\bf 1_{\Anti}},1) + h.c.]$ & $(0,0,0,\pm 2,0)$\\
 $ee'$ & $[(\1,\1,\1, \1, {\bf 3_{\Sym}}) + h.c.]$ & $(0,0,0, 0, \pm 2)$\\
 & $[(\1,\1,\1, \1, {\bf 1_{\Anti}}) + h.c.]$& $(0,0,0,0,\pm 2)$\\
 $ad$ &   $2 \times [ ({\bf 4}, \1,\1, {\bf \bar 2},\1) + h.c.] $ & $(\pm 1, 0, 0,\mp 1,0)$ \\
 $de'$ &  $(\1, \1, \1, {\bf 2}, {\bf 2}) + h.c. $  & $(0,0,0,\pm1,\pm 1)$ \\
 \hline
\end{tabular}
\caption{Non-chiral spectrum for the globally defined Pati-Salam D6-brane configuration in table~\ref{Tab:decentPatiSalam1}.\label{Tab:NCSdecentPatiSalam1}}
\end{center}
\end{table}

While the amount of exotic matter charged under $SU(4)$ is minimal compared to any of the toroidal orbifold models of 
Type IIA/$\OR$ on $T^6/\Z_{2N}$ or $T^6/\Z_2 \times \Z_{2M}$ without discrete torsion, e.g. in~\cite{Honecker:2003vq,Gmeiner:2007zz,Cvetic:2011vz},
it is necessary to either find field theoretical couplings which render the exotic matter charged under $SU(4)_a \times SU(2)_L \times SU(2)_R$
massive or show that its couplings to visible matter and the electroweak gauge bosons are weak enough to have evaded experimental detection
to date.

\subsubsection{Yukawa and other Three-Point Interactions for the Pati-Salam Model}\label{Sss:PS-Yukawas}

Charge selection rules allow for the perturbative Yukawa coupling~(\ref{Eq:PS-Yukawa}) of one particle generation.
Similarly, three-point couplings among the non-Standard Model particle in table~\ref{Tab:PS_per-Sector_1} and with some of the non-chiral states in 
table~\ref{Tab:NCSdecentPatiSalam1} can potentially provide mass terms for matter charged under the `hidden' gauge group $SU(2)_d \times SU(2)_e$ 
if some scalar in the non-chiral sector (including states with $U(1)^5_{\text{massive}}$ chirality) acquires a {\it vev}. 

To determine the value of the Yukawa coupling in equation~(\ref{Eq:PS-Yukawa}) and of other three-point couplings, it is a priori necessary to 
determine the localisation of each matter and Higgs state per intersection sector, just as for the `local' MSSM and left-right
symmetric models. The result is collected in tables~\ref{Tab:PS_per-Sector_1} and~\ref{Tab:PS_per-Sector_2}. 
\begin{table}[h]
\begin{center}
\begin{tabular}{|c||c|c|c|c|c|}
\hline \multicolumn{6}{|c|}{\bf Counting of states for a five-stack Pati-Salam model, Part I} \\
\hline \hline $(\chi^{x y}, \chi^{x (\omega y)},\chi^{x (\omega^2 y)})$ & $y=a$ & $y=b$ & $y=c$ & $y=d$&$y=e$\\
\hline\hline $x=a$& (0,0,0) & (0,1,0) & (0,0,-1) &  (0,2,-2)&(0,0,0)\\
$x=b$& & (0,0,0) & (0,1,0) & (1,0,0) & (0,0,0) \\
$x=c$& & & (0,0,0) & (-1,0,0) &(0,0,0)\\
$x=d$ &&&& (0,1,-1)&(0,0,0)\\
$x=e$ &&&&& (0,0,0)\\
\hline
\end{tabular}
\caption{Counting $\chi^{x (\omega^k y)}$ of chiral states per sector $x(\omega^k y)$ for the D6-brane configuration in table~\protect\ref{Tab:decentPatiSalam1}.
For each sector in this table, the total amount of matter is given by $\varphi^{x (\omega^k y)}=|\chi^{x (\omega^k y)}|$.}
\label{Tab:PS_per-Sector_1}
\end{center}
\end{table}
%
\begin{table}[h]
\begin{center}
\begin{tabular}{|c||c|c|c|c|c|}
\hline \multicolumn{6}{|c|}{\bf Counting of states for a five-stack Pati-Salam, Part II} \\
\hline \hline $(\chi^{x y'}, \chi^{x (\omega y)'},\chi^{x (\omega^2 y)'})$ & $y=a$ & $y=b$ & $y=c$ & $y=d$&$y=e$\\
\hline\hline $x=a$& (0,0,0) & (0,1,1) & (-1,0,-1) &  (0,0,0)&(0,0,0)\\
$x=b$& & (0,0,0) & (0,0,0) & (0,-1,-2)&(-1,0,0) \\
$x=c$& & & (0,0,0) & (1,0,2)&(0,0,1)\\
$x=d$ &&&& (-2,2,$|4|$)&(1,0,-1)\\
$x=e$ &&&&&($|2|$,-1,1) \\
\hline
\end{tabular}
\caption{Counting of states per sector for the D6-brane configuration in table~\protect\ref{Tab:decentPatiSalam1}. 
If  the net-chirality $\chi^{x(\omega^k y)'} =0$ vanishes but the total amount of matter $\varphi^{x(\omega^k y)'} \neq 0$ is non-zero,
 the net-chirality is replaced by the (absolute value of the) total amount of matter $|\varphi^{x(\omega^k y)'}|$. 
 The net-chirality $\chi^{x(\omega^k x)'}$ counts the amount of symmetric + antisymmetric matter as introduced in equation~(\ref{Eq:ChiralMatter-Sym+Anti}).}
\label{Tab:PS_per-Sector_2}
\end{center}
\end{table}
Table~\ref{Tab:PS_per-Sector_1} shows that the chiral matter state $\left(Q_L , L\right)_{ab}$ arises from the sector $a (\omega b)$, confirmed by the fact that the three-cycles $a$ and $(\omega b)$ intersect only 
one time on each torus $T^2_{(i)}$. Similarly, the chiral state $\left( \bar u_R, \bar d_R, \bar\nu_R, \bar e_R \right)_{ac}$ can be traced back to the $a (\omega^2 c)$ sector and the chiral state $ \left(H_d, H_u\right)$ to the $b (\omega c)$ sector. The sequence $[a, (\omega^2 c), (\omega b)]$ then represents the corresponding closed sequence built from these sectors.  On each of the three two-tori $T^2_{(i)}$ the triangle enclosed by the intersecting three-cycles has the $\Z_2$ fixed points $4$, $5$ and $6$ as apexes with a non-zero area $v_i/8$. Hence, the Yukawa interaction in equation (\ref{Eq:PS-Yukawa}) is allowed and exponentially suppressed by a factor ${\cal O}(e^{-\sum_{i=1}^3v_i/8})$.

\begin{sidewaystable}[h]
\begin{center}
\begin{tabular}{|c|c||c|c||c|}
\hline \multicolumn{5}{|c|}{\bf Yukawa-couplings and mass generating three-point couplings for a 5-stack Pati-Salam model} \\
\hline \hline \bf sequence $[x,y,z]$ & \bf coupling & \bf triangle & \bf enclosed area& \bf K\"ahler factor $(K_{xy} K_{yz} K_{zx} )^{-1/2}$\\
\hline \hline $[a,(\omega^2 c), (\omega b)]$ & $Q_{ab}\, U_{ac}\, H_{bc}$ &$\{ [4,5,6], [6,4,5], [5,6,4] \}$ & $\frac{1}{8} \sum_{i=1}^3 v_i $ & $\left(\frac{v_1 v_2 v_3}{ \sqrt{8}\, g_{\text{string}} }\right)^{3/2}$ \\
\hline$[b,d,e']$ & $H_{bd}\, \bar h_{be'}\, h_{de'}$ & $\{ 6, 6, 4 \}$ & 0 & $ \left(\frac{v_1 v_2 v_3}{2\pi \,g_{\text{string}}}\right)^{3/2} \left( \frac{\ell_{\text{s}}^2}{\sqrt{10}\,r_2 r_3}\right)^{1/2}$ \\
\hline$[b, (\omega c), (\omega d)']$ & 
$\bar h^{(3)}_{bd'}\, h_{cd'}^{(3)} \, H_{bc}$ & $\{ 4,6,5 \}$ & 0 & \multirow{2}{*}{$\left(\frac{v_1 v_2 v_3}{g_{\text{string}}}\right)^{3/2} \left(\frac{2 }{25\sqrt{8}}\right)^{1/2}  $} \\
&$\bar h^{(3)}_{bd'}\, h_{cd'}^{(2)} \, H_{bc}$& $\{ [4,5,4],6,5 \} $ & $\frac{v_1}{4}$ &\\
\hline$[c,d,(\omega^2 e)']$& $\bar H_{cd}\, h_{ce'} \, \bar h_{de'}$ & $\{ 6,6,5 \}$ & 0 & $ (2\pi)^{-3/4} \left(\frac{v_1 v_2 v_3}{2\pi \,g_{\text{string}}} \frac{ \ell_{\text{s}}}{r_3}\right)^{3/2}  $ \\
\hline$[a, (\omega d), (\omega^2 d)]$ & $Q^{(1)}_{ad}\,  A_{d}\, U^{(1)}_{ad} $& $\{ 4,5, [5,6,4] \}$ & $\frac{v_3}{8}$ &\multirow{4}{*}{$(10 \sqrt{8})^{-1/2} \left(\frac{v_1 v_2 v_3}{g_{\text{string}}}\right)^{3/2} $  } \\
&$Q^{(2)}_{ad}\,  A_{d}\, U^{(1)}_{ad} $& $\{[6,(R,R'),4], 5, [5,6,4]  \}$ & $\frac{v_1}{12}+\frac{v_3}{8}$ &  \\
&$Q^{(1)}_{ad}\,  A_{d}\, U^{(2)}_{ad} $& $\{ 4,[5,(S,S'),6], [5,6,4]  \}$ & $\frac{v_2}{12}+\frac{v_3}{8}$ & \\
&$Q^{(2)}_{ad}\,  A_{d}\, U^{(2)}_{ad} $& $\{ [6,(R,R'),4],[5,(S,S'),6], [5,6,4]  \}$ & $\frac{v_1+v_2}{12}+\frac{v_3}{8}$ & \\
\hline$[d,(\omega d), d']$ & $T^{(1)}_d\, A_d\, \bar T^{(1)}_d$ & $\{ 6,4,[5,6,4] \}$ & $\frac{v_3}{8}$ & \multirow{4}{*}{$ \left(\frac{\ell^2_{\text{s}}}{3 \sqrt{8}À, r_1 r_2 }\right)^{1/2} \left(\frac{v_1 v_2 v_3}{2\pi\,g_{\text{string}}}\right)^{3/2}$} \\
 & $T^{(2)}_d\, A_d\, \bar T^{(1)}_d$& $\{ (R,R'),4,[5,6,4] \}$ & $\frac{v_3}{8}$ & \\
  &$T^{(1)}_d\, A_d\, \bar T^{(2)}_d$ & $\{ 6,(S,S'),[5,6,4] \}$ & $\frac{v_3}{8}$ & \\
   &$T^{(2)}_d\, A_d\, \bar T^{(2)}_d$ & $\{ (R,R'),(S,S'),[5,6,4] \}$ & $\frac{v_3}{8}$ & \\
\hline$[d, (\omega^2 d), e']$ & $h_{de'}\, A_d\, \bar h_{de'}$ & $\{ 5, 6, 4 \}$ & 0 & $(10 \sqrt{8})^{-1/2} \left(\frac{v_1 v_2 v_3}{g_{\text{string}}}\right)^{3/2} $ \\
\hline
\end{tabular}
\caption{Overview of Yukawa interactions and mass generating three-point interactions for the Pati-Salam model presented in table \ref{Tab:decentPatiSalam1}. The triangles formed by the sequence $[x,y,z]$ are split over the three tori and listed in the middle column. A triple intersection in a single point is indicated by the $\Z_2$ fixed point or by a $\Z_2$ invariant tuple $(R,R')$. A non-trivial triangle is denoted by $[i,j,k]$ where $i$, $j$ and $k$ denote the apices of the triangle. The expressions for the K\"ahler metrics $K_{xy}$ used in the K\"ahler factor, 
which enters the physical Yukawa couplings $Y_{ijk} = (K_{xy} K_{yz} K_{zx} )^{-1/2} \, e^{\kappa^2_4 {\cal K}/2} \, W_{ijk}$,
 are given in table~\ref{tab:beta_coeffs_Kaehler_metrics}.  \label{Tab:YukPS1Full}}
\end{center}
\end{sidewaystable}
Table \ref{Tab:YukPS1Full} provides a full list of perturbative three-point interactions generating mass terms for the chiral and non-chiral matter states of the Pati-Salam model presented above. For each three-point interaction,
 the corresponding triangle on each torus $T^2_{(i)}$ is indicated and the enclosed area ${\cal A}_{ijk}$ in terms of the areas $v_i$ of the two-tori $T^2_{(i)}$ is given. 
 To simplify the expressions for the couplings, the following notation for the matter states is introduced,
\begin{equation}\label{Eq:PS1NotationStates}
\begin{array}{crl@{\hspace{0.4in}}crl}
({\bf 4}, {\bf\bar 2}_y, \1,\1,\1)^{(i)}_{xy}& \rightarrow &Q^{(i)}_{xy}, &
({\bf\bar 4}, {\bf 2}_y,\1,\1,\1)^{(i)}_{xy}&\rightarrow & U^{(i)}_{xy},\\
(\1, {\bf2}_x, {\bf\bar 2}_y,\1,\1)_{xy}^{(i)}&\rightarrow &H^{(i)}_{xy},&
(\1, {\bf2}_x, {\bf 2}_y,\1,\1)_{xy'}^{(i)}&\rightarrow &h^{(i)}_{xy'},\\
\Sym^{(i)}_x &\rightarrow & S^{(i)}_x, &
\Adj_x&\rightarrow & A_x,\\
\Anti^{(i)}_x&\rightarrow & T^{(i)}_x, &&&
\end{array}
\end{equation}
where the parameter $i$ counts the multiplicity of the state and is omitted when the respective state only occurs once. 

The one of the two non-chiral matter pairs charged under $SU(4)$ can be made massive by turning on a non-zero {\it vev} for the multiplet in the 
adjoint representation of $U(2)_d$, causing the gauge group to break spontaneously if the {\it vev} is chosen in the irreducible 
$\3$ of $SU(2)_d$ or without breaking the gauge group if the {\it vev} is switched on for the singlet in the decomposition ${\bf 4}=\3 + \1$. 
The same reducible multiplet in the adjoint representation of $U(2)_d=SU(2)_d \times U(1)_d$ can be used in a three-point coupling with the 
non-chiral  matter in the antisymmetric representation of $SU(2)_d$ to make one pair of the latter states massive. Also the bifundamental
$de'$ states in the last line of table~\ref{Tab:PS_per-Sector_2} acquire a mass by the same {\it vev}.

One combination of bifundamental states $\bar{h}^{(3)}_{bd'}=(\1,\ov{\2},\1,\ov{\2},\1)$ and $h^{(i)}_{cd'}=(\1,\1,\2,\2,\1)$, which are 
chiral with respect to the anomalous $U(1)^5_{\text{massive}}$ symmetries, receives a mass by coupling to the standard Higgs particle, 
and four more bifundamental states in the lower half of table~\ref{Tab:CSdecentPatiSalam1} couple perturbatively to bifundamental states in
the last line of table~\ref{Tab:NCSdecentPatiSalam1}, which provides mass terms if some scalar in a bifundamental representation $(\2_x,\2_y)$
or $(\2_x,\ov{\2}_y)$ or its conjugate receives a {\it vev}, breaking $SU(2)_x \times SU(2)_y$ to its diagonal subgroup.
 
The non-chiral matter in the symmetric representation of $SU(2)_d$ arises in a single sector $d (\omega^2 d)'$ and cannot be made massive 
through a three-point interaction. Similarly due to the rigidness of cycle $e$ there are no multiplets in the adjoint representation of 
$U(2)_e$ at our  disposal to make the non-chiral matter in the symmetric or antisymmetric representation of  $SU(2)_e$ massive via a 
three-point coupling. 

The remaining exotic matter states have to acquire masses through higher order couplings or non-perturbative effects such as instanton 
corrections.

\clearpage
\section{The U(1)s: Potential $Z'$ Bosons or Dark Photons?}\label{S:U1s}

The article so far focused on the construction of realistic matter spectra and their pattern of perturbative Yukawa interactions.
In this section, the non-Abelian gauge couplings, unification up to one-loop order, consistency with a low string scale $M_{\text{string}}$  
and kinetic mixing parameters for the Abelian gauge factors of the global five-stack Pati-Salam model 
of section~\ref{Sss:PS-Spectrum} are analysed. 
The gauge couplings of $SU(4)_a \times SU(2)_b \times SU(2)_c \times SU(2)_e$ unify at tree level at the string scale since the three-cycle 
lengths in equation~(\ref{Eq:tree-gauge-coupling}) are equal,
\begin{equation}\label{Eq:tree-gauge-repeated}
\frac{4 \pi}{g_{SU(N_x),{\rm tree}}^2} = \frac{1}{4 \, g_{\text{string}}} \frac{r_1 r_2 r_3}{\ell_s^3}
\qquad
\text{for}
\quad
x \in \{a,b,c\}
.
\end{equation}
At one-loop, threshold corrections lead to deviations, which depend on the 
individual two-torus volumes $v_i$ in units of $\alpha'$ as discussed below.

The dimensional reduction of the ten dimensional Einstein-Hilbert term in string frame to four dimensions,
\begin{equation}
{\cal S} = \frac{1}{2\kappa_{10}^2} \int_{\mathbb{R}^{1,9}} d^{10}x \sqrt{-g_{10}} \, e^{-2\phi_{10}} \; {\cal R}_{10} \to
\frac{1}{2\kappa_{4}^2}  \int_{\mathbb{R}^{1,3}}  d^{4}x \sqrt{-g_{4}} \; {\cal R}_{4}
\quad 
\text{with}
\quad
\kappa_{10}^2 = \frac{\ell_s^8}{4\pi}
,\quad 
\kappa_{4}^2 =M^{-2}_{\text{Planck}}
,
\end{equation}
leads to the relation
\begin{equation}\label{Eq:mass-ratios}
\frac{M^{2}_{\text{Planck}}}{M_{\text{string}}^2} = \frac{4\pi}{g^2_{\text{string}}} \, v_1v_2v_3
\end{equation}
with $g_{\text{string}} \equiv e^{\phi_{10}}$ and $v_1v_2v_3 = \left( 2\pi^2 \sqrt{3} \right)^3
\left(\frac{r_1r_2r_3}{\ell_s^3}\right)^2 $
the compact six-dimensional volume  in units of $\ell_s^6\equiv (2\pi \sqrt{\alpha'})^6 $.
With $M_{\text{Planck}} \sim 10^{19} \text{ GeV}$, the size of the compact volume is constrained by the weakness of four-dimensional 
gravity for given choices of the string scale $M_{\text{string}}$ and string coupling constant $g_{\text{string}}$
as displayed for some examples in table~\ref{tab:various-scales}.
\mathtabfix{
\begin{array}{|c||c|c|c|c||c|c|c|c||c|c|c|c|c|c|}\hline
\muc{13}{|c|}{\text{\bf Mass scales and values of the string coupling}}
\\\hline\hline
M_{\text{string}} & \muc{4}{|c||}{\text{1 TeV}} & \muc{4}{|c||}{10^{12} \text{ GeV}}  & \muc{4}{|c|}{10^{16} \text{ GeV}}
\\\hline
g_{\text{string}} & 10^{-3} & 0.01 & 0.1 & 0.5 & 10^{-3} & 0.01 & 0.1 & 0.5 &10^{-3} & 0.01 & 0.1 & 0.5 
\\\hline
v_1v_2v_3 & 8 \cdot 10^{24} & 8 \cdot 10^{26} &  8 \cdot 10^{28} &  2 \cdot 10^{30} & 8 \cdot 10^6 &  8 \cdot 10^8 &  8 \cdot 10^{10}  &  2 \cdot 10^{12} & 0.08 & 8 & 800 & 2 \cdot 10^4
\\\hline
4\pi/g^2_{a,\text{tree}} & \muc{4}{|c||}{4 \cdot 10^{12}} & \muc{4}{|c||}{4 \cdot 10^{3}} & \muc{4}{|c|}{4 \cdot 10^{-1}}
\\\hline
\end{array}
}{various-scales}{Size of the compact six-dimensional volume $v_1v_2v_3 >1 $ in the geometric regime in units of the string length.
The tree-level value of $\alpha^{-1}_{a,\text{tree}} \equiv 4\pi/g_{a,\text{tree}}^2 \approx  4 \cdot 10^{-4} \, M_{\text{Planck}}/M_{\text{string}}$ is by equation~(\protect\ref{Eq:tree-gauge-repeated}) set by the 
ratio of Planck to string scale.
}

The tree level value of the gauge couplings of $SU(4)_a \times SU(2)_b \times SU(2)_c$ is determined 
through the relation~(\ref{Eq:tree-gauge-coupling}) in terms of the string coupling and square root of the 
six-dimensional volume which can by equation~(\ref{Eq:mass-ratios}) be expressed in terms of the ratio
of the Planck to string mass, and numerical values for sample values of the former are displayed in table~\ref{tab:various-scales}.

The tree-level value for low string mass clearly deviates from the measured values at the electro-weak scale,
${\alpha_{\text{strong}}}^{-1} (M_Z) \approx 9 $, $\alpha_{\text{weak}}^{-1} (M_Z) \approx 29$.
However, as will be shown below, one-loop corrections can significantly change the values of the gauge couplings 
at $M_{\text{string}}$. Before discussing this issue, one-loop corrections are briefly introduced in the context of kinetic mixing.

All five Abelian gauge factors of the globally defined five-stack Pati-Salam model are anomalous and receive string scale masses through 
the generalised Green-Schwarz mechanism as follows, see e.g.~\cite{Ibanez:2001nd,MarchesanoBuznego:2003hp}. 
The dual two-forms to the RR scalars in the $\Z_2^{(k)}$  twisted closed string sectors,
\begin{equation}
B_2^{(k),\alpha} \sim \int_{\chi^{(k),\alpha}} C_5^{\text{RR}}
\qquad 
\text{with}
\qquad
\chi^{(k),\alpha} =\left\{\begin{array}{cr} -\rho_1 + 2 \rho_2 & \alpha=0 \\
 -\varepsilon^{(k)}_{\alpha} + 2 \, \tilde{\varepsilon}^{(k)}_{\alpha} & 1,2,3\\
\varepsilon^{(k)}_{4} - \varepsilon^{(k)}_{5} & 4,5\\ 
\tilde{\varepsilon}^{(k)}_{4} + \tilde{\varepsilon}^{(k)}_{5} - \frac{\varepsilon^{(k)}_{4} + \varepsilon^{(k)}_{5}}{2} & 4,5
\end{array}\right.
,
\end{equation} 
with $\chi^{(k),\alpha}$ the $\OR$ odd three-cycles on the {\bf AAA} lattice and the choice of exotic O6-plane $\eta_{\OR}=-1$,
couple to the Abelian gauge fields by
\begin{equation}
\sum_{k=1}^3 \sum_{\alpha=1}^5 N_a c^{(k),\alpha}_a \int_{\mathbb{R}^{1,3}} B_2^{(k),\alpha}  \wedge F_a
\quad 
\text{with}
\quad
c^{(k),\alpha}_a = \left\{\begin{array}{cr} Y_a & \alpha=0 \\ y^{(k)}_{a,\alpha} & 1,2,3 \\
x^{(k)}_{a,4} -x^{(k)}_{a,5} + \frac{1}{2}(y^{(k)}_{a,4} -y^{(k)}_{a,5}) & 4,5\\
 y^{(k)}_{a,4} + y^{(k)}_{a,5} & 4,5
\end{array}\right.
.
\end{equation} 
For the five-stack Pati-Salam model, the matrix  in equation (\ref{Eq:GGS-matrix}) has maximal rank
and thus all five $U(1)$ factors receive string scale masses,
\begin{equation} \label{Eq:GGS-matrix}
\left(\begin{array}{c} N_x c^{(1),\alpha}_x \\ N_x c^{(2),\alpha}_x \\ N_x c^{(3),\alpha}_x \end{array}\right)=\left(\begin{array}{ccccc}  
0 & 0 & 0 & 0 & 2 N_e \\
0 & 0 & 0 & 0 & 0 \\
0 & 0 & 0 & - 2 N_d &0 \\
2 N_a & 0 & -2 N_c &-3 N_d  & - N_e \\
0& 0 & 0 & 2 N_d & -2 N_e \\  
0&0 & 0 &0 & -2 N_e \\
0 & 0 & 0 & 0 & 0 \\
0 & 0 & 0 & 2 N_d & 0\\
-2 N_a & 2 N_b & 0 & 3 N_d & N_e \\
0 & 0 & 0 & -2 N_d & 2 N_e \\
 0 &0 & 0 &0 & 0 \\
0 & 0 & 0 & 0 & 0 \\
0 & 0 & 0 & 0 & 0\\
 N_a & 2 N_b & -N_c & - N_d & 0 \\
-2 N_a & 0 & -2 N_c & 2 N_d & 0 \\
\end{array}\right)
.
\end{equation}

The one-loop mixing parameters are given in terms of the holomorphic gauge kinetic functions,
\begin{equation}\label{Eq:1-loop-mixings}
\begin{aligned}
\Re \left(\delta^{\text{1-loop}} {\rm f}_{U(1)_aU(1)_y} \right) &= -\frac{6}{\pi^2} \, \ln \frac{\vartheta_1(\frac{1}{2}, i v_k)}{\eta (i v_k)} 
-\sum_{i \neq k} \frac{2}{\pi^2} \ln(i v_i) + \frac{\ln (2)}{3\pi^2} 
\qquad 
\text{for}
\quad 
\left\{\begin{array}{rr} k=2 & y=b \\ 1 & c \end{array}\right.
,
\\
\Re \left(\delta^{\text{1-loop}} {\rm f}_{U(1)_aU(1)_d}  \right) &= \frac{2 \; \ln (2)}{3\pi^2}
,
\\
\Re \left(\delta^{\text{1-loop}} {\rm f}_{U(1)_aU(1)_e}  \right) &= -\frac{3}{\pi^2} \, \ln \left( e^{-\pi v_3/4}   
\frac{\vartheta_1(\frac{-iv_3}{2}, i v_3)}{\eta (i v_3)} \right)   -\sum_{i=1,2} \frac{2}{\pi^2} \, \ln \frac{\vartheta_1(\frac{1}{2}, i v_i)}{\eta (i v_i)}
,
\\
\Re \left(\delta^{\text{1-loop}} {\rm f}_{U(1)_bU(1)_c}  \right) &=  0
,
\\
\Re \left(\delta^{\text{1-loop}} {\rm f}_{U(1)_bU(1)_d}  \right) &=\Re \left(\delta^{\text{1-loop}} {\rm f}_{U(1)_cU(1)_d}  \right)
= -\frac{3}{\pi^2} \ln(i v_3) - \frac{\ln (2)}{3\pi^2}
,
\\
\Re \left(\delta^{\text{1-loop}} {\rm f}_{U(1)_yU(1)_e}  \right) &= -\frac{2}{\pi^2} \, \ln \left( e^{-\pi v_3/4}   \frac{\vartheta_1(\frac{-iv_3}{2}, i v_3)}{\eta (i v_3)} \right)  -\frac{1}{\pi^2} \, \ln \frac{\vartheta_1(\frac{1}{2}, i v_k)}{\eta (i v_k)} + \frac{\ln (2)}{6\pi^2}
\qquad 
\left\{\begin{array}{rr} k=1 & y=b \\  2 & c \end{array}\right.
,
\\
,
\end{aligned}
\end{equation}
which lead to deviations from gauge coupling unification of the $SU(N_x)_{x \in \{a,b,c,e\}}$ gauge factors
for generic values of the two-torus volume parameters $v_i \equiv \frac{\sqrt{3}}{2} \frac{r_i^2}{\alpha'}$,
as can be seen from the relation
\begin{equation}
\Re \left(\delta^{\text{1-loop}} {\rm f}_{U(1)_xU(1)_{y\neq x}} \right) = 2 N_x \, \Re \left(\delta^{\text{1-loop}}_{y \neq x} {\rm f}_{SU(N_x)}\right)
\end{equation}
and the additional contributions from sectors with adjoint and symmetric plus antisymmetric matter,
\begin{equation}
\begin{aligned}
\Re \left(\delta^{\text{1-loop}, {\cal A}}_{SU(N_x)} {\rm f}_{SU(N_x), x \in \{a,b,c\}} \right) &=
\sum_{i=1}^3 \frac{1}{\pi^2} \ln \eta(iv_i) - \left\{\begin{array}{cc}  \frac{\ln (2)}{6 \pi^2} & a \\ 0 & b,c
\end{array}\right.
,
\\
\Re \left(\delta^{\text{1-loop}, {\cal A}}_{SU(N_d)} {\rm f}_{SU(N_d)} \right) &=  - \frac{1}{2\pi^2} \ln \eta(iv_3) - \frac{\ln (2)}{4\pi^2}
,
\\
\Re \left(\delta^{\text{1-loop}, {\cal A}}_{SU(N_e)} {\rm f}_{SU(N_e)} \right) &= \frac{1}{2\pi^2} \ln \eta(iv_3) + \frac{\ln (2)}{12 \pi^2}
.
\end{aligned}
\end{equation}
For stacks $x \in \{a,b,c\}$ moreover, there is no matter in any $x(\omega^k x')$ sector and therefore it is reasonable to assume that 
no additional one-loop contribution arises from the amplitudes with M\"obius strip topology since all known corrections in 
table~\ref{tab:hol-gauge-kin} of appendix~\ref{A:intersections+betas} are proportional 
to the contributions to beta function coefficients from the corresponding sectors. The one-loop corrections to the observable gauge couplings are thus
\begin{equation}\label{Eq:one-loops}
\begin{aligned}
\Re \left( \delta^{\text{1-loop}} {\rm f}_{SU(N_a)}  \right) =& \frac{1}{2 N_a} \sum_{x \in\{b,c,d,e\}} \Re \left(\delta^{\text{1-loop}} {\rm f}_{U(1)_aU(1)_x} \right) + \Re \left( \delta^{\text{1-loop}, {\cal A}}_{SU(N_a)} {\rm f}_{SU(N_a)}   \right)
\\
=&  \sum_{i=1,2} \left(- \frac{1}{\pi^2} \, \ln \frac{\vartheta_1(\frac{1}{2}, i v_i)}{\eta (i v_i)} + \frac{3}{4\pi^2} \ln \eta (i v_i) \right)
\\
&   -\frac{3}{8\pi^2} \, \ln \left( e^{-\pi v_3/4}   \frac{\vartheta_1(\frac{-iv_3}{2}, i v_3)}{\eta (i v_3)} \right) + \frac{1}{2\pi^2} \ln \eta(iv_3)
\\
\stackrel{v_i \to \infty}{\longrightarrow}& \, \frac{5}{48 \pi} ( v_1 +  v_2)  - \frac{7}{96 \pi}  v_3 - \frac{2}{\pi^2} \ln (2)
,
\\
\Re \left( \delta^{\text{1-loop}} {\rm f}_{SU(N_x),x \in \{b,c\}}  \right) =& 
 -\frac{1}{4\pi^2} \, \ln \frac{\vartheta_1(\frac{1}{2}, i v_m)}{\eta (i v_m)}  + \frac{1}{2\pi} \ln \eta (i v_m)  
-\frac{3}{2 \pi^2} \, \ln \frac{\vartheta_1(\frac{1}{2}, i v_n)}{\eta (i v_n)} + \frac{1}{\pi^2} \ln \eta (i v_n)  \\
& -\frac{1}{2 \pi^2} \, \ln \left( e^{-\pi v_3/4}   \frac{\vartheta_1(\frac{-iv_3}{2}, i v_3)}{\eta (i v_3)} \right) -\frac{1}{4\pi^2} \ln \eta (i v_3)
   + \frac{\ln (2)}{24 \pi^2}
\\
\stackrel{v_i \to \infty}{\longrightarrow} &\, \frac{1}{6 \pi} v_n  -\frac{1}{16\pi}  v_3 -\frac{41}{24 \pi^2} \ln (2)
\qquad
\text{with} \qquad (m,n) = \left\{\begin{array}{cr} (1,2) & x=b \\ (2,1) & c \end{array}\right.
,
 \end{aligned}
\end{equation}
which differ for generic choices of two-torus volume moduli $v_i$.

\begin{figure}[h]
\begin{tabular}{c@{\hspace{0.4in}}c}
\includegraphics[width=7.8cm]{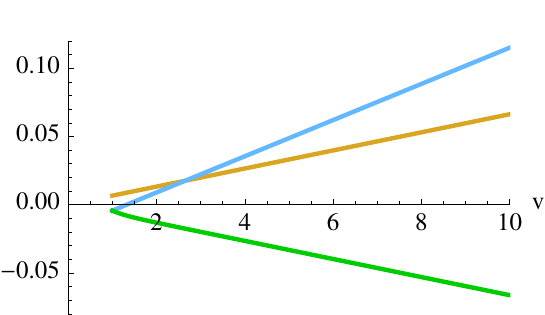} &
\includegraphics[width=7.8cm]{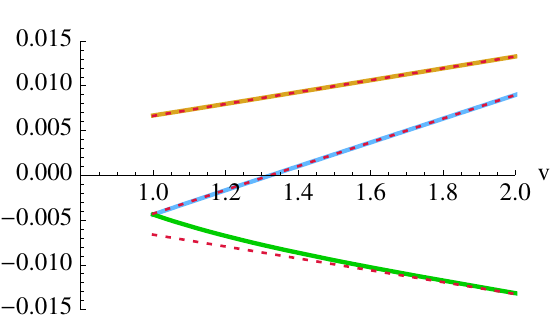}
\end{tabular}
\caption{The three types of one-loop contributions to the gauge coupling as a function of the K\"ahler modulus $v$ : $-\frac{1}{4\pi^2} \ln(\eta(iv))$ (orange), $-\frac{1}{4\pi^2}  \ln \left(\frac{\vartheta_1(\frac{1}{2}, iv)}{\eta (i v)}\right) $ (blue), $-\frac{1}{4\pi^2} \ln \left(e^{-\pi v /4}\frac{|\vartheta_1(- i\frac{v}{2}, iv)|}{\eta (i v)}\right) $ (green). The left plot shows the regime where the geometric approximation is valid ($v>1$), while the right plot gives a closer look at the region $1<v<2$.  The red, dotted curves in the right plot correspond to the large $v$-behaviour of the one-loop contributions, showing that the contributions asymptote very fast to linear contributions in terms of the K\"ahler modulus $v$.\label{plot9}}
\end{figure}

The asymptotic behaviour, which is linear in the two-torus volumes is reached already for values very close to $v_i=1$ as displayed in figure~\ref{plot9}.
From the one-loop contributions~(\ref{Eq:one-loops}), one can thus deduce that gauge coupling unification occurs for the special choice of 
nearly isotropic two-torus volumes,
\begin{equation}
v_1=v_2 =4 v_3 + \frac{168 \ln (2)}{\pi}
,
\end{equation}
which is of interest for model building with a high string scale $M_{\text{string}} \sim M_{\text{GUT}}$.

On the other hand, the asymptotic behaviour in~(\ref{Eq:one-loops}) with positive sign for the K\"ahler moduli $v_1$ and $v_2$ and negative 
contribution from $v_3$ indicates that highly unisotropic choices and large volumes such as e.g. $v_3 \sim 10^{13}$ and $v_1=v_2 \sim 10^6$ 
are consistent with \mbox{$M_{\text{string}} \sim $1 TeV} and very weak string coupling 
$g_{\text{string}} \sim 10^{-3}$ according to the required value for the six-dimensional volume in table~\ref{tab:various-scales}.
This regime of low string mass is not only of interest for the possible detection of the first massive string excitations at the LHC, but also 
in view of the search for potential $Z'$ bosons and dark photons.

In~\cite{Camara:2011jg}, it was argued that $U(1)$ factors from the closed string RR sector can kinetically mix among each other and with
the open string $U(1)$s discussed above. However, the five-stack Pati-Salam model possesses a single RR photon, which stems from the $\Z_3$ 
sector twisted by $\omega^2$ and does therefore not contribute to the Chern-Simons couplings of fractional D6-branes of the 
type~(\ref{Eq:def_frac_cycle}) composed solely of bulk and $\Z_2$ twisted sector contributions.

\section{Conclusions}\label{S:Conclusions}

In this article, an exhaustive study of D6-brane model building on the $T^6/(\Z_2\times \Z_6'\times \OR)$ orientifold with discrete torsion has been performed, taking fully advantage of the existence of rigid fractional three-cycles 
- which are linear combinations of some orbifold invariant bulk cycle inherited from the covering torus $T^6$ and all orbifold invariant exceptional cycles stuck at the $\Z_2$ fixed points traversed by the bulk cycle.  
As the orbifold group \mbox{$\Z_2\times \Z_6'$} contains three different $\Z_2^{(i)}$ subgroups ($i \in \{ 1,2,3 \}$), an exceptional three-cycle in each subsector wraps a one-cycle on the two-torus $T^2_{(i)}$ and an exceptional divisor on
the remaining four-torus \mbox{$T^2_{(j)} \times T^2_{(k)}$}. Only in the presence of discrete torsion $(\eta = -1)$ are these exceptional divisors not projected out by the orientifold projection and does the orientifold $T^6/(\Z_2\times \Z_6'\times \OR)$ support two times five  exceptional three-cycles per $\Z_2^{(i)}$ twisted sector. Each fractional cycle is fully characterised by the wrapping numbers of the bulk cycle, by the $\Z_2^{(1)}$ and $\Z_2^{(2)}$ eigenvalues, by three discrete displacements $(\vec \sigma)$ from the origin of each two-torus and by three discrete Wilson lines $(\vec\tau)$.   

The orientifold projection $\OR$ allows some freedom on the lattice orientation of a two-torus $T^2_{(i)}$, leading at first sight to four distinct lattice orientations for  a factorisable six-torus: {\bf AAA}, {\bf AAB}, {\bf ABB} and {\bf BBB}. However, the first accomplishment of this paper consists in relating the lattices pairwise and thereby reducing the number of inequivalent models by a factor of two. More explicitly, the naive relation between the {\bf AAA} and the {\bf ABB} lattice on the one hand, and between the {\bf AAB} and the {\bf BBB} lattice on the other hand has been promoted to a full equivalence at the level of the massless matter spectra and perturbative field theoretic results. By counting the massless closed string states, or comparing the bulk RR tadpole cancellation and the supersymmetry condition one already obtains some intuition for the equivalence ${\bf AAA} \leftrightarrow {\bf ABB}$ and ${\bf AAB} \leftrightarrow {\bf BBB}$. A thorough and precise analysis on the level of global consistency conditions, massless spectra and their K\"ahler metrics, and the one-loop corrections to the holomorphic gauge kinetic functions confirms this intuition completely. As a warm-up exercise for the identification of the lattices on the $T^6/(\Z_2\times \Z_6'\times \OR)$ orientifold,  the equivalence of the {\bf AAA} and {\bf ABB} lattices on the $T^6/(\Z_6\times \OR)$ orientifold was also proven.

The virtue of considering the $T^6/(\Z_2\times \Z_6'\times \OR)$ orientifold with discrete torsion lies in the presence of rigid fractional three-cycles on the one hand and the simplicity of the global bulk consistency conditions on the other hand. 
The moduli describing the position of the D6-branes along the transverse compact directions are projected out by the $\Z_2 \times \Z_2$ subsymmetry. Nevertheless, this does not imply the complete absence of chiral multiplets in the adjoint representation. One should be aware that the intersections between a cycle $a$ and its orbifold images $(\omega^k a)$ could yield new matter in the adjoint representation, corresponding to the deformation moduli of the three-cycle at the self-intersection point. It turns out that one can write down clear conditions for the total absence of matter in the adjoint representation depending on the one-cycle lengths of the factorisable three-cycle, its discrete Wilson lines and its displacements
but independent of its $\Z_2$ eigenvalues. Requiring the complete rigidity of a three-cycle therefore forms a first step to finding cycles suitable for the `QCD stack' of the strong interaction. Together with other constraints (such as the absence of matter in the symmetric and antisymmetric representation and the occurrence of three quark generations) the amount of fractional three-cycles suitable for the `QCD stack' and the $SU(2)_L$ stack can be severely reduced.

By constraining the `QCD stack' and the $SU(2)_L$ stack in this way, the search for a consistent intersecting D-brane model of particle physics is greatly simplified. The constraints on the non-chiral spectrum (no adjoints, no symmetrics and no antisymmetrics) together with the required chiral spectrum tend to deliver only `local' MSSM and left-right symmetric models. A  supersymmetric global completion of these types of models by adding `hidden' gauge groups is obstructed by 
large non-vanishing twisted RR tadpoles. Pati-Salam models on the other hand do allow for a supersymmetric global completion with additional `hidden' gauge groups. The largest hidden gauge groups found for a global Pati-Salam model is $U(2)_d \times U(2)_e$, assuming that the `QCD stack' is completely rigid and free of matter in symmetric and antisymmetric representations. The only exotic matter with $SU(4)$ charge in this model consists of two pairs of non-chiral matter states in the bifundamental representation of the `QCD stack' and the `hidden' gauge group $SU(2)_d$. By allowing non-chiral matter in the antisymmetric representation of the `QCD stack'  six- stack global Pati-Salam models can be constructed with a larger hidden gauge group $U(4)_d\times U(2)_e\times U(2)_f$ and all gauge factors wrapped on rigid fractional three-cycles. Besides the non-chiral exotics in the antisymmetric representation, this second model also comes with non-chiral exotic matter in the bifundamental representation of the `QCD stack'  and a `hidden' $SU(2)$ gauge factor. 
Both Pati-Salam models also contain some amount of matter charged under $SU(2)_L \times SU(2)_R$, some of which can receive masses through perturbative three-point couplings.
For the five-stack model it was furthermore shown that the heaviest particle generation is rendered massive through area-suppressed perturbative Yukawa couplings. 

Last but not least, the masses of Abelian gauge factors and the possibility of gauge coupling unification or a low string scale $M_{\text{string}} \sim $ 1 TeV were discussed with the result that gauge coupling unification is possible with
approximately isotropic two-torus volumes, whereas a low string scale can be achieved for highly unisotropic two-torus volumes through cancellations among tree- and one-loop  contributions to the gauge couplings. The latter is of great interest in view of a potential detection of either higher string excitations of Standard Model
particles or of new massive Abelian gauge bosons at the LHC. These aspects clearly deserve further investigation.

From a more technical point of view, it is interesting to investigate the existence of discrete global symmetries along the lines~\cite{Ibanez:2012wg,BerasaluceGonzalez:2012vb}.
Due to the incomplete factorisation of intersection numbers on toroidal orbifolds, it is expected that the pattern can differ from the one recently derived on the six-torus~\cite{BerasaluceGonzalez:2012vb}.

Besides the $T^6/(\Z_2\times \Z_6'\times \OR)$ orientifold of this article, the $T^6/(\Z_2\times \Z_6\times \OR)$ orientifold, for which the consistency conditions have been derived in~\cite{Forste:2010gw},
is of great phenomenological interest. It is expected that also on the latter orbifold exact pairwise relations among compactification lattices can be derived, leading to a considerable reduction of effort in
the model search. This search is expected to be even more involved than the present one due to the existence of one complex structure modulus, and supersymmetric bulk three-cycles need to be classified per value of this
complex structure. The investigation is, however, expected to provide a rich family of particle physics models since the net-chiralities of matter in symmetric and antisymmetric representations are no 
longer identical and $SU(5)$ GUTs as well as Standard Model spectra with some particles in these representations could be found. 
We plan to investigate these options in the future.

\noindent{\bf Acknowledgements.} G.H. is grateful to the program {\it Mathematics and Applications of Branes in String and M-theory}
at the Isaac Newton Institute for Mathematical Sciences in Cambridge, UK, for kind hospitality during various stages of this project. 
The work of G.H. and W.S. is partially supported by the Research Center {\it Elementary Forces and Mathematical Foundations (EMG)} at JGU Mainz.

\clearpage

\begin{appendix}

\section{Tables of Three-cycle Wrapping Numbers}\label{A:tables_bulk-cycles}

This appendix contains the exhaustive lists of supersymmetric bulk three-cycles not overshooting the RR tadpole cancellation conditions
(cf. table~\ref{tab:Bulk-RR+SUSY-Z2Z6p})
on the {\bf AAA} and {\bf BBB} background lattice orientations.
The corresponding complete lists for {\bf ABB} and {\bf AAB} lattices are obtained applying the matching relations~(\ref{Eq:identify_nm_AAA+ABB}) and~(\ref{Eq:identify_nm_AAB+BBB}),
respectively, on the toroidal wrapping numbers in tables~\ref{Tab:all-cycles-AAA} and~\ref{Tab:all-cycles-BBB}.
Furthermore, table~\ref{tab:exceptional-wrappings-Z2Z6} concisely provides all exceptional wrapping numbers $(x^{(i)}_{\alpha},y^{(i)}_{\alpha})$ in dependence of 
the discrete parameters specifying each fractional three-cycle.
\begin{table}[h]
\begin{center}
$\begin{array}{|c|c|c||c|c|c|}
\hline
 \multicolumn{6}{|c|}{\text{\bf Orbits $(n^1, m^1; n^2, m^2; n^3, m^3 )$ of the SUSY bulk cycles on the {\color{red}AAA} lattice}} \\
\hline \hline \text{\bf shortest}  & X & Y &  \text{\bf next-to-shortest}  & X & Y \\
\hline
\Omega {\cal R}: (1,0;1,0;1,0) & 1 & 0 & \Omega {\cal R} \Z^{(1)}_2:  (1,0;-1,2;1,-2) & 3&0\\
(0,1;1,0;1,-1) & 1 & 0 &  \Omega {\cal R} \Z^{(2)}_2: (-1,2;1,0;1,-2) & 3&0\\
  &&&  \Omega {\cal R} \Z^{(3)}_2: (-1,2;1,-2;1,0) & 3&0\\
   &&& (-1,2;2,-1;1,-1) & 3&0 \\
   &&& (1,0;2,-1;1,1) & 3& 0\\
   &&&  (2,-1;1,0;1,1) & 3& 0\\
   \hline \hline \text{\bf next-to-next-to-shortest}  & X & Y &  \text{\bf next-to-next-to-next-to-shortest}  & X & Y \\
   \hline
 (-2,3;1,0;1,-3)&7&0 & (-3,4;4,-3;1,-1) & 13 & 0 \\
 (-2,3;3,-2;1,-1)&7&0 & (-1,4;4,-1;1,-1) & 13 & 0 \\
 (-1,1;3,-1;1,-3)&7&0 & (1,-1;3,1;1,3) & 13 & 0 \\
 (-1,3;3,-1;1,-1)&7&0 &  (1,0;4,-3;1,3) & 13 & 0 \\
 (0,1;3,-2;3,-1)&7&0 &  (1,0;4,-1;3,1) & 13 & 0 \\ 
 (1,0;2,1;3,-1)&7&0 & (1,3;3,1;1,-1)  & 13 & 0 \\
 (1,2;2,1;1,-1)&7&0 & (4,-3;1,0;1,3) & 13 & 0 \\
 (2,1;1,0;3,-1)&7&0 & (4,-1;1,0;3,1) & 13 & 0 \\
 (-1,3;1,-1;3,-1)&7&0 &(1,3;1,-1;3,1) & 13 & 0 \\
 \hline
\end{array}$
\caption{Complete list of independent bulk three-cycles on the {\bf AAA} lattice, which do not overshoot the bulk RR tadpole cancellation
condition in table~\protect\ref{tab:Bulk-RR+SUSY-Z2Z6p}. The representant with \mbox{$(n^3,m^3)=\text{(odd,odd)}$} and $n^3 >0$
is given - except for the four orbits containing the $\OR$ and $\OR\Z_2^{(k)}$ invariant planes as detailed at the end of section~\protect\ref{Sss:Geo_bulk}.
}
\label{Tab:all-cycles-AAA}
\end{center}
\end{table}

\begin{table}[h]
\begin{center}
$\begin{array}{|c|c|c||c|c|c|}
\hline
 \multicolumn{6}{|c|}{\text{\bf Orbits $(n^1, m^1; n^2, m^2; n^3, m^3 )$ of the SUSY cycles on the {\color{blue}BBB} lattice}} \\
\hline \hline \text{\bf shortest}  & X & Y &  \text{\bf next-to-shortest}  & X & Y \\
\hline
  \Omega {\cal R} \Z^{(1)}_2:  (1,1;-1,1;1,-1) & -1 & 2  & \Omega {\cal R}: (1,1;1,1;1,1) &-3 &6\\ 
 \Omega {\cal R} \Z^{(2)}_2:  (-1,1;1,1;1,-1) & -1&2 &  (1,1;2,-1;-1,2) &-3 &6\\
 \Omega {\cal R} \Z^{(3)}_2:   (-1,1;1,-1;1,1) & -1 & 2 &&&\\
  (1,-1;1,0;-2,1) & -1 & 2 &&&\\
  (1,-1;2,-1;-1,0) & -1 & 2 &&&\\
  (1,1;1,0;0,1) & -1 & 2 &&&\\
     \hline \hline \text{\bf next-to-next-to-shortest}  & X & Y &  \text{\bf next-to-next-to-shortest}  & X & Y \\
   \hline
(1,-5;1,0;-2,-1)& -7 & 14 & (1,1;2,1;1,2) & -7 & 14 \\
(1,-5;2,1;-1,0)& -7 & 14 & (1,1;3,-2;-2,3)  & -7 & 14 \\
(1,-5;3,-2;0,-1)& -7 & 14 & (1,1;3,-1;-1,3) & -7 & 14 \\
(1,-3;1,0;-4,-1)& -7 & 14 &(3,-1;1,0;-4,5) & -7 & 14 \\
(1,-3;2,-1;-1,-2)& -7 & 14 & (3,-1;2,-1;-3,2) & -7 & 14 \\
(1,-3;3,-1;-1,-1)& -7 & 14 & (3,-1;2,1;-1,2) & -7 & 14 \\
(1,-3;4,1;-1,0)& -7 & 14 & (3,-1;3,-2;-2,1) & -7 & 14 \\
(1,-3;5,-4;0,-1)& -7 & 14 & (3,-1;5,-1;-1,1)  & -7 & 14 \\
(1,-1;2,1;-5,4)& -7 & 14 &(5,-1;1,0;-2,3)  & -7 & 14 \\
(1,-1;3,-2;-4,-1)& -7 & 14 & (5,-1;3,-1;-1,1) & -7 & 14 \\
(1,-1;3,-1;-5,1)& -7 & 14 & (3,-1;1,-1;-5,1)  & -7 & 14 \\
(1,-1;4,1;-3,2)& -7 & 14 & (3,-1;1,1;-1,3) & -7 & 14 \\
(1,-1;5,-4;-2,-1)& -7 & 14 & (5,-1;1,-1;-3,1)  & -7 & 14 \\
(1,-1;5,-1;-3,1)& -7 & 14 &  & &\\
\hline
\end{array}$
\caption{Complete list of independent bulk cycles on the {\bf BBB} lattice, which do not overshoot the bulk RR tadpole cancellation condition
in table~\protect\ref{tab:Bulk-RR+SUSY-Z2Z6p}.  The representant with $(n^1_a,m^1_a)=$(odd,odd) and $n^1_a>0$  is chose for each orbifold
and orientifold orbit.}
\label{Tab:all-cycles-BBB}
\end{center}
\end{table}


\mathsidetabfix{
\begin{array}{|c|c||c|c||c|c||c|c|}\hline
& \muc{4}{|c|}{\text{\bf Exceptional wrapping numbers $(x^{(i)}_{\alpha}, y^{(i)}_{\alpha})$ in dependence of torus-wrappings $(n^i,m^i)$, 
$\Z_2$ eigenvalues $(-1)^{\tau^{\Z_2^{(i)}}}$, displacements  $(\vec{\sigma})$ and Wilson lines $(\vec{\tau})$}}
\\
(n^j,m^j;n^k,m^k) & (\sigma^j;\sigma^k)=(0;0) & (1;0) & (0;1) & (1;1)
\\\hline\hline
\text{(odd,odd;odd,odd)} & (z^{(i)}_{\alpha} m^i \,, \, -z^{(i)}_{\alpha} (n^i+m^i) )_{\alpha=1,2,3}
&  \begin{array}{c}  \left( (1-z^{(i)}_{\alpha}) n^i -z^{(i)}_{\alpha}m^i \, , \, m^i+z^{(i)}_{\alpha}n^i \right)_{\alpha=1} \\   
(z^{(i)}_{\alpha} n^i \,, \, z^{(i)}_{\alpha} m^i)_{\alpha=5}  \\
(-z^{(i)}_{\alpha} (n^i+m^i)  \, , \, z^{(i)}_{\alpha} n^i)_{\alpha=4} \end{array}
&  \begin{array}{c}  (z^{(i)}_{\alpha} m^i \,, \, -z^{(i)}_{\alpha} (n^i+m^i) )_{\alpha=4,5}  \\  
\left( (1-z^{(i)}_{\alpha}) n^i -z^{(i)}_{\alpha}m^i \, , \, m^i+z^{(i)}_{\alpha}n^i\right)_{\alpha=2}  
\end{array}
&  \begin{array}{c}  (-z^{(i)}_{\alpha} (n^i+m^i)  \, , \, z^{(i)}_{\alpha} n^i)_{\alpha=5}   \\  (z^{(i)}_{\alpha} n^i \,, \, z^{(i)}_{\alpha} m^i)_{\alpha=4}  
\\ \left( (1-z^{(i)}_{\alpha}) n^i -z^{(i)}_{\alpha}m^i \, , \, m^i+z^{(i)}_{\alpha}n^i\right)_{\alpha=3}   \end{array}
\\\hline
\text{(odd,even;odd,even)} &  (z^{(i)}_{\alpha} n^i \,, \, z^{(i)}_{\alpha} m^i)_{\alpha=1,2,3}
&   \begin{array}{c}  \left(-n^i+ (z^{(i)}_{\alpha}-1)m^i \, , \, (1-z^{(i)}_{\alpha})n^i -z^{(i)}_{\alpha}m^i \right)_{\alpha=1} \\  
(-z^{(i)}_{\alpha} (n^i+m^i)  \, , \, z^{(i)}_{\alpha} n^i)_{\alpha=5} \\    (z^{(i)}_{\alpha} m^i \,, \, -z^{(i)}_{\alpha} (n^i+m^i) )_{\alpha=4} \end{array}
&  \begin{array}{c}   (z^{(i)}_{\alpha} n^i \,, \, z^{(i)}_{\alpha} m^i)_{\alpha=4,5}  \\  
\left(-n^i+ (z^{(i)}_{\alpha}-1)m^i \, , \, (1-z^{(i)}_{\alpha})n^i -z^{(i)}_{\alpha}m^i \right)_{\alpha=2}  
\end{array}
&  \begin{array}{c}  (z^{(i)}_{\alpha} m^i \,, \, -z^{(i)}_{\alpha} (n^i+m^i) )_{\alpha=5} \\  (-z^{(i)}_{\alpha} (n^i+m^i)  \, , \, z^{(i)}_{\alpha} n^i)_{\alpha=4}  \\  
\left(-n^i+ (z^{(i)}_{\alpha}-1)m^i \, , \, (1-z^{(i)}_{\alpha})n^i -z^{(i)}_{\alpha}m^i \right)_{\alpha=3}  \end{array}
\\\hline
\text{(even,odd;even,odd)} & (-z^{(i)}_{\alpha} (n^i+m^i)  \, , \, z^{(i)}_{\alpha} n^i)_{\alpha=1,2,3}
&   \begin{array}{c}  \left( z^{(i)}_{\alpha} n^i+m^i \, , \, -n^i +(z^{(i)}_{\alpha} -1)m^i \right)_{\alpha=1} \\ (z^{(i)}_{\alpha} m^i \,, \, -z^{(i)}_{\alpha} (n^i+m^i) )_{\alpha=5} 
 \\   (z^{(i)}_{\alpha} n^i \,, \, z^{(i)}_{\alpha} m^i)_{\alpha=4}  \end{array}
&  \begin{array}{c}  (-z^{(i)}_{\alpha} (n^i+m^i)  \, , \, z^{(i)}_{\alpha} n^i)_{\alpha=4,5}  \\  
\left( z^{(i)}_{\alpha} n^i+m^i \, , \, -n^i +(z^{(i)}_{\alpha} -1)m^i \right)_{\alpha=2}   \end{array}
&  \begin{array}{c}  (z^{(i)}_{\alpha} n^i \,, \, z^{(i)}_{\alpha} m^i)_{\alpha=5}  \\  (z^{(i)}_{\alpha} m^i \,, \, -z^{(i)}_{\alpha} (n^i+m^i) )_{\alpha=4}  \\ 
 \left( z^{(i)}_{\alpha} n^i+m^i \, , \, -n^i +(z^{(i)}_{\alpha} -1)m^i \right)_{\alpha=3}  \end{array}
\\\hline
\muc{5}{|c|}{z^{(i)}_{\alpha} = (-1)^{\tau^{\Z_2^{(i)}}} \times \left\{\begin{array}{c}
1 \\ (-1)^{\tau^j} \\ (-1)^{\tau^k} \\ (-1)^{\tau^j+\tau^k}  \end{array}\right.
\qquad
\begin{array}{c} (\sigma^j;\sigma^k)=(0;0) \to (1;0) \to (0;1) \to (1;1)
\\1 \to 1 \to 4 \to 5
\\ 2 \to 5 \to 2 \to 4
\\ 3 \to 4 \to 5 \to 3
\end{array}
\qquad
\alpha=
\left\{
\begin{array}{c}
\alpha=0
 \\ 1 +  \sigma^k (3 + \sigma^j)
\\  2 + \sigma^j (3 - \sigma^k)
\\ 3+\sigma^j + 2 \sigma^k  - 3\sigma^j\sigma^k
\end{array}
\right.
}
\\\hline\hline
\text{(odd,even;odd,odd)} & \begin{array}{c}    (z^{(i)}_{\alpha} n^i \,, \, z^{(i)}_{\alpha} m^i)_{\alpha=1,5} \\  (z^{(i)}_{\alpha} m^i \,, \, -z^{(i)}_{\alpha} (n^i+m^i) )_{\alpha=2} \end{array}
&   \begin{array}{c}    \left(-n^i+ (z^{(i)}_{\alpha}-1)m^i \, , \, (1-z^{(i)}_{\alpha})n^i -z^{(i)}_{\alpha}m^i \right)_{\alpha=1} \\   (-z^{(i)}_{\alpha} (n^i+m^i)  \, , \, z^{(i)}_{\alpha} n^i)_{\alpha=4} \\  (z^{(i)}_{\alpha} m^i \,, \, -z^{(i)}_{\alpha} (n^i+m^i) )_{\alpha=3}  \end{array}
&  \begin{array}{c}    (z^{(i)}_{\alpha} n^i \,, \, z^{(i)}_{\alpha} m^i)_{\alpha=3,4}  \\ \left( (1-z^{(i)}_{\alpha}) n^i -z^{(i)}_{\alpha}m^i \, , \, m^i+z^{(i)}_{\alpha}n^i \right)_{\alpha=2}  \end{array}
&  \begin{array}{c}    (z^{(i)}_{\alpha} m^i \,, \, -z^{(i)}_{\alpha} (n^i+m^i) )_{\alpha=4} \\  (-z^{(i)}_{\alpha} (n^i+m^i)  \, , \, z^{(i)}_{\alpha} n^i)_{\alpha=3}   \\ \left(-n^i+ (z^{(i)}_{\alpha}-1)m^i \, , \, (1-z^{(i)}_{\alpha})n^i -z^{(i)}_{\alpha}m^i \right)_{\alpha=5}   \end{array}
\\\hline
\text{(even,odd;odd,even)} &  \begin{array}{c}     (-z^{(i)}_{\alpha} (n^i+m^i)  \, , \, z^{(i)}_{\alpha} n^i)_{\alpha=1,5}  \\ (z^{(i)}_{\alpha} n^i \,, \, z^{(i)}_{\alpha} m^i)_{\alpha=2}   \end{array}
&   \begin{array}{c}   \left( z^{(i)}_{\alpha} n^i+m^i \, , \, -n^i +(z^{(i)}_{\alpha} -1)m^i \right)_{\alpha=1} \\ (z^{(i)}_{\alpha} m^i \,, \, -z^{(i)}_{\alpha} (n^i+m^i) )_{\alpha=4} \\  (z^{(i)}_{\alpha} n^i \,, \, z^{(i)}_{\alpha} m^i)_{\alpha=3}  \end{array}
&  \begin{array}{c}  (-z^{(i)}_{\alpha} (n^i+m^i)  \, , \, z^{(i)}_{\alpha} n^i)_{\alpha=3,4} \\ \left(-n^i+ (z^{(i)}_{\alpha}-1)m^i \, , \, (1-z^{(i)}_{\alpha})n^i -z^{(i)}_{\alpha}m^i \right)_{\alpha=2}   \end{array}
&  \begin{array}{c}   (z^{(i)}_{\alpha} n^i \,, \, z^{(i)}_{\alpha} m^i)_{\alpha=4}  \\  (z^{(i)}_{\alpha} m^i \,, \, -z^{(i)}_{\alpha} (n^i+m^i) )_{\alpha=3}    \\ \left( z^{(i)}_{\alpha} n^i+m^i \, , \, -n^i +(z^{(i)}_{\alpha} -1)m^i \right)_{\alpha=5} \end{array}
\\\hline
\text{(odd,odd;even,odd)} &  \begin{array}{c}   (z^{(i)}_{\alpha} m^i \,, \, -z^{(i)}_{\alpha} (n^i+m^i) )_{\alpha=1,5}   \\   (-z^{(i)}_{\alpha} (n^i+m^i)  \, , \, z_{\alpha} n^i)_{\alpha=2}  \end{array}
&   \begin{array}{c}  \left( (1-z_{\alpha}) n^i -z_{\alpha}m^i \, , \, m^i+z_{\alpha}n^i \right)_{\alpha=1} \\   (z_{\alpha} n^i \,, \, z_{\alpha} m^i)_{\alpha=4} \\ (-z_{\alpha} (n^i+m^i)  \, , \, z_{\alpha} n^i)_{\alpha=3}  \end{array}
&  \begin{array}{c}   (z_{\alpha} m^i \,, \, -z_{\alpha} (n^i+m^i) )_{\alpha=3,4} \\  \left( z_{\alpha} n^i+m^i \, , \, -n^i +(z_{\alpha} -1)m^i \right)_{\alpha=2}  \end{array}
&  \begin{array}{c} (-z_{\alpha} (n^i+m^i)  \, , \, z_{\alpha} n^i)_{\alpha=4} \\ (z_{\alpha} n^i \,, \, z_{\alpha} m^i)_{\alpha=3}   \\    \left( (1-z_{\alpha}) n^i -z_{\alpha}m^i \, , \, m^i+z_{\alpha}n^i \right)_{\alpha=5} \end{array}
\\\hline
&  \muc{4}{|c|}{\begin{array}{c} 1 \to 1 \to 3 \to 4
\\ 2 \to 4 \to 2 \to 3
\\ 5 \to 3 \to 4 \to 5 \end{array}
\qquad
\alpha=
\left\{ \begin{array}{c} 1 +  \sigma^k (2 + \sigma^j)
\\   2 + \sigma^j (2 - \sigma^k)
\\  5 -2 \sigma^j - \sigma^k + 3 \sigma^j\sigma^k
\end{array} \right.   }
\\\hline\hline
\text{(even,odd;odd,odd)} &  \begin{array}{c}  (-z_{\alpha} (n^i+m^i)  \, , \, z_{\alpha} n^i)_{\alpha=1,4} \\   (z_{\alpha} m^i \,, \, -z_{\alpha} (n^i+m^i) )_{\alpha=2}   \end{array}
&   \begin{array}{c}  \left( z_{\alpha} n^i+m^i \, , \, -n^i +(z_{\alpha} -1)m^i \right)_{\alpha=1} \\  (z_{\alpha} m^i \,, \, -z_{\alpha} (n^i+m^i) )_{\alpha=3}  \\  (z_{\alpha} n^i \,, \, z_{\alpha} m^i)_{\alpha=5} \end{array}
&  \begin{array}{c}   (-z_{\alpha} (n^i+m^i)  \, , \, z_{\alpha} n^i)_{\alpha=5,3}  \\ \left( (1-z_{\alpha}) n^i -z_{\alpha}m^i \, , \, m^i+z_{\alpha}n^i \right)_{\alpha=2}   \end{array}
&  \begin{array}{c} (z_{\alpha} n^i \,, \, z_{\alpha} m^i)_{\alpha=3}   \\ (z_{\alpha} m^i \,, \, -z_{\alpha} (n^i+m^i) )_{\alpha=5} \\ \left( z_{\alpha} n^i+m^i \, , \, -n^i +(z_{\alpha} -1)m^i \right)_{\alpha=4}  \end{array}
\\\hline
\text{(odd,odd;odd,even)} &  \begin{array}{c}  (z_{\alpha} m^i \,, \, -z_{\alpha} (n^i+m^i) )_{\alpha=1,4}  \\  (z_{\alpha} n^i \,, \, z_{\alpha} m^i)_{\alpha=2}    \end{array}
&   \begin{array}{c}  \left( (1-z_{\alpha}) n^i -z_{\alpha}m^i \, , \, m^i+z_{\alpha}n^i \right)_{\alpha=1}  \\ (z_{\alpha} n^i \,, \, z_{\alpha} m^i)_{\alpha=3}  \\ (-z_{\alpha} (n^i+m^i)  \, , \, z_{\alpha} n^i)_{\alpha=5} \end{array}
&  \begin{array}{c}  (z_{\alpha} m^i \,, \, -z_{\alpha} (n^i+m^i) )_{\alpha=5,3}  \\   \left(-n^i+ (z_{\alpha}-1)m^i \, , \, (1-z_{\alpha})n^i -z_{\alpha}m^i \right)_{\alpha=2}   \end{array}
&  \begin{array}{c}   (-z_{\alpha} (n^i+m^i)  \, , \, z_{\alpha} n^i)_{\alpha=3}  \\ (z_{\alpha} n^i \,, \, z_{\alpha} m^i)_{\alpha=5}  \\  \left( (1-z_{\alpha}) n^i -z_{\alpha}m^i \, , \, m^i+z_{\alpha}n^i \right)_{\alpha=4} \end{array}
\\\hline
\text{(odd,even;even,odd)} &  \begin{array}{c}   (z_{\alpha} n^i \,, \, z_{\alpha} m^i)_{\alpha=1,4} \\   (-z_{\alpha} (n^i+m^i)  \, , \, z_{\alpha} n^i)_{\alpha=2}  \end{array}
&   \begin{array}{c}   \left(-n^i+ (z_{\alpha}-1)m^i \, , \, (1-z_{\alpha})n^i -z_{\alpha}m^i \right)_{\alpha=1}\\  (-z_{\alpha} (n^i+m^i)  \, , \, z_{\alpha} n^i)_{\alpha=3}  \\ (z_{\alpha} m^i \,, \, -z_{\alpha} (n^i+m^i) )_{\alpha=5} \end{array}
&  \begin{array}{c}  (z_{\alpha} n^i \,, \, z_{\alpha} m^i)_{\alpha=5,3}  \\   \left( z_{\alpha} n^i+m^i \, , \, -n^i +(z_{\alpha} -1)m^i \right)_{\alpha=2}   \end{array}
&  \begin{array}{c}  (z_{\alpha} m^i \,, \, -z_{\alpha} (n^i+m^i) )_{\alpha=3}  \\  (-z_{\alpha} (n^i+m^i)  \, , \, z_{\alpha} n^i)_{\alpha=5} \\   \left(-n^i+ (z_{\alpha}-1)m^i \, , \, (1-z_{\alpha})n^i -z_{\alpha}m^i \right)_{\alpha=4}\end{array}
\\\hline
&  \muc{4}{|c|}{\begin{array}{c} 1 \to 1 \to 5 \to 3
\\ 2 \to 3 \to 2 \to 5
\\ 4 \to 5 \to 3 \to 4 \end{array}
\qquad
\alpha=
\left\{
\begin{array}{c} 1 + \sigma^k(4-2\sigma^k)
\\  2 + \sigma^j( 1+ 2 \sigma^k)
\\ 4 + \sigma^j-\sigma^k
\end{array} \right. }
\\\hline
\end{array}
}{exceptional-wrappings-Z2Z6}{Complete list of exceptional wrapping numbers $(x^{(i)}_{\alpha}, y^{(i)}_{\alpha})$ in dependence 
of the discrete parameters per D6-brane in extension of schematic form in table~\protect\ref{Tab:xy-entries-Z2Z6p} of 
section~\protect\ref{Sss:Geo_exceptional}. 
Under the matching ${\bf ABB} \to {\bf AAA}$ in section~\ref{Sss:T6Z2Z6p_identifications}, the three blocks of rows are 
permuted as $1^{\rm st} \to 3^{\rm rd} \to 2^{\rm nd} \to 1^{\rm st}$ while preserving the ordering of the three rows per block.
For the matching ${\bf BBB} \to {\bf AAB}$ in appendix~\ref{A:identifications_AAB+BBB}, the permutation of blocks is as before,
by the rows in each block are permuted likewise.}

\clearpage

\section{Intersection Numbers, Beta Function Coefficients, K\"ahler Metrics and One-loop Vacuum Amplitudes}\label{A:intersections+betas}

In section~\ref{Sss:Geo_exceptional}, exceptional three-cycles at $\Z_2^{(k)}$ fixed points have been introduced with relative sign
factors due to $\Z_2^{(k)}$ eigenvalues $(-1)^{\tau^{\Z_2^{(k)}}}$
and discrete Wilson lines $(-1)^{\tau^k}$ as displayed in table~\ref{Tab:Reference_Point+Signs}, 
and in section~\ref{Sss:T6Z2Z6p_identifications} it was argued that the pairwise matching of lattice orientations
is extended to the full spectrum, which is computed from intersection numbers,
\begin{equation}
\begin{aligned}
I_{ab} & \equiv \prod_{i=1}^3 I_{ab}^{(i)} =  \prod_{i=1}^3  \left( n^i_a m^i_b - m^i_a n^i_b \right)
,
\\
I_{ab}^{\Z_2^{(i)}} &\equiv   I_{ab}^{(i)} \cdot   I_{ab}^{\Z_2^{(j)},(j)}  \cdot I_{ab}^{\Z_2^{(k)},(k)}  
\qquad \text{ with 
$(ijk)$ a permutation of (123)}
,
\end{aligned}
\end{equation}
where $I_{ab}^{(i)}$ are toroidal one-cycle intersection numbers and $I_{ab}^{\Z_2^{(j)},(j)}$ counts the $\Z_2$ invariant intersection points
along $T^2_{(j)}$ with signs as displayed in table~\ref{tab:Wilson_signs}. Apart from the slight change of notation by 
including the $\Z_2^{(i)}$ eigenvalue $(-1)^{\tau^{\Z_2^{(i)}}_a } =(-1)^{\tau^{\Z_2^{(j)}}_a + \tau^{\Z_2^{(k)}}_a }$ in the definition of $I_{ab}^{\Z_2^{(j)},(j)}$
and $I_{ab}^{\Z_2^{(k)},(k)}$, the formulas agree with those given in appendix~A of~\cite{Gmeiner:2009fb}.
\mathtabfix{
\begin{array}{|c||c|c|c|}\hline
\muc{4}{|c|}{\text{\bf  Intersection numbers $I^{\Z_2^{(i)},(i)}_{ab}$ in dependence of discrete parameters on the two-torus $T^2_{(i)}$}}
\\\hline\hline
& \muc{3}{|c|}{ (n^i_a,m^i_a \, ; \, n^i_b,m^i_b)}
\\\hline
(\sigma^i_a;\sigma^i_b) & \begin{array}{c}  (1) \quad \text{(odd,odd \, ; \, odd,odd)}\\ \stackrel{\omega}{\to}  \text{(odd,even \, ; \, odd,even)}\\ \stackrel{\omega}{\to}  \text{(even,odd \, ; \, even,odd)}\end{array}
 & \begin{array}{c}  (2) \quad \text{(odd,odd \, ; \, odd,even)}\\ \stackrel{\omega}{\to}  \text{(odd,even \, ; \, even,odd)}\\ \stackrel{\omega}{\to}  \text{(even,odd \, ; \, odd,odd)}\end{array}
  & \begin{array}{c}  (3) \quad \text{(odd,odd \, ; \, even,odd)}\\ \stackrel{\omega}{\to}  \text{(odd,even \, ; \, odd,odd)}\\ \stackrel{\omega}{\to}  \text{(even,odd \, ; \, odd,even)}\end{array}
\\\hline
(0;0) & (-1)^{\tau^{\Z_2^{(i)}}_{ab}} \, \left[1+ (-1)^{\tau^i_{ab}} \right] 
&  (-1)^{\tau^{\Z_2^{(i)}}_{ab} } &  (-1)^{\tau^{\Z_2^{(i)}}_{ab} }
\\
(0;1) & 0 &  (-1)^{\tau^{\Z_2^{(i)}}_{ab}} \,(-1)^{\tau^i_{ab}} &  (-1)^{\tau^{\Z_2^{(i)}}_{ab}} \,(-1)^{\tau^i_a}
\\
(1;0) & 0 &   (-1)^{\tau^{\Z_2^{(i)}}_{ab}} \,(-1)^{\tau^i_b} & (-1)^{\tau^{\Z_2^{(i)}}_{ab}} \,(-1)^{\tau^i_{ab}} 
\\
(1;1) &  (-1)^{\tau^{\Z_2^{(i)}}_{ab} } \, \left[1+ (-1)^{\tau^i_{ab}} \right] &  (-1)^{\tau^{\Z_2^{(i)}}_{ab}} \,(-1)^{\tau^i_a} &  (-1)^{\tau^{\Z_2^{(i)}}_{ab}} \,(-1)^{\tau^i_b}
\\\hline
\end{array}
}{Wilson_signs}{$\Z_2^{(i)}$ invariant intersection numbers  among D6-branes $a$ and $b$
in dependence of the even- or oddness of one-cycle wrapping numbers $(n^i_a,m^i_a \, ; \, n^i_b,m^i_b)$,
displacement parameters \mbox{$\sigma^i_a,\sigma^i_b \in \{0,1\}$}
and of the relative $\Z_2$ eigenvalue $(-1)^{\tau^{\Z_2^{(i)}}_{ab}} \equiv (-1)^{\tau^{\Z_2^{(i)}}_{a} + \tau^{\Z_2^{(i)}}_{b}}$ and
discrete  Wilson lines $(-1)^{\tau^i_a}$ or 
$(-1)^{\tau^i_b}$ with the notation $(-1)^{\tau^i_{ab}} \equiv (-1)^{\tau^i_{a} + \tau^i_{b}}$ for a relative Wilson line.
The list is compatible with appendix~A of~\cite{Gmeiner:2009fb}.
$\omega$ denotes the $\Z_6'$ transformation of one-cycles in equation~(\ref{Eq:1-cycle-orbits}).}

The number of adjoint representations at intersections of orbifold images is determined from the case (2) and (3) in table~\ref{tab:Wilson_signs}
for $a(\omega \, a)$ and $a(\omega^2 \, a)$ intersections, respectively, as can be read off 
from the transformation of one-cycle wrapping numbers in equation~(\ref{Eq:1-cycle-orbits}) with $(-1)^{\tau^{\Z_2^{(i)}}_{aa}} \equiv 1$ for all
$i \in \{1,2,3\}$
and $(\sigma^i_a;\sigma^i_a), (\tau^i_a;\tau^i_a) \in \{(0;0),(1;1)\}$.
 The existence of fractional D6-branes without any  matter in the adjoint representation 
at self-intersections is studied in section~\ref{Ss:no-adjoints}.

To determine the number of symmetric plus antisymmetric representations, it is important to notice that the relative $\Z_2^{(i)}$ eigenvalue when using table~\ref{tab:Wilson_signs} is
 $(-1)^{\tau^{\Z_2^{(i)}}_{aa'}} = - \eta_{(i)} (-1)^{2b_i \, \sigma^i_a \tau^i_a}$ with $2b_i \equiv 1$ for the $SU(3)^3$ root lattice
as can be checked by performing the ${\cal R}$ image on the entries of table~\ref{Tab:Reference_Point+Signs}, 
 and the even- or oddness of one-cycle wrapping numbers $(n^i_{a'},m^i_{a'})$ is determined by equation~(\ref{Eq:OR_on_n+m}). On an {\bf A}-type lattice,  if $(n^i_a,m^i_a)=\text{(odd,odd)}$
 the $aa'$ sector is of the type (3), the $(\omega a)(\omega a)'$ sector of type (1) and the $(\omega^2 a)(\omega^2 a)'$ sector of type (2), whereas on a {\bf B}-type lattice the ordering is
 (1),(2) and (3), in other words $\left[(\omega^k a)(\omega^k a)'\right]_{\bf A} = \left[(\omega^{k-1} a)(\omega^{k-1} a)'\right]_{\bf B}$
for \mbox{$k=0,1,2$ mod 3}. 
The relevant one-cycle wrapping numbers are collected in table~\ref{tab:aaP-intersections}.
\mathtabfix{
\begin{array}{|c||c|c|c||c|c|c|}\hline
\muc{7}{|c|}{\text{\bf One-cycle intersections involving orientifold images and O6-planes}}
\\\hline\hline
 & \muc{3}{|c|}{\text{{\bf A} lattice }}  & \muc{3}{|c|}{\text{{\bf B} lattice}} 
\\\cline{2-7}
x & I_{xx'}^{(i)} & \tilde{I}_x^{\OR,(i)} & \tilde{I}_x^{\OR\Z_2,(i)}  
& I_{xx'}^{(i)} & \tilde{I}_x^{\OR,(i)} & \tilde{I}_x^{\OR\Z_2,(i)} 
\\\hline\hline
a & -m^i_a \, \left[ 2 \, n^i_a + m^i_a \right] & -m^i_a & 2 \, n^i_a + m^i_a  
& \left(n^i_a\right)^2 - \left(m^i_a\right)^2 & n^i_a - m^i_a & n^i_a + m^i_a  
\\\hline
(\omega a) & \left(m^i_a\right)^2 - \left(n^i_a\right)^2 & n^i_a + m^i_a  & m^i_a - n^i_a
& -n^i_a \,  \left[ n^i_a + 2 \, m^i_a \right] & n^i_a + 2 \, m^i_a & - n^i_a 
\\\hline
(\omega^2 a) & n^i_a \,  \left[ n^i_a + 2 \, m^i_a \right] & -n^i_a & -n^i_a - 2 \, m^i_a
& m^i_a \, \left[ 2 \, n^i_a + m^i_a \right] & - 2 \, n^i_a - m^i_a & - m^i_a
\\\hline
\end{array}
}{aaP-intersections}{One-cycle intersection numbers for toroidal D6-branes $(\omega^k a)_{k\in \{0,1,2\}}$ and their orientifold images
in dependence of the lattice orientation {\bf A} or {\bf B}. Intersections with O6-plane orbits $\OR\Z_2^{(k)}$
are computed from $\tilde{I}_x^{\OR\Z_2^{(k)}} \equiv \tilde{I}_x^{\OR\Z_2,(i)} \cdot \left( - \tilde{I}_x^{\OR\Z_2,(j)} \right)\cdot \tilde{I}_x^{\OR,(k)}$
due to the relative angles $\frac{\pi}{2}( 1_i,-1_j,0_k)$ with the $\OR$-invariant plane.
}
The relative sign factor $(-1)^2=+1$ in table~\ref{tab:aaP-intersections} arises on $T^2_{(2)} \times T^2_{(3)}$ in the mapping ${\bf ABB} \to {\bf AAA}$,
while the non-trivial transformation of one-cycle wrapping numbers on $T^2_{(1)}$ in equation~(\ref{Eq:identify_nm_AAA+ABB}) leads to 
identical results (with same sign) on both sides of the map. 
To complete the proof that the configurations involving some orientifold images on {\bf AAA} and  {\bf ABB} are identical under the identification~(\ref{Eq:identify_nm_AAA+ABB}),
the fixed points of table~\ref{Tab:Reference_Point+Signs} and corresponding $\Z_2^{(i)}$ invariant intersections,
\begin{equation}\label{Eq:IaapZ2-values}
I_{(\omega^k a)(\omega^k a)'}^{\Z_2^{(i)},(i)} = - \eta_{(i)} \times \left\{\begin{array}{cr}
2 \,(-1)^{\sigma^i_a \tau^i_a} &  \text{for } \; I_{(\omega^k a)(\omega^k a)'}^{(i)} \;  \text{ even}\\
1 & \text{ odd}
\end{array}\right.
\end{equation}
have to be taken into account as well.

The survey of the existence of fractional D6-branes without symmetric or antisymmetric matter is given in section~\ref{Ss:no-anti+sym}.

In section~\ref{Ss:identifications} and appendix~\ref{A:identifications_AAB+BBB}, it is shown that configurations on the {\bf AAA} and {\bf ABB}
as well as {\bf AAB} and {\bf BBB} lattices are pairwise equivalent at the level of global consistency conditions and massless spectra.

It remains to be shown that also the complete tower of massive string excitations
and the effective field theory - to our current level of understanding - are identical under the matching~(\ref{Eq:identify_nm_AAA+ABB})
among {\bf AAA} and {\bf ABB} background  orientations. 
To this aim, {\bf string vacuum amplitudes} with Annulus and M\"obius strip topology are briefly reviewed here.
The amplitudes are of the form
\begin{equation}\label{Eq:general_amplitudes}
\begin{aligned}
{\cal A}^{\text{topology}} & \sim \sum_{\text{sectors}} \int_0^{\infty}  dl \sum_{\alpha,\beta \in \{0,\frac{1}{2}\}} (-1)^{2(\alpha+\beta)}  
\frac{\vartheta \targ{\alpha}{\beta} (\nu,2 i l - \frac{c_{\text{top}}}{2})}{\eta^3 (2il - \frac{c_{\text{top}}}{2})}
\prod_{k=1}^3 A^{\text{top},\text{sector}}_{T^2_{(k)}} (\alpha,\beta; \phi^{(k)};  2 i l - \frac{c_{\text{top}}}{2})
,
\end{aligned}
\end{equation}
with $c_{\text{top}} = 0$ for the Annulus and $c_{\text{top}} =1$ for the M\"obius strip. $\alpha,\beta$ parameterises the different spin 
structures and $\frac{\vartheta \targ{\alpha}{\beta} (\nu,2 i l - \frac{c_{\text{top}}}{2})}{\eta^3 (2il - \frac{c_{\text{top}}}{2})}$ accounts
for all non-compact string excitations in light cone gauge with gauging $\nu$.
The RR and NS-NS tadpole cancellation conditions are obtained by setting $\nu=0$ and considering $\alpha=\frac{1}{2}$  and $\alpha=0$,
respectively. The gauge threshold amplitudes - which encode the beta function coefficients~(\ref{Eq:beta-coeffs}), K\"ahler metrics to leading
order and one-loop corrections to the holomorphic gauge kinetic function - 
 are obtained via the magnetic background field by considering variations in the gauging $\nu$,
which amounts to replacing $\frac{\vartheta \targ{\alpha}{\beta} (\nu,2 i l - \frac{c_{\text{top}}}{2})}{\eta^3 (2il - \frac{c_{\text{top}}}{2})}
\to \frac{\vartheta^{\prime\prime} \targ{\alpha}{\beta} (0,2 i l - \frac{c_{\text{top}}}{2})}{\eta^3 (2il - \frac{c_{\text{top}}}{2})}$ 
in equation~(\ref{Eq:general_amplitudes}).
The contributions of the full tower of string excitations originating from the compact dimensions can be factorised into contributions per two-torus,
\begin{equation}\label{Eq:amplitudes_per_two-torus}
\begin{aligned}
 A^{\text{Annulus},\text{sector}}_{T^2_{(k)}} (\alpha,\beta; \phi^{(k)};  2 i l) =&  \left\{\begin{array}{l}
I_{ab}^{(i)} \; \frac{\vartheta \targ{\alpha}{\beta} } {\vartheta \targ{1/2}{1/2}}(\phi^{(i)}_{ab}, 2il) \\
\stackrel{\phi^{(i)}_{ab} \to 0}{\longrightarrow}  \frac{(L_a^{(i)})^2}{\text{Vol}(T^2_{(i)})}  
\tilde{\cal L}^{(i)}_{\text{Annulus}} (\sigma^i_{ab},\tau^i_{ab}, L^{(i)}_a; r_i, l) 
\frac{\vartheta \targ{\alpha}{\beta} (0,2 i l)}{\eta^3 (2il)} \\
I_{ab}^{\Z_2,(i)} \; \frac{\vartheta \targ{\alpha+1/2}{\beta} } {\vartheta \targ{0}{1/2}}(\phi^{(i)}_{ab}, 2il)
 \end{array}\right.
,
\\
 A^{\text{M\"obius},\text{sector}}_{T^2_{(k)}} (\alpha,\beta; \phi^{(k)};  2 i l - \frac{1}{2}) =&  \left\{\begin{array}{l}
\tilde{I}_{a}^{\OR\Z_2^{(k)},(i)} \; \frac{\vartheta \targ{\alpha}{\beta} } {\vartheta \targ{1/2}{1/2}}(\phi^{(i)}_{ab}, 2il - \frac{1}{2})  \\
\stackrel{\phi^{(i)}_{a,\OR\Z_2^{(k)}} \to 0}{\longrightarrow} \frac{(\tilde{L}_a^{(i)})^2}{\text{Vol}(T^2_{(i)})} 
\tilde{\cal L}^{(i)}_{\text{M\"obius}} (\tilde{\sigma}^i_{aa'},\tilde{\tau}^i_{aa'},\ \tilde{L}^{(i)}_a; \tilde{r}_i, l) 
\frac{\vartheta \targ{\alpha}{\beta} (0,2 i l  - \frac{c_{\text{top}}}{2})}{\eta^3 (2il  - \frac{1}{2})}
 \end{array}\right.
\end{aligned}
\end{equation}
where $\phi^{(i)}_{ab}$, $\sigma^i_{ab}$ and $\tau^i_{ab}$ are the {\it relative} angle, displacement and Wilson line among D6-branes $a$ and $b$. 
$I_{ab}^{(i)}$ and $I_{ab}^{\Z_2,(i)}$ are the total and $\Z_2$ invariant intersections numbers as detailed above, 
and $L_a^{(i)}$ is the length of the one-cycle.
The quantities with tilde in the M\"obius strip contributions depend on the shape of the lattice, $\tilde{I}_{a}^{\OR\Z_2^{(k)},(i)} \equiv 
2(1-b_i) \, I_{a}^{\OR\Z_2^{(k)},(i)}$, $ (\tilde{L}^{(i)}_a)^2 \equiv 2(1-b_i) \, (L_a^{(i)})^2$ and $\tilde{r}_i^2 \equiv \frac{r_i^2}{1-b_i}$
with $b_i =\frac{1}{2}$ for the $SU(3)$ root lattice.

The Klein bottle amplitude is obtained in analogy to the amplitudes with Annulus and M\"obius strip topology 
for $c_{\text{top}} = 0$ by replacing the angles $\phi^{(i)}_{ab} \to m v_i + n w_i'$ by (multiples of) the 
shift vectors generating the orbifold.

Since each of the ingredients, such as {\it relative} angles and intersection numbers, is preserved (up to permutation of 
sectors $a(\omega^k b') \leftrightarrow a(\omega^{k-1} b')$ involving orientifold images)
under an overall rotation by $\pm \pi/3$ of the 
two-torus $T^2_{(1)}$, the vacuum amplitudes are identical for pairs of D6-brane models on {\bf AAA} and {\bf ABB} related 
by~(\ref{Eq:identify_nm_AAA+ABB}). The same is true for a rotation by $\pm \pi/3$ along $T^2_{(3)}$ among the {\bf AAB} and {\bf BBB}
lattices according to equation~(\ref{Eq:identify_nm_AAB+BBB}).
This shows that the complete tower of massive string states as well as field theoretical data extracted 
from gauge threshold computations agree. 
To our knowledge, amplitudes with vertex operator insertions are only known for the six-torus. It thus remains to be seen if new subtleties
arise for open string vertex operators of matter  on D6-branes intersecting at orbifold singularities.

The expressions for beta function coefficients, which are used in this article to determine the non-chiral matter spectrum,  
and leading order K\"ahler metrics have been derived for $T^6/(\Z_2 \times \Z_{2M} \times \OR)$ with discrete torsion 
in~\cite{Honecker:2011sm} based on partial results in~\cite{Blumenhagen:2007ip}. They are summarised in table~\ref{tab:beta_coeffs_Kaehler_metrics}.
\mathtabfix{
\begin{array}{|c||c|c|c|}\hline
\muc{4}{|c|}{\text{\bf  K\"ahler metrics and beta function coefficients on $T^6/(\Z_2 \times \Z_{2M} \times \OR)$ with discrete torsion}}
\\\hline\hline
(\phi^{(1)}_{ab},\phi^{(2)}_{ab},\phi^{(3)}_{ab}) & K_{{\bf R}_a  \in \{ (\N_a,\ov{\N}_b), (\N_a,\N_b), (\Anti_a), (\Sym_a) \}}
& b^{\cal A}_{ab} & b^{\cal M}_{aa'} \; \text{ (only for $b=a'$)} 
\\\hline\hline
(0,0,0) & \frac{g_{\text{string}}}{v_1v_2v_3} \, \sqrt{2\pi}^3 \frac{L_a^{(i)}}{\ell_s} &  -\frac{N_b}{4} \sum_{i=1}^3 \delta_{\sigma^i_a}^{\sigma^i_b} \; \delta_{\tau^i_a}^{\tau^i_b} I_{ab}^{\Z_2^{(i)},(j \cdot k)} 
&  - \frac{1}{2} \sum_{j<k} \sum_{m=0}^3 \eta_{\OR\Z_2^{(m)}} \; (-1)^{2 b_i \, \sigma^i_a \tau^i_a} \,| \tilde{I}_a^{\OR\Z_2^{(m)},(j \cdot k)}| 
\\\hline
(0^{(i)},\phi,-\phi) &   \frac{g_{\text{string}}}{v_1v_2v_3} \, \sqrt{2\pi}^3 \frac{L_a^{(i)}}{\ell_s}
& \frac{N_b \; \delta_{\sigma^i_a}^{\sigma^i_b} \; \delta_{\tau^i_a}^{\tau^i_b}}{4} \left( |I_{ab}^{(j \cdot k)}| - I_{ab}^{\Z_2^{(i)},(j \cdot k)}\right) & 
 - \frac{1}{2}  \sum_{ \text{\tiny $\begin{array}{c} m \in \{0 \ldots 3\} \text{ with }  \\ a \pp \OR\Z_2^{(m)} \text{ on } T^2_{(i)} \end{array} $}} \eta_{\OR\Z_2^{(m)}} \; (-1)^{2 b_i \, \sigma^i_a \tau^i_a} \,| \tilde{I}_a^{\OR\Z_2^{(m)},(j \cdot k)}| 
\\\hline
\begin{array}{c} (\phi^{(1)}_{ab},\phi^{(2)}_{ab},\phi^{(3)}_{ab}) \\ \text{\tiny ${\sum_i \phi^{(i)}_{ab}=0}$ }\end{array} 
&  \frac{g_{\text{string}}}{v_1v_2v_3}
\, \sqrt{ \prod_{i=1}^3 
\frac{ \Gamma (|\phi^{(i)}|)}{\Gamma (1-|\phi^{(i)}|)}^{-\frac{\sgn(\phi^{(i)})}{\sgn(I)}} }
& \frac{N_b}{8} \left( |I_{ab}| + \sgn(I_{ab}) \sum_{i=1}^3 I_{ab}^{\Z_2^{(i)}} \right) 
& \frac{1}{4} \sum_{m=0}^3 c_a^{\OR\Z_2^{(m)}} \eta_{\OR\Z_2^{(m)}} |\tilde{I}_a^{\OR\Z_2^{(m)}}|
\\\hline
\end{array}
}{beta_coeffs_Kaehler_metrics}{K\"ahler metrics $ K_{{\bf R}_a}$ and contributions to the beta function coefficients
$b^{\cal A}_{ab}$ , $b^{\cal M}_{aa'}$. 
The universally appearing function of the four-dimensional dilaton $S$ and bulk K\"ahler and complex structure moduli $T$ and $U$
is  $f(S,T,U) \sim e^{\phi_{10}} /{\rm Vol}(T^6)$~\cite{Blumenhagen:2007ip,Honecker:2011sm}.
The factor $(-1)^{2 b_i \, \sigma^i_a \tau^i_a}$ appearing in the M\"obius strip contribution to the beta function coefficient was first noted in the 
caption of table 49 in~\cite{Forste:2010gw}. Its necessity for the consistency of matter spectra derived on the one hand from 
intersection numbers and $\OR$ invariance properties and on the other hand using the relation~(\ref{Eq:beta-coeffs}) to beta function coefficients
is demonstrated in section~\protect\ref{Aa:sublety_Moebius}.
}
The one-loop corrections to the holomorphic gauge kinetic function have likewise been 
computed~\cite{Lust:2003ky,Akerblom:2007uc,Blumenhagen:2007ip,Gmeiner:2009fb,Honecker:2011sm,Honecker:2011hm} 
from open string vacuum amplitudes
using the magnetic background field method with the result summarised in table~\ref{tab:hol-gauge-kin}.  
\mathtabfix{
\begin{array}{|c||c|c|}\hline
\muc{3}{|c|}{\text{\bf One-loop corrections to holomorphic gauge kinetic functions on $T^6/\Z_2 \times \Z_{2M}$ with discrete torsion}}
\\\hline\hline
(\phi^{(1)}_{ab},\phi^{(2)}_{ab},\phi^{(3)}_{ab}) & \Re \left( \delta^{\text{1-loop},{\cal A}}_b {\rm f}_{SU(N_a)} \right)
 &  \Re \left( \delta^{\text{1-loop},{\cal M}}_{b=a'} {\rm f}_{SU(N_a)}  \right)
\\\hline\hline
(0,0,0) & - \sum_{i=1}^3 \frac{b^{{\cal A},(i)}_{ab}}{4 \pi^2} \, \ln \eta(i v_i) 
&  - \sum_{i=1}^3 \frac{b^{{\cal M},(i)}_{aa'}}{4 \pi^2} \, \ln \eta(i \tilde{v}_i)  - \frac{b^{{\cal M}}_{aa'} \, \ln(2)}{8 \pi^2} 
\\
& - \sum_{i=1}^3 \frac{\tilde{b}^{{\cal A},(i)}_{ab} \left(1- \delta_{\sigma^i_a}^{\sigma^i_b} \; \delta_{\tau^i_a}^{\tau^i_b}\right)}{4 \pi^2} 
\ln \left( e^{-\pi (\sigma^i_{ab})^2 v_i/4} 
\frac{\vartheta_1 (\frac{\tau^i_{ab} - i \sigma^i_{ab} v_i}{2}, iv_i)}{\eta(iv_i)}  \right)
& \text{(for $b_i=0$ or $(\sigma^i_a,\tau^i_a)=(0,0)$)}
\\\hline
(0^{(i)},\phi^{(j)}_{ab},\phi^{(k)}_{ab}) &  -\frac{b^{{\cal A}}_{ab}}{4 \pi^2} \, \ln \eta(i v_i) 
& -  \frac{b^{{\cal M}}_{aa'}}{4 \pi^2} \, \ln \eta(i \tilde{v}_i) 
\\ 
& - \frac{\tilde{b}^{{\cal A},(i)}_{ab} \left(1-\delta_{\sigma^i_a}^{\sigma^i_b} \; \delta_{\tau^i_a}^{\tau^i_b} \right)}{4 \pi^2} 
\ln \left( e^{-\pi (\sigma^i_{ab})^2 v_i/4} 
\frac{\vartheta_1 (\frac{\tau^i_{ab} - i \sigma^i_{ab} v_i}{2}, iv_i)}{\eta(iv_i)}  \right) 
&  \text{(for $b_i=0$ or $(\sigma^i_a,\tau^i_a)=(0,0)$)}
\\
& + \sum_{l=j,k}  \frac{N_b \, I^{\Z_2^{(l)}}}{32 \pi^2} \left(\frac{\sgn(\phi^{(l)}_{ab})}{2} - \phi^{(l)}_{ab} \right) 
& + \frac{\ln(2)}{32 \pi^2} \sum_{[m \text{ with } a \perp \OR\Z_2^{(m)} \text{ on } T^2_{(i)}]} \eta_{\OR\Z_2^{(m)}} |\tilde{I}_a^{\OR\Z_2^{(m)}}|
\\\hline
(\phi^{(1)},\phi^{(2)},\phi^{(3)})  & 
 \sum_{l=1}^3  \frac{N_b \, I^{\Z_2^{(l)}}}{32 \pi^2} \left(\frac{\sgn(\phi^{(l)}_{ab}) + \sgn(I_{ab})}{2} - \phi^{(l)}_{ab} \right)
&
\frac{\ln(2)}{32 \pi^2} \sum_{m=0}^3 \eta_{\OR\Z_2^{(m)}} |\tilde{I}_a^{\OR\Z_2^{(m)}}|
\\\hline
\end{array}
}{hol-gauge-kin}{One-loop contributions to the holomorphic gauge kinetic function with the two-torus volume modulus 
$v_i \equiv \frac{\sqrt{3}}{2} \frac{r_i^2}{\alpha'}$ and $\tilde{v}_i \equiv \frac{v_i}{1-b_i}$. The beta function coefficients are 
given in table~\protect\ref{tab:beta_coeffs_Kaehler_metrics}. Contributions from M\"obius strip topologies for some vanishing angle
are only known for square tori $(b_i=0)$ or vanishing displacements and Wilson lines $(\sigma^i_a,\tau^i_a)=(0,0)$ as argued in the text.
}

The K\"ahler metrics in table~\ref{tab:beta_coeffs_Kaehler_metrics} and one-loop corrections to the holomorphic gauge kinetic functions 
in table~\ref{tab:hol-gauge-kin} contain the angles between D6-branes.
The invariance of these field theoretical quantities under the identifications ${\bf AAA} \leftrightarrow {\bf ABB}$
is explained in section~\ref{Sss:T6Z2Z6p_identifications}, see in particular equation~(\ref{Eq:trafo-angles_AAA+ABB}) for the 
permutation of $a(\omega^k \, b)$ sectors and corresponding angles

\subsection{A Subtlety in M\"obius Strip Contributions to the Beta Function Coefficient}\label{Aa:sublety_Moebius}

In section~\ref{Ss:Gaugeenhanc}, orientifold invariant D6-branes are classified, and the corresponding gauge groups $USp(2N)$ or $SO(2N)$
are derived by means of the beta function coefficients. Furthermore, in section~\ref{Ss:no-anti+sym} the existence of D6-branes without any
matter in the symmetric or antisymmetric representation is investigated.
In both cases, contributions to the beta function coefficients for some vanishing angle need to be computed. The factor 
$(-1)^{2 b_i \, \sigma^i_a \tau^i_a}$ in the first two rows of the M\"obius strip contributions in 
table~\ref{tab:beta_coeffs_Kaehler_metrics} has been conjectured on the basis of consistency with interpretations in terms of the topology
of three-cycles and counting of massless open string states using Chan-Paton matrices.

The origin of  the uncertainty stems from the shape of the lattice sum 
$\tilde{\cal L}^{(i)}_{\text{M\"obius}} (\tilde{\sigma}^i_{aa'},\tilde{\tau}^i_{aa'},\ \tilde{L}^{(i)}_a; \tilde{r}_i, l)$ with 
$\sigma^i_{a},\tau^i_{a} \neq 0$ in equation~(\ref{Eq:amplitudes_per_two-torus}). The analogous sum 
$\tilde{\cal L}^{(i)}_{\text{Annulus}} (\sigma^i_{ab},\tau^i_{ab}, L^{(i)}_a; r_i, l)$ for Annulus amplitudes has been derived in~\cite{Gmeiner:2009fb}
for arbitrary $\sigma^i_{ab},\tau^i_{ab}$. For the M\"obius strip, the lattice sum $\tilde{\cal L}^{(i)}_{\text{M\"obius}}$ for $\sigma^i_{a}=\tau^i_{a}=0$
is well-known, see e.g.~\cite{Forste:2001gb,Gmeiner:2009fb} and references therein. It can be recovered from $\tilde{\cal L}^{(i)}_{\text{Annulus}}$
by replacing $(r_i^2, (L^{(i)}_a)^2) \to (\tilde{r}_i^2, (\tilde{L}^{(i)}_a)^2)$. The generalisation to $\sigma^i_{a} \neq 0$ but $\tau^i_{a}=0$ relies
on purely geometric considerations and can also be implemented by $\sigma^i_{aa'}= \tilde{\sigma}^i_{aa'}= 2\, \sigma^i_{a}$. For {\it untilted} tori,
T-duality arguments put Wilson lines on an equal footing with displacements. However, T-duality on {\it tilted} tori exchanges $\Z_2$ fixed points
in a different manner and has to our knowledge not been worked out so far.

The factor $(-1)^{2 b_i \, \sigma^i_a \tau^i_a}$ will substantiated in two examples below.

\subsubsection{Orientifold Invariant D6-branes Parallel to Ordinary O6-planes}

Orientifold invariant three-cycles have been classified in table~\ref{Tab:OR-inv-branes}. 
As an example, a D6-brane $a$ wrapping the orientifold invariant three-cycle parallel to the $\OR$-invariant O6-plane orbit with
$\sigma^1_a=\tau^1_a=1$ will be considered in the presence of the $\OR\Z_2^{(1)}$ the exotic O6-plane orbit.

The beta function coefficient for an $USp(2N_a)$ or $SO(2N_a)$ gauge group is given by 
\begin{equation}\label{Eq:Def-beta-SO+Sp}
\begin{aligned}
b_{USp/SO(2N_a)} &= \underbrace{N_a \left( -3+\varphi^{\Sym_a} + \varphi^{\Anti_a} \right) } 
+ \underbrace{ \left(  \varphi^{\Sym_a} - \varphi^{\Anti_a} -3 \, \xi_a \right)}
+ \underbrace{\sum_{b \neq a} \frac{N_b}{2}  \varphi^{ab}}
\\
&\equiv \qquad\qquad\qquad   b_{aa}^{\cal A}  \qquad\qquad +  \qquad\qquad  b_{aa}^{\cal M}  \qquad\qquad  +  \quad\quad  \sum_{b \neq a} b_{ab}^{\cal A}
,
\end{aligned}
\end{equation}
with $\xi_a=+1$ for an $USp(2N_a)$ and $\xi_a=-1$ for a $SO(2N_a)$ gauge group.

For the above specified D6-brane configuration and a tilted two-torus $T^2_{(1)}$, i.e. $b_1=\frac{1}{2}$, the first line of 
table~\ref{tab:beta_coeffs_Kaehler_metrics} gives
\begin{equation}
 b_{aa}^{\cal A} + b_{aa}^{\cal M} = -3 \, N_a - \left( (-1)^{\sigma^1_a\tau^1_a} \, \eta_{\OR\Z_2^{(1)}} + \eta_{\OR\Z_2^{(2)}} +   \eta_{\OR\Z_2^{(3)}} \right)
= -3 \, N_a - 3
,
\end{equation}
in accord with the interpretation that the Chan-Paton factor associated to the massless states $\psi^{\mu}_{-1/2}|0\rangle_{\text{NS}}$
generates the gauge group $USp(2N_a)$. Due to $\eta_{\OR\Z_2^{(1)}}=-1$ and $\eta_{\OR\Z_2^{(2)}}=\eta_{\OR\Z_2^{(3)}}=+1$, the interpretation of 
the contribution $b_{aa}^{\cal A} + b_{aa}^{\cal M}$ to the total beta function coefficient in terms of a vector multiplet in the adjoint
representation is only possible if the sign factor $(-1)^{\sigma^1_a\tau^1_a}$ is inserted.

\subsubsection{Orientifold Image D6-branes at one Vanishing Angle}

It is instructive to use the bulk cycles of the $T^6/(\Z_2 \times \Z_6' \times \OR)$ example with discrete torsion in~\cite{Forste:2010gw}
but switch on displacements and Wilson lines. For the {\bf AAB}, the bulk three-cycle with wrapping numbers
\begin{equation}
(n^1_a,m^1_a;n^2_a,m^2_a;n^3_a,m^3_a)=(0,1;1,-1;1,1)  
\end{equation}
is at angle $\pi (\frac{1}{3},-\frac{1}{3},0)$ with respect to the $\OR$-invariant plane. 

For the choice $\eta_{\OR}=-1$ of an exotic O6-plane, the intersection number of any D6-brane $a$ with its orientifold image $a'$
can be written as 
\begin{equation}\label{Eq:Ex-one-angle-inters}
\Pi_a^{\text{frac}} \circ \Pi_{a'}^{\text{frac}} = \left\{\begin{array}{cr} 0 & (\sigma^1_a\tau^1_a,\sigma^2_a\tau^2_a)=(0,0),(1,1) \\ +2 & (1,0)
\\ -2 & (0,1) \end{array}\right.  
,
\qquad 
\Pi_a^{\text{frac}} \circ \Pi_{O6}=0
.
\end{equation}
The contributions to the beta function coefficients are read off from the second line in table~\ref{tab:beta_coeffs_Kaehler_metrics}
with $(ijk)$ permutations of $(123)$,
\begin{equation}
\begin{aligned}
b_{(\omega^k \, a)(\omega^k \, a)'}^{\cal A} + b_{(\omega^k \, a)(\omega^k \, a)'}^{\cal M} = &  
\frac{N_a}{4} \left( |I_{(\omega^k \, a)(\omega^k \, a)'}^{(i \cdot j)}| -I_{(\omega^k \, a)(\omega^k \, a)'}^{\Z_2^{(3-k)},(i \cdot j)} \right)
\\& + \frac{(-1)^{\sigma^{3-k}_a \tau^{3-k}_a}}{2} \left( \eta_{\OR} \, \tilde{I}_{(\omega^k \, a)}^{\OR,(i \cdot j)} + 
\eta_{\OR\Z_2^{(3-k)}} \, \tilde{I}_{(\omega^k \, a)}^{\OR\Z_2^{(3-k)},(i \cdot j)} \right)
\\
=&\left\{\begin{array}{cr} 0 & k=0 \\ \frac{N_a}{2} + (-1)^{\sigma^2_a\tau^2_a} & 1 \\ \frac{N_a}{2} + (-1)^{\sigma^1_a\tau^1_a} & 2 \end{array}\right. 
\end{aligned}
\end{equation}
leading to
\begin{equation}
\sum_{k=0}^2 \left(b_{(\omega^k \, a)(\omega^k \, a)'}^{\cal A} + b_{(\omega^k \, a)(\omega^k \, a)'}^{\cal M} \right)
= N_a + (-1)^{\sigma^1_a\tau^1_a} + (-1)^{\sigma^2_a\tau^2_a} 
\end{equation}
and finally using~(\ref{Eq:beta-coeffs}) to the amount of matter states in symmetric and antisymmetric representations
\begin{equation}
(\varphi^{\Sym_a}, \varphi^{\Anti_a}) = \left\{\begin{array}{cr} (2,0) &  (\sigma^1_a\tau^1_a,\sigma^2_a\tau^2_a)=(0,0) \\ (1,1) & (1,0),(0,1) \\
(0,2) & (1,1)
 \end{array}\right.
\Leftrightarrow
\begin{array}{c}
(\Sym_a) + h.c.\\
{} [(\Sym_a) + (\Anti_a)] \text{ or } [(\ov{\Sym}_a) + (\ov{\Anti}_a)]\\
(\Anti_a ) + h.c.
\end{array}
, 
\end{equation}
which is consistent with the net-chiralities $\chi^{\Anti_a/\Sym_a}=\frac{\Pi_a^{\text{frac}} \circ \Pi_{a'}^{\text{frac}} \pm\Pi_a^{\text{frac}} \circ \Pi_{O6}}{2}$
and intersection numbers in equation~(\ref{Eq:Ex-one-angle-inters}). 
The matching clearly requires the conjectured sign factors $(-1)^{\sigma^i_a\tau^i_a}$ in the M\"obius strip contribution to the beta function coefficients.

\section{Identification of Compactifications on the {\bf AAB} and {\bf BBB} Lattices}\label{A:identifications_AAB+BBB}

The map between ${\bf AAA}$ and ${\bf ABB}$ lattice backgrounds has been proven in section~\ref{Ss:identifications}. 
The action on toroidal one-cycle wrapping numbers is, up to permutation of two-torus indices, identical for the map between {\bf AAB} and {\bf BBB} 
lattices,
\begin{equation}\label{Eq:identify_nm_AAB+BBB}
\begin{aligned}
\left(\begin{array}{cc} n^1_a & m^1_a  \\ n^2_a  & m^2_a  \\ n^3_a  & m^3_a   \end{array}\right)_{\bf AAB}
&=\left(\begin{array}{cc} \ov{n}^1_a  & \ov{m}^1_a  \\ \ov{n}^2_a  & \ov{m}^2_a  \\ \ov{n}^3_a +\ov{m}^3_a   & -\ov{n}^3_a   \end{array}\right)_{\bf BBB}
,
\end{aligned}
\end{equation}
where again all other discrete data ($\Z_2$ eigenvalues, Wilson lines, displacements) are preserved,
\begin{equation}\label{Eq:identify_sigma+tau_AAB+BBB}
\begin{aligned}
\left(\tau^{\Z_2^{(i)}}_a \, ; \,  \tau^i_a \, ; \, \sigma^i_a \right)_{\bf AAB} &=  \quad
\left(\ov{\tau}^{\Z_2^{(i)}}_a \, ; \,  \ov{\tau}^i_a \, ; \, \ov{\sigma}^i_a \right)_{\bf BBB} 
\qquad \text{ for } \quad i=1,2,3
.
\end{aligned}
\end{equation}
Since for the present pair of lattice orientations, the wrapping numbers on the first two-torus are preserved under the identification~(\ref{Eq:identify_nm_AAB+BBB}),
it is convenient to use the condition $(n^1_a,m^1_a)=\text{(odd,odd)}$ with $n^1_a >0$  to single out one orbifold image and one-cycle orientation
instead of the same constraint on the third two-torus as stated at the end of section~\ref{Sss:Geo_bulk} for the {\bf AAA} and {\bf ABB} lattices.

The invariance of the RR tadpole cancellation and supersymmetry conditions under this map is analogously to the discussion for the first
pair of lattices in section~\ref{Sss:T6Z6_identifications}
at first most easily checked for the 
$T^6/(\Z_6 \times \OR)$ orbifold with $\Z_2^{(3)} \subset \Z_6$, for which only the bulk and $\Z_2^{(3)}$ twisted sectors need to be considered.
The identifications~(\ref{Eq:identify_nm_AAB+BBB}) and~(\ref{Eq:identify_sigma+tau_AAB+BBB}) imply
\begin{equation}\label{Eq:identify_xy3_AAB+BBB}
\begin{aligned}
\left(X_a\, , \, Y_a \right)_{\bf AAB} &= \left( \ov{X}_a + \ov{Y}_a \, ,\,  -\ov{X}_a \right)_{\bf BBB} ,
\\
\left(x^{(3)}_{\alpha,a} \, ,\, y^{(3)}_{\alpha,a} \right)_{\bf AAB} &= \left( \ov{x}^{(3)}_{\alpha,a} + \ov{y}^{(3)}_{\alpha,a} \, , \,  - \ov{x}^{(3)}_{\alpha,a} \right)_{\bf BBB}
\qquad \text{ for each } \quad 
\alpha \in \{1 \ldots 5\}
.
\end{aligned}
\end{equation}
The global bulk consistency conditions for $T^6/(\Z_6 \times \OR)$ in table~\ref{tab:Bulk-RR+SUSY-Z6} obviously match, and so do
the $\Z_2^{(3)}$ twisted tadpoles, which can be read off from table~\ref{tab:twistedRR-AAB+BBB-Z2Z6p} with $\eta_{(3)} \equiv 1$
on both lattices since the $\Z_2^{(3)}$ fixed points do not transform under~(\ref{Eq:identify_nm_AAB+BBB}) and~(\ref{Eq:identify_sigma+tau_AAB+BBB}).

For $T^6/(\Z_2 \times \Z_6' \times \OR)$ with discrete torsion, the bulk RR tadpole cancellation and supersymmetry
conditions are identical to  $T^6/(\Z_6 \times \OR)$, 
and $\Z_2^{(3)}$ twisted consistency conditions in table~\ref{tab:twistedRR-AAB+BBB-Z2Z6p} are likewise satisfied for both 
choices $(\eta_{(3)} )_{\bf AAB} =(\eta_{(3)} )_{\bf BBB} = \pm 1$ as can be explicitly checked by inserting the second line of relations in
equation~(\ref{Eq:identify_xy3_AAB+BBB}). 

\mathtabfix{
\begin{array}{|c||c|c|c|}\hline
\multicolumn{4}{|c|}{\text{\bf Twisted RR tadpole cancellation conditions on $T^6/(\Z_2 \times \Z_6' \times \OR)$ with discrete torsion, Part II}}
\\\hline\hline
i &  & {\bf AAB} & {\bf BBB}
\\\hline\hline
3 & \alpha=1,2,3 & \sum_a N_a \left[x^{(3)}_{\alpha,a} -\eta_{(3)} \;  y^{(3)}_{\alpha,a}) \right]=0
& \sum_a N_a \left[    x^{(3)}_{\alpha,a} + \eta_{(3)} \; \left( x^{(3)}_{\alpha,a} + y^{(3)}_{\alpha,a} \right)\right]=0 = \sum_a N_a (1 - \eta_{(3)}) \;  y^{(3)}_{\alpha,a}
\\\hline
& \alpha=4,5 & \sum_a N_a \left[ x^{(3)}_{4,a}\! -\!\eta_{(3)}  \; y^{(3)}_{5,a} \right]=0 =  \sum_a N_a  \left[ x^{(3)}_{5,a} -\eta_{(3)} \;  y^{(3)}_{4,a} \right]
&\sum_a N_a \left[ \left( 2\, x^{(3)}_{4,a} +y^{(3)}_{4,a} \right) +\eta_{(3)} \;  \left( 2 \, x^{(3)}_{5,a} + y^{(3)}_{5,a} \right) \right]=0 =\sum_a N_a \left[ y^{(3)}_{4,a} - \eta_{(3)} \; y^{(3)}_{5,a} \right]
\\\hline\hline
1,2 & \alpha=1,5 & \sum_a N_a \left[   x^{(i)}_{\alpha,a}  -\eta_{(i)} \; \left(x^{(i)}_{\alpha,a} + y^{(i)}_{\alpha,a} \right) \right]=0 = \sum_a N_a  (1 +\eta_{(i)}) \;  y^{(i)}_{\alpha,a}
& \text{analogous to } i=3
\\\hline
& \alpha=2 & \sum_a N_a \left[  y^{(i)}_{2,a} -\eta_{(i)} \; \left(x^{(i)}_{2,a} +  y^{(i)}_{2,a} \right) \right]=0 =  \sum_a N_a (1+\eta_{(i)}) \; x^{(k)}_{2,a} &
\\\hline
&  \alpha=3,4 & \sum_a N_a \left[ \left( 2\, x^{(i)}_{3,a} +y^{(i)}_{3,a}  \right) -\eta_{(i)} \; \left( 2 \, x^{(i)}_{4,a} +  y^{(i)}_{4,a} \right) \right]=0 =  \sum_a N_a \left[y^{(i)}_{3,a} + \eta_{(i)} \; y^{(i)}_{4,a} \right] &
\\\hline
\end{array}
}{twistedRR-AAB+BBB-Z2Z6p}{Twisted RR tadpole cancellation conditions for the {\bf AAB} and {\bf BBB} lattice orientations, which are related by the identifications~(\protect\ref{Eq:identify_nm_AAB+BBB})
and~(\protect\ref{Eq:identify_sigma+tau_AAB+BBB}) of wrapping numbers and sign factors. The twisted RR tadpole cancellation conditions for 
$T^6/(\Z_6 \times \OR)$ with $\Z_2^{(3)} \subset \Z_6$ are obtained by setting $\eta_{(3)} \equiv 1$ and 
truncating the remaining two $\Z_2^{(1)}$ and $\Z_2^{(2)}$ twisted sectors.}

The exceptional wrapping numbers in the remaining two $\Z_2^{(1)}$ and $\Z_2^{(2)}$ twisted sectors are related by 
$\left(\eta_{(i)}\right)_{\bf AAB} \leftrightarrow \left(- \eta_{(i)}\right)_{\bf BBB}$ for $i \in \{1,2\}$
and 
\begin{equation}\label{Eq:identify_xy_12_AAB+BBB}
\begin{aligned}
\left(x^{(i)}_{\alpha,a}\, , \, y^{(i)}_{\alpha,a} \right)^{i =1,2}_{\bf AAB}  &= \left( - (\ov{x}^{(i)}_{\alpha,a} + \ov{y}^{(i)}_{\alpha,a}) \, , \, \ov{x}^{(i)}_{\alpha,a} \right)^{i =1,2}_{\bf BBB}  
 \qquad 
 \alpha = (1,3,4,5)
; \quad
 \ov{\alpha} = (1,5,3,4)
 ,
\\ 
\\
\left( x^{(i)}_{2,a}\, ,\, y^{(i)}_{2,a} \right)^{i =1,2}_{\bf AAB}  &= \left( \ov{y}^{(i)}_{2,a}, - ( \ov{x}^{(i)}_{2,a} +\ov{y}^{(i)}_{2,a} ) \right)^{i =1,2}_{\bf BBB}  
,
\end{aligned}
\end{equation}
as can be verified on a case-by-case basis using the complete list in table~\ref{tab:exceptional-wrappings-Z2Z6}.
For example, the transformation of one-cycle wrapping numbers~(\ref{Eq:identify_nm_AAB+BBB}) leads to
\begin{equation}
\begin{aligned}
{\bf AAB} & \quad \leftrightarrow \quad {\bf BBB}\\
(n^2_a,m^2_a;n^3_a,m^3_a)=\text{(odd,even;odd,odd)} & \quad \leftrightarrow \quad (\ov{n}^2_a,\ov{m}^2_a;\ov{n}^3_a,\ov{m}^3_a)=\text{(odd,even;odd,even)}
,
\end{aligned}
\end{equation}
and for $(\sigma^2_a;\sigma^3_a)=(0;1)$ the three-cycle traverses  on $T^2_{(2)} \times T^2_{(3)}$
the fixed points 14, 15, 44, 45 for the {\bf AAB} lattice 
and 15, 16, 45, 46 for the {\bf BBB} lattice. The first two fixed points contribute to $\Z_2^{(1)}$ twisted cycles 
$(\varepsilon^{(1)}_{\alpha}, \tilde{\varepsilon}^{(1)}_{\alpha})$ with $\alpha=\ov{\alpha}=2$ and prefactors $(-1)^{\tau^{\Z_2^{(1)}}_a}$ and 
$(-1)^{\tau^{\Z_2^{(1)}}_a+\tau^3_a}$,
respectively, while the remaining two fixed points contribute to $\alpha=3,4$ on {\bf AAB} and $\ov{\alpha}=4,5$ on {\bf BBB} with
prefactors $(-1)^{\tau^{\Z_2^{(1)}}_a+\tau^1_a}$ and $(-1)^{\tau^{\Z_2^{(1)}}_a+\tau^1_a+\tau^3_a}$.
This constitutes one example of the blockwise permutation $1^{\rm st} \to 3^{\rm rd} \to 2^{\rm nd} \to 1^{\rm st}$ in 
table~\ref{tab:exceptional-wrappings-Z2Z6} under ${\bf BBB} \to {\bf AAB}$ with a simultaneous permutation 
$1^{\rm st} \to 3^{\rm rd} \to 2^{\rm nd} \to 1^{\rm st}$ of the three lines within each block.

The discussion is straightforwardly extended to intersection numbers and angles as in section~\ref{Sss:T6Z2Z6p_identifications} 
upon permutation of two-torus indices $T^2_{(1)} \leftrightarrow T^2_{(3)}$, e.g. for the angles with respect to the $\OR$-invariant plane
and relative intersection angles, one obtains
\begin{equation}\label{Eq:trafo-angles_AAB+BBB}
\begin{aligned}
\pi \left(\phi^{(1)}_a, \phi^{(2)}_a, \phi^{(3)}_a \right)_{\bf AAB}  =& 
 \pi  \, \bigl(\ov{\phi}^{(1)}_{\ov{a}},  \ov{\phi}^{(2)}_{\ov{a}}, \ov{\phi}^{(3)}_{\ov{a}} \bigr)_{\bf BBB} 
+ \pi \bigl( \frac{1}{6},\frac{1}{6} , -\frac{1}{3}\bigr)
,
\\
\pi \bigl( \vec{\phi}_{a(\omega^k \, b)} \bigr)_{\bf AAB} = & \pi \bigl( \vec{\ov{\phi}}_{\ov{a}(\omega^k \, \ov{b})} \bigr)_{\bf BBB}
,
\\
\pi \bigl( \vec{\phi}_{a(\omega^k \, b')} \bigr)_{\bf AAB} =&\pi \bigl( \vec{\ov{\phi}}_{\ov{a}(\omega^{k-1} \, \ov{b}')} \bigr)_{\bf BBB}
,
\end{aligned}
\end{equation}
in agreement with equations~(\ref{Eq:trafo-angles_AAA+ABB}) to~(\ref{Eq:abP-angles}).
The full towers of massless and massive string excitations and field theoretical results at our current state of knowledge 
are thus identical for pairs of D6-brane configurations on the {\bf AAB} and {\bf BBB} lattices under the identifications of
D6-brane data~(\ref{Eq:identify_nm_AAB+BBB}) and~(\ref{Eq:identify_sigma+tau_AAB+BBB}).

\section{Rigid D6-Branes with Three Generations of Quarks}
\label{A:RigidD63Gen}

The three different sets of completely rigid D6-branes, for which three generations of quarks are found, are given . All these D6-branes wrap the cycle (0,1;1,0;1,-1)
on the {\bf AAA} lattice. The $\Z_2$ eigenvalues for the $QCD$ stack can without loss of generality be as chosen $(-1)^{\tau_a^{\Z_2^{(i)}}} =(+,+,+)$, since all physical quantities only 
depend on the {\it relative} $\Z_2$ eigenvalues.
\begin{table}[h]
\begin{center}
\begin{tabular}{|c||c|c||c|c|c||c|c|c|}
\hline \multicolumn{9}{|c|}{\bf Rigid cycles on (0,1;1,0;1,-1) with three generations (set 1)}\\
\hline
\hline&\multicolumn{2}{|c||}{\bf $D6_a$ stack} &  \multicolumn{3}{|c||}{\bf $D6_b$ stack} &  \multicolumn{3}{|c|}{\bf $D6_c$ stack} \\
\hline\hline& $(\vec{\tau_a})$ & $(\vec{\sigma_a})$ & $\tau_{ab}^{\Z^{(i)}_2}$ & $(\vec{\tau_b})$& $(\vec{\sigma_b})$ & $\tau_{ac}^{\Z^{(i)}_2}$ & $(\vec{\tau_c})$& $(\vec{\sigma_c})$\\
\hline  \hline
1&(0,0,1) &(0,0,1)&   (0,1,1)& (0,1,1)& (0,1,1) &   (1,1,0) & (1,0,1) & (1,0,1) \\
2&(0,0,1) &(0,1,1)&   (1,1,0)& (0,1,1)& (0,1,1) &   (1,1,0) & (1,0,1) & (1,1,1) \\
3&(0,0,1) &(1,0,1)&   (0,1,1)& (0,1,1)& (1,1,1) &   (1,0,1) & (1,0,1) & (1,0,1) \\
4&(0,0,1) &(1,1,1)&   (1,1,0)& (0,1,1)& (1,1,1) &   (1,0,1) & (1,0,1) & (1,1,1) \\
5&(0,1,1) &(0,0,1)&   (1,1,0)& (0,1,1)& (0,1,1) &   (1,1,0) & (1,1,1) & (1,0,1) \\
6&(0,1,1) &(1,0,1)&   (1,1,0)& (0,1,1)& (1,1,1) &   (1,0,1) & (1,1,1) & (1,0,1) \\
7&(1,0,1) &(0,0,1)&   (0,1,1)& (1,1,1)& (0,1,1) &   (1,0,1) & (1,0,1) & (1,0,1) \\
8&(1,0,1) &(0,1,1)&   (1,1,0)& (1,1,1)& (0,1,1) &   (1,0,1) & (1,0,1) & (1,1,1) \\
9&(1,1,1) &(0,0,1)&   (1,1,0)& (1,1,1)& (0,1,1) &   (1,0,1) & (1,1,1) & (1,0,1) \\
\hline
\end{tabular}
\caption{First list of rigid fractional cycles on bulk orbit (0,1;1,0;1,-1) on the {\bf AAA} lattice
with net-chiralities $(\chi^{ab},\chi^{ab'}) = (1,2)$, $(\chi^{ac},\chi^{ac'}) = (-1,-2)$ and $(\chi^{bc},\chi^{bc'}) = (1,0)$. \label{Tab:RigFreeSet1}}
\end{center}
\end{table}

\begin{table}[h]
\begin{center}
\begin{tabular}{|c||c|c||c|c|c||c|c|c|}
\hline \multicolumn{9}{|c|}{\bf Rigid cycles on (0,1;1,0;1,-1) with three generations (set 2)}\\
\hline
\hline &\multicolumn{2}{|c||}{\bf $D6_a$ stack} &  \multicolumn{3}{|c||}{\bf $D6_b$ stack} &  \multicolumn{3}{|c|}{\bf $D6_c$ stack} \\
\hline\hline &$(\vec{\tau_a})$ & $(\vec{\sigma_a})$ & $\tau_{ab}^{\Z^{(i)}_2}$ & $(\vec{\tau_b})$& $(\vec{\sigma_b})$ & $\tau_{ac}^{\Z^{(i)}_2}$ & $(\vec{\tau_c})$& $(\vec{\sigma_c})$\\
\hline \hline
1&(0,1,0) &(0,1,0) &   (1,1,0) &(1,1,0) &(1,1,0)&   (1,0,1) & (0,1,1) & (0,1,1)\\
2&(0,1,0) &(0,1,1) &   (1,1,0) &(1,1,0) &(1,1,1)&   (0,1,1) & (0,1,1) & (0,1,1)\\
3&(0,1,0) &(1,1,0) &   (1,0,1) &(1,1,0) &(1,1,0)&   (1,0,1) & (0,1,1) & (1,1,1)\\
4&(0,1,0) &(1,1,1) &   (1,0,1) &(1,1,0) &(1,1,1)&   (0,1,1) & (0,1,1) & (1,1,1)\\
5&(0,1,1) &(0,1,0) &   (1,1,0) &(1,1,1) &(1,1,0)&   (0,1,1) & (0,1,1) & (0,1,1)\\
6&(0,1,1) &(1,1,0) &   (1,0,1) &(1,1,1) &(1,1,0)&   (0,1,1) & (0,1,1) & (1,1,1)\\
7&(1,1,0) &(0,1,0) &   (1,0,1) &(1,1,0) &(1,1,0)&   (1,0,1) & (1,1,1) & (0,1,1)\\
8&(1,1,0) &(0,1,1) &   (1,0,1) &(1,1,0) &(1,1,1)&   (0,1,1) & (1,1,1) & (0,1,1)\\
9&(1,1,1) &(0,1,0) &   (1,0,1) &(1,1,1) &(1,1,0)&   (0,1,1) & (1,1,1) & (0,1,1)\\
\hline
\end{tabular}
\caption{Second list of rigid fractional cycles on bulk orbit (0,1;1,0;1,-1)  on the {\bf AAA} lattice
with net-chiralities $(\chi^{ab},\chi^{ab'}) = (1,2)$, $(\chi^{ac},\chi^{ac'}) = (-1,-2)$ and $(\chi^{bc},\chi^{bc'}) = (1,0)$.\label{Tab:RigFreeSet2}}
\end{center}
\end{table}

\begin{table}[h]
\begin{center}
\begin{tabular}{|c||c|c||c|c|c||c|c|c|}
\hline \multicolumn{9}{|c|}{\bf Rigid cycles on (0,1;1,0;1,-1) with three generations (set 3)}\\
\hline
\hline &\multicolumn{2}{|c||}{\bf $D6_a$ stack} &  \multicolumn{3}{|c||}{\bf $D6_b$ stack} &  \multicolumn{3}{|c|}{\bf $D6_c$ stack} \\
\hline\hline &$(\vec{\tau_a})$ & $(\vec{\sigma_a})$ & $\tau_{ab}^{\Z^{(i)}_2}$ & $(\vec{\tau_b})$& $(\vec{\sigma_b})$ & $\tau_{ac}^{\Z^{(i)}_2}$ & $(\vec{\tau_c})$& $(\vec{\sigma_c})$\\
\hline \hline
1&(1,0,0) & (1,0,0) &   (1,0,1)& (1,0,1) & (1,0,1) &    (0,1,1) & (1,1,0) & (1,1,0)\\
2&(1,0,0) & (1,0,1) &   (0,1,1)& (1,0,1) & (1,0,1) &    (0,1,1) & (1,1,0) & (1,1,1)\\
3&(1,0,0) & (1,1,0) &   (1,0,1)& (1,1,1) & (1,0,1) &    (1,1,0) & (1,1,0) & (1,1,0)\\
4&(1,0,0) & (1,1,1) &   (0,1,1)& (1,0,1) & (1,1,1) &    (1,1,0) & (1,1,0) & (1,1,1)\\
5&(1,0,1) & (1,0,0) &   (0,1,1)& (1,0,1) & (1,0,1) &    (0,1,1) & (1,1,1) & (1,1,0)\\
6&(1,0,1) & (1,1,0) &   (0,1,1)& (1,0,1) & (1,1,1) &    (1,1,0) & (1,1,1) & (1,1,0)\\
7&(1,1,0) & (1,0,0) &   (1,0,1)& (1,1,1) & (1,0,1) &    (1,1,0) & (1,1,0) & (1,1,0)\\
8&(1,1,0) & (1,0,1) &   (0,1,1)& (1,1,1) & (1,0,1) &    (1,1,0) & (1,1,0) & (1,1,1)\\
9&(1,1,1) & (1,0,0) &   (0,1,1)& (1,1,1) & (1,0,1) &    (1,1,0) & (1,1,1) & (1,1,0)\\
\hline
\end{tabular}
\caption{Third list of rigid fractional cycles on bulk orbit (0,1;1,0;1,-1)  on the {\bf AAA} lattice
with net-chiralities $(\chi^{ab},\chi^{ab'}) = (1,2)$, $(\chi^{ac},\chi^{ac'}) = (-1,-2)$ and $(\chi^{bc},\chi^{bc'}) = (1,0)$.\label{Tab:RigFreeSet3}}
\end{center}
\end{table}

\clearpage

\section{A Six-stack Pati-Salam Model}\label{A:6stack_PS}
In subsection \ref{Ss:PS-model} we presented a five-stack Pati-Salam model where the `QCD stack' was taken to be free of non-chiral matter in the symmetric or antisymmetric representation. If this assumption is dropped - while maintaining the rigidness of the `QCD stack' - other types of Pati-Salam models can be constructed. Here we present a global six-stack Pati-Salam model  with two non-chiral matter pairs in the antisymmetric representation charged under the `QCD stack' and with `hidden' gauge groups $U(4)_d \times U(2)_e \times U(2)_f$. The full D-brane configuration is given in table \ref{Tab:OtherDecentPatiSalam}. The corresponding chiral and non-chiral spectrum are given in tables \ref{Tab:OtherDecentPatiSalamC} and \ref{Tab:OtherDecentPatiSalamNC} respectively. 

\begin{table}[h]
\begin{center}
\begin{tabular}{|c|c|c||c|c|c|c|}
\hline &\bf wrapping numbers& $\frac{\rm Angle}{\pi}$ &\bf $\Z_2^{(i)}$ eigenvalues  & ($\vec \tau$) & ($\vec \sigma$)& \bf gauge group\\
\hline \hline
 $a$&(1,0;1,0,1,0)&$(0,0,0)$&$(+++)$&$(0,1,0)$ & $(1,1,0)$& $U(4)$\\
 $b$&(0,1;1,0,1,-1)&$(\frac{1}{3},0,-\frac{1}{3})$&$(--+)$&$(1,1,0)$ & $(1,1,0)$&$U(2)_L$\\
 $c$&(0,1;1,0,1,-1)&$(\frac{1}{3},0,-\frac{1}{3})$&$(+--)$&$(1,1,0)$ & $(1,1,0)$&$U(2)_R$\\
 \hline $d$ & (1,0;1,0;1,0) &$(0,0,0)$& $(-+-)$ &$(0,1,0)$&$(1,1,1)$&$U(4)_d$\\
 $e$ & (1,0;1,0;1,0)&$(0,0,0)$ &$(+++)$&$(0,1,0)$& $(0,1,1)$&$U(2)_e$\\
 $f$ & (1,0;1,0;1,0)&$(0,0,0)$  &$(+--)$&$(0,1,1)$& $(0,1,0)$&$U(2)_f$\\
 \hline
\end{tabular}
\caption{D-brane configuration with six D-brane stacks yielding a Pati-Salam model with gauge group $SU(4)_a\times SU(2)_b\times SU(2)_c \times SU(4)_d \times SU(2)_e\times SU(2)_f\times U(1)^6$\label{Tab:OtherDecentPatiSalam}}
\end{center}
\end{table}

In this model the Higgs-sector is no longer minimal: the $bc'$ sector yields two doublets $(H_d, H_u)$ and from the $bc$ sector we obtain another non-chiral pair. 
This plethora of Higgses allows us to write down Yukawa interactions for all three generations: 
\begin{equation}\label{Eq:PS2Yukawa}
\begin{array}{ll}
i = 2, 3: &\left( \bar u_R, \bar d_R, \bar\nu_R, \bar e_R \right)_{ac}  \left(H_d, H_u\right)_{bc} \left(Q_L^{(i)} , L^{(i)}\right)_{ab},\\
i = 2, 3: &\left( \bar u_R^{(i)}, \bar d_R^{(i)}, \bar\nu_R^{(i)}, \bar e_R^{(i)} \right)_{ac'}  \left(H_d, H_u\right)_{bc} \left(Q_L , L\right)_{ab'},\\
i, k = 2,3; j= 1,2: & \left( \bar u_R^{(i)}, \bar d_R^{(i)}, \bar\nu_R^{(i)}, \bar e_R^{(i)} \right)_{ac}  \left(H_d^{(j)}, H_u^{(j)}\right)_{bc'} \left(Q_L^{(k)} , L^{(k)}\right)_{ab},
\end{array}
\end{equation} 
such that all three generations can be made massive according to the charge selection rule. In table \ref{Tab:PS2Yukawa} we take a closer look at these Yukawa coupling and investigate which ones correspond to a closed sequence of intersecting three-cycles. It turns out that not all Yukawa couplings written down above survive the sequence selection rule. Therefore, not all generations of quarks and leptons acquire mass through a Yukawa interaction with the Higgs fields responsible for the spontaneous symmetry breaking of the gauge group $SU(2)_b\times SU(2)_c$. The Yukawa couplings corresponding to a closed triangle are of order ${\cal O}(1)$, while the others merely vanish.

Looking at the non-chiral spectrum one notices the absence of adjoint matter, indicating that the six- stack Pati-Salam model has been constructed on completely rigid D-branes. On the other hand, the model also comes with some additional chiral bifundamental matter charged under $U(2)_L$ or $U(2)_R$ and one of the hidden gauge groups, and with non-chiral bifundamental matter charged under the `QCD stack' $U(4)$ and one of the hidden gauge groups. 
\begin{table}[h]
\begin{center}
\begin{tabular}{|c||c|c|c|}
\hline \multicolumn{4}{|c|}{\bf Chiral spectrum of a 6-stack Pati-Salam model}\\
\hline \hline
Matter & Sector & $U(4)\times U(2)_L \times U(2)_R \times U(4)_d \times U(2)_e\times U(2)_f$& $(Q_a, Q_b, Q_c, Q_d, Q_e, Q_f)$ \\
\hline $\left(Q_L , L\right)  $&$ab$&$2 \times ({\bf 4}, {\bf\bar 2}, \1,\1,\1,\1)$ &(1,-1,0,0,0,0)\\
$\left( Q_L, L \right)$&$ab'$&  $({\bf4}, {\bf 2}, \1,\1,\1,\1)$&(1,1,0,0,0,0)\\
$\left( \bar u_R, \bar d_R, \bar\nu_R, \bar e_R \right)  $ & $ac$ & $({\bf\bar 4}, \1,{\bf 2},\1,\1,\1)$ &(-1,0,1,0,0,0) \\
$\left( \bar u_R, \bar d_R, \bar\nu_R, \bar e_R \right)  $& $ac'$ &  $2 \times ({\bf\bar 4}, \1,{\bf\bar 2},\1,\1,\1)$&(-1,0,-1,0,0,0) \\
$\left(H_d, H_u\right)$&$bc'$&$2 \times (\1, {\bf2}, {\bf 2},\1,\1,\1)$&(0,1,1,0,0,0) \\
& $bd$ & $(\1, {\bf 2}, \1, {\bf \bar 2},\1,\1) $&(0,1,0,-1,0,0)\\
&$bd'$&  $(\1, {\bf \bar  2}, \1, {\bf \bar 2},\1,\1) $ &(0,-1,0,-1,0,0)\\
&$bf$&$(\1, {\bf 2}, \1, \1,\1, {\bf \bar 2}) $&(0,1,0,0,0,-1)\\
&$bf'$&$(\1, {\bf \bar  2}, \1, \1, \1, {\bf \bar 2}) $&(0,-1,0,0,0,-1)\\
&$cd$& $(\1, \1, {\bf \bar 2}, {\bf 2},\1,\1) $   &(0,0,-1,1,0,0) \\
&$cd'$&  $(\1, \1, {\bf  2}, {\bf 2},\1,\1) $ & (0,0,1,1,0,0)\\
&$cf$&  $(\1, \1, {\bf 2}, \1, \1, {\bf \bar 2}) $ & (0,0,1,0,0,-1)\\
&$cf'$&  $(\1, \1, {\bf \bar 2}, \1, \1, {\bf \bar 2}) $ & (0,0,-1,0,0,-1)\\
\hline
\end{tabular}
\caption{Chiral spectrum for the D-brane configuration in table \ref{Tab:OtherDecentPatiSalam}.\label{Tab:OtherDecentPatiSalamC}}
\end{center}
\end{table}

\begin{table}[h]
\begin{center}
\begin{tabular}{|c|c|c|}
\hline \multicolumn{3}{|c|}{\bf Non-chiral spectrum  of a 6-stack Pati-Salam model}\\
\hline \hline sector  & $U(4)\times U(2)_L \times U(2)_R \times U(4)_d \times U(2)_e \times U(2)_f$& $(Q_a, Q_b, Q_c, Q_d, Q_e)$ \\
\hline
$aa'$& $2 \times [({\bf {6}_\Anti}, \1,\1,\1,\1,\1) + h.c.]$ & ($\pm 2$,0,0,0,0,0) \\
$dd'$& $2 \times [(\1,\1,\1,{\bf {6}_\Anti}, \1,\1) + h.c.]$ & (0,0,0,$\pm 2$,0,0) \\
 $ee'$ & $2 \times [(\1,\1,\1, \1, {\bf 1_{\Anti}},\1) + h.c.]$ & (0,0,0,0,$\pm 2$,0)\\
 $ff'$ &  $2 \times [(\1,\1,\1, \1, \1,{\bf 1_{\Anti}}) + h.c.]$  & (0,0,0,0,0,$\pm 2$)\\
 $ad$ &   $ ({\bf 4}, \1,\1, {\bf \bar 2},\1,\1) + h.c.$ & $(\pm 1, 0, 0, \mp 1, 0 , 0)$ \\
  $ae'$ &  $ ({\bf 4}, \1,\1, \1, {\bf  2},\1) + h.c.$ &$(\pm 1, 0, 0, 0, \pm 1, 0)$\\
 $bc$ & $ (\1, {\bf 2}, {\bf \bar 2},\1,\1,\1) + h.c. $  &$(0, \pm 1, \mp 1, 0, 0, 0)$ \\
 $de$ &  $(\1, \1, \1, {\bf 2}, {\bf \bar 2},\1) + h.c. $  &$(0, 0, 0, \pm 1, \mp 1 , 0)$ \\
 $df'$ &  $(\1,\1,\1, {\bf 2},\1, {\bf 2}) + h. c. $&$(0, 0, 0, \pm 1, 0, \pm 1)$\\
 $ef$ & $ (\1,\1,\1,\1, {\bf 2}, {\bf \bar 2}) + h.c.  $&$(0, 0, 0, 0,\pm 1, \mp 1)$\\
 \hline
\end{tabular}
\caption{Non-chiral spectrum for the D-brane configuration in table \ref{Tab:OtherDecentPatiSalam}.\label{Tab:OtherDecentPatiSalamNC}}
\end{center}
\end{table}

\begin{table}[h]
\begin{center}
\begin{tabular}{|c||c|c|c|c|c|c|}
\hline \multicolumn{7}{|c|}{\bf Counting of states for a six-stack Pati-Salam model, part I} \\
\hline \hline $(\chi^{x y}, \chi^{x (\omega y)},\chi^{x (\omega^2 y)})$ & $y=a$ & $y=b$ & $y=c$ & $y=d$&$y=e$&$y=f$\\
\hline\hline $x=a$& (0,0,0) & (1,1,0) & (-1,0,0) &  (0,1,-1)&(0,0,0)&(0,0,0)\\
$x=b$& & (0,0,0) & ($|2|$,0,0) & (1,0,0) & (0,0,0) &(-1,0,0)\\
$x=c$& & & (0,0,0) & (-1,0,0) &(0,0,0)&(1,0,0)\\
$x=d$ &&&& (0,0,0)&(0,1,-1)&(0,0,0)\\
$x=e$ &&&&& (0,0,0)&(0,1,-1)\\
$x=f$&&&&&&(0,0,0) \\
\hline
\end{tabular}
\caption{Counting $\chi^{x (\omega^k y)}$ of chiral states per sector $x(\omega^k y)$ for the D6-brane configuration in table \ref{Tab:OtherDecentPatiSalam}. As the net-chirality $\chi^{bc}$ vanishes, we indicate the total amount of matter $|\varphi^{bc}|$.}
\end{center}
\end{table}

\begin{table}[h]
\begin{center}
\begin{tabular}{|c||c|c|c|c|c|c|}
\hline \multicolumn{7}{|c|}{\bf Counting of states for a six-stack Pati-Salam model, part II} \\
\hline \hline $(\chi^{x y'}, \chi^{x (\omega y)'},\chi^{x (\omega^2 y)'})$ & $y=a$ & $y=b$ & $y=c$ & $y=d$&$y=e$&$y=f$\\
\hline\hline $x=a$& ($|2|$,-1,1) & (1,0,0) & (-1,-1,0) &  (0,0,0)&(0,-1,1)&(0,0,0)\\
$x=b$& & (0,0,0) & (01,1) & (-1,0,0)&(0,0,0)&(1,0,0) \\
$x=c$& & & (0,0,0) & (1,0,0)&(0,0,0)&(-1,0,0)\\
$x=d$ &&&& ($|2|$,-1,1)&(0,0,0)&(0,-1,1)\\
$x=e$ &&&&&($|2|$,-1,1) &(0,0,0)\\
$x=f$ &&&&&&($|2|$,-1,1)\\
\hline
\end{tabular}
\caption{Counting of states per sector for the D6-brane configuration in table \ref{Tab:OtherDecentPatiSalam}. If  the net-chirality $\chi^{x(\omega^k y)'} =0$ but the total amount of matter $\varphi^{x(\omega^k y)'} \neq 0$, the net-chirality is replaced by the absolute value of the total amount of matter $|\varphi^{x(\omega^k y)'}|$. The net-chirality $\chi^{x(\omega^k x)'}$ counts the amount of symmetric + antisymmetric matter as introduced in equation~(\ref{Eq:ChiralMatter-Sym+Anti}).}
\end{center}
\end{table}

\begin{table}[h]
\begin{center}
\begin{tabular}{|c|c|c|c|}
\hline \multicolumn{4}{|c|}{\bf Yukawa couplings for a six-stack Pati-Salam Model} \\
\hline \hline \bf sequence & \bf coupling & \bf triangle & \bf enclosed\\
 $[x,y,z]$ & & & \bf area  \\
 \hline \hline $[a,c,b]$ & $Q^{(3)}_{ab}\, H_{bc}\, U_{ac}$ & $\{6, \|=[5,6], 1\}$ & 0 \\
 \hline $[a,b',c']$ & $Q_{ab'}\, H_{bc}\, U^{(3)}_{ac'}$& $\{5,\|=[5,6],1\}$ & 0\\
 \hline $[a,b,(\omega c)']$ & $Q^{(2)}_{ab}\, h^{(1)}_{bc'}\, U^{(3)}_{ac'} $ &$\{6,5,1\}$ &0\\
  $[a, (\omega b),c']$ & $Q^{(3)}_{ab}\, h^{(1)}_{bc'}\, U^{(2)}_{ac'} $ &$\{5,6,1\}$ &0\\
\hline
\end{tabular}
\caption{Overview of the surviving Yukawa couplings from equation (\ref{Eq:PS2Yukawa}). The three-point interactions are written down with a generalised notation introduced in equation (\ref{Eq:PS1NotationStates}). The symbol $\|=[5,6]$ represents three parallel one-cycles going through the $\Z_2$ fixed points $5$ and $6$. \label{Tab:PS2Yukawa}}
\end{center}
\end{table}

\end{appendix}

\clearpage

\addcontentsline{toc}{section}{References}
\bibliographystyle{ieeetr}
\bibliography{refs_Z2Z6p}

\end{document}